\documentclass[lettersize,journal]{IEEEtran}
\usepackage{amsmath,amsfonts}
\usepackage[ruled,vlined,linesnumbered,noresetcount]{algorithm2e} % http://www.ctan.org/pkg/algorithm2e
\usepackage{xcolor} % http://www.ctan.org/tex-archive/macros/latex/contrib/xcolor
\usepackage{tikz}
\usetikzlibrary{fit,calc}
\newcommand*{\tikzmk}[1]{\tikz[remember picture,overlay,] \node (#1) {};\ignorespaces}
\newcommand{\boxit}[1]{\tikz[remember picture,overlay]{\node[xshift = 3pt, yshift=1pt,fill=#1,opacity=.23,fit={(A)($(B)+(.9\linewidth,.7\baselineskip)$)}] {};}\ignorespaces}
\colorlet{pink}{pink!40}
\colorlet{cyan}{cyan!20}
\colorlet{orange}{orange!20}
\colorlet{yellow}{yellow!20}
\usepackage{multicol}
\usepackage{array}
\SetArgSty{textnormal}

\usepackage{cuted}
\usepackage{textcomp}
\usepackage{stfloats}
\usepackage{url}
\usepackage{verbatim}
\usepackage{graphicx}
\usepackage{cite}
\usepackage{xcolor}
\usepackage{adjustbox}
\usepackage{subfigure}
\usepackage{capt-of}
\usepackage{graphics}

\newtheorem{theorem}{Theorem}
\usepackage{blkarray}

\newtheorem{remark}{Remark}
\newtheorem{definition}{Definition}
\newcommand\given[1][]{\:#1\vert\:}
\newcommand{\floor}[1]{\lfloor #1 \rfloor}
\begin{document}

\title{Basil: A Fast and  Byzantine-Resilient Approach for Decentralized Training}
 \author{Ahmed Roushdy Elkordy, Saurav Prakash, Salman Avestimehr
         \thanks{The authors are with the Electrical Engineering Department, University of Southern California, Los Angeles, CA 90089, USA (e-mail: aelkordy@usc.edu sauravpr@usc.edu,  avestimehr@ee.usc.edu).}}
        % \thanks{Preprint, under review.}

%\author{IEEE Publication Technology,~\IEEEmembership{Staff,~IEEE,}
        % <-this % stops a space
%\thanks{This paper was produced by the IEEE Publication Technology Group. They are in Piscataway, NJ.}% <-this % stops a space
%\thanks{Manuscript received April 19, 2021; revised August 16, 2021.}}

\iffalse \markboth{Journal of \LaTeX\ Class Files,~Vol.~14, No.~8, August~2021}%
{Shell \MakeLowercase{\textit{et al.}}: A Sample Article Using IEEEtran.cls for IEEE Journals}\fi


\maketitle

\begin{abstract}
Decentralized (i.e., serverless) training across edge nodes can suffer substantially from potential Byzantine nodes that can degrade the training performance. However, detection and mitigation of Byzantine behaviors in a decentralized learning setting is a daunting task, especially when the data distribution at the users is heterogeneous. As our main contribution, we propose \texttt{Basil}, a fast and computationally efficient Byzantine-robust algorithm  for decentralized training systems, which leverages a novel sequential, memory-assisted and performance-based criteria for training over a logical ring while filtering the Byzantine users. In the IID dataset setting, we provide the theoretical convergence guarantees of \texttt{Basil}, demonstrating its linear convergence rate. Furthermore, for the IID setting, we experimentally demonstrate that \texttt{Basil} is robust to various  Byzantine attacks,  including the strong  Hidden attack, while providing up to absolute ${\sim}16 \%$ higher test accuracy over the state-of-the-art Byzantine-resilient decentralized learning approach. Additionally, we generalize \texttt{Basil} to the non-IID   setting by proposing Anonymous Cyclic Data Sharing (ACDS), a technique that allows each node to anonymously share a random fraction of its  local non-sensitive dataset (e.g., landmarks images)   with all other nodes. Finally, to reduce the overall latency of  \texttt{Basil} resulting from its sequential implementation over the logical ring, we propose \texttt{Basil+} that  enables  Byzantine-robust parallel training across groups of logical rings, and at the same time, it retains the performance gains of  \texttt{Basil} due to sequential training within each group. Furthermore, we experimentally demonstrate the scalability gains of \texttt{Basil+} through different sets of experiments. 
\end{abstract}

\section{Introduction}

 Thanks to the large amounts of  data generated on and held by the edge devices,   machine learning (ML) applications can achieve significant performance  \cite{devlin2018bert,dosovitskiy2020image}. However,  privacy concerns and regulations \cite{feld2020united} make it extremely difficult to pool clients' datasets for a centralized ML training procedure. As a result, distributed  machine learning methods are  gaining a surge of recent interests.  The  key underlying goal in distributed machine learning at the edge is to learn a  global model  using the data stored across the edge devices.

Federated learning (FL) has emerged as a promising framework \cite{kairouz2019advances} for distributed  machine learning. In federated learning, the training process is facilitated by a central server. In an FL architecture, the task of training is federated among the clients themselves. Specifically, each participating client trains a local model based on its own (private) training dataset and shares only the trained local model with the central entity, which appropriately aggregates the clients’ local models. \iffalse At a high level,  the server maintains a global model which  is iteratively shared with the mobile users who improve this shared model by using their own datasets before sending it back to the central server. \fi While the existence of the parameter server in FL is advantageous for   orchestrating the training process, it brings new security and efficiency drawbacks \cite{lian2017decentralized, kairouz2019advances}.  Particularly, the  parameter server in FL is a single point of failure, as the entire learning process could fail when the server crashes or gets hacked.  Additionally, the parameter server can become a performance bottleneck itself due to the large number of the mobile devices that need to be handled simultaneously.  

Training using a decentralized setup is another approach for distributed  machine learning without having to rely on a central coordinator (e.g., parameter server), thus avoiding the aforementioned limitations of FL. Instead, it only requires on-device computations on the edge nodes and peer-to-peer communications. In fact, many decentralized training algorithms have been proposed for   the decentralized training  setup.  In particular, a class of  gossip-based algorithms over random graphs has been proposed, e.g.,  \cite{koloskova2019decentralized, dobbe2017fully,5399485},  in which  all the nodes participate in each training round. During training, each node  maintains a local model and communicates with others over a graph-based decentralized network. More specifically, every node updates its local model using its local dataset, as well as the models received from the nodes in its neighborhood. For example, a simple aggregation rule at each node is to average the locally updated model with the models from the neighboring nodes. Thus, each node performs both model training and model aggregation.

% Decentralized  training setup  is another distributed training framework that  does not rely on a central  coordinator (e.g. parameter server) for solving the optimization problem in \eqref{ma}. Instead, it only requires on-device computations on the edge nodes and peer-to-peer communications. Many decentralized training algorithms have been proposed for solving the collaborative learning problem in \eqref{ma}.  In particular, a class of  gossip-based algorithms over random graphs has been proposed, e.g.,  \cite{koloskova2019decentralized, dobbe2017fully,5399485},  in which  all the nodes participate in each training round. During training, each node  maintains a local model and communicates with others over a graph-based decentralized network. More specifically, every node updates its local model using its local dataset, as well as the models received from the nodes in its neighborhood. For example, a simple aggregation rule at each node is to average the locally updated model with the models from the neighboring nodes. Thus, each node performs both model training and model aggregation. 

Although decentralized training provides many benefits, its decentralized nature makes it vulnerable to performance degradation due to system failures, malicious nodes, and data heterogeneity \cite{kairouz2019advances}. Specifically, one of the key challenges in decentralized training  is the presence of  different threats that can alter the learning process, such as the software/hardware errors and adversarial attacks. Particularly, some clients can become faulty due to software bugs, hardware components which may behave arbitrarily, or even get hacked during training, sending arbitrary or malicious values to  other  clients, thus severely degrading the overall convergence performance. Such faults, where client nodes arbitrarily deviate from the agreed-upon protocol, are called Byzantine faults \cite{Byz}. To mitigate Byzantine nodes in a graph-based  decentralized setup where nodes are randomly connected  to each other,  some Byzantine-robust optimization algorithms have been introduced recently, e.g., \cite{guo2020byzantineresilient,yang2019bridge}. In these algorithms, each node combines the  set of models received from its  neighbors   by using robust aggregation rules, to ensure that the training is not impacted by the Byzantine nodes. However, to the best of our knowledge, none of these algorithms have considered the scenario when the data distribution at the nodes is heterogeneous.      Data  heterogeneity makes the detection of   Byzantine nodes a daunting task, since it becomes  unclear whether the  model drift can be attributed
to a Byzantine node, or to the very heterogeneous nature of the data.  Even in the absence of Byzantine nodes, data heterogeneity can degrade the convergence rate \cite{kairouz2019advances}.

The limited  computation and communication resources of edge devices (e.g., IoT devices)   are another important consideration in the decentralized training setting. In fact, the resources at the edge devices are considered as a critical bottleneck for performing distributed training for large models \cite{kairouz2019advances, devlin2018bert}.  In prior Byzantine-robust decentralized algorithms (e.g., \cite{guo2020byzantineresilient,yang2019bridge}), which are based on parallel training over a random graph, all nodes need to be always active and perform training during the entire training process. Therefore, they might not be suitable for the resource constrained edge devices, as the parallel training nature of their algorithms  requires  them to be perpetually active which could  drain their resources. In contrast to parallel training over a random graph, our work takes the view that sequential training  over a logical ring is better suited for decentralized training in resource constrained edge setting. Specifically, sequential training over a logical ring allows each node to become active and perform model training \textit{only} when it receives the  model from its counterclockwise neighbor.  Since nodes need not be active during the whole training time, the sequential training nature makes it  suitable for IoT devices with limited computational resources. 
\subsection{Contributions}
To overcome the aforementioned limitations of prior graph-based Byzantine-robust algorithms, we propose \texttt{Basil}, an efficient Byzantine mitigation algorithm, that achieves Byzantine robustness in a decentralized setting by leveraging the  sequential training over a logical ring. To highlight  some of the benefits of \texttt{Basil}, Fig. 1(a) illustrates a sample result that demonstrates the performance gains and the cost reduction compared to the state-of-the-art Byzantine-robust algorithm  UBAR that leverages the parallel training over   a graph-based setting. We observe that \texttt{Basil} retains a higher accuracy than UBAR with an absolute value of $ \sim 16\%$. Additionally, we note that while UBAR achieves its highest accuracy in $\sim 500$ rounds, \texttt{Basil} achieves   UBAR's highest accuracy in just $\sim 100$ rounds. This implies that for achieving UBAR's highest accuracy, each client  in \texttt{Basil} uses $ 5\times$ lesser computation and communication resources compared to that in UBAR confirming the gain attained from the sequential training nature of \texttt{Basil}.

In the following, we further highlight the key aspects and performance gains of \texttt{Basil}:
\begin{itemize}
\item  In \texttt{Basil}, the defense technique to filter out Byzantine nodes is a performance-based strategy, wherein each node evaluates a received  set of models from its counterclockwise neighbors  by using its own local dataset to select the best candidate.
\item We  theoretically show  that    \texttt{Basil}  for convex loss  functions in the IID data setting has a linear convergence rate with respect to the product of the number of  benign nodes and the total number of training rounds over the ring.  Thus, our theoretical result demonstrates scalable performance for  \texttt{Basil} with respect to the number of nodes.

%\item We provide the theoretical guarantees of \texttt{Basil}  showing  that    \texttt{Basil}  for convex loss  functions in the IID data setting has a linear convergence rate with respect to the product of the number of  benign nodes and the total number of training rounds over the ring.  Thus, our theoretical result demonstrates scalable performance for  \texttt{Basil} with respect to the number of nodes.

\iffalse\item { For our  theoretical analysis, we  would like to highlight that the aggregation rule of  \texttt{Basil}  is based on performance-based and memory-assisted criteria, which makes the convergence analysis a challenging task. We overcome this challenge and   show that  \texttt{Basil}  for convex loss functions has a linear convergence rate with respect to the product of the number of  benign nodes and the total training rounds over the ring.    } \fi   
 \item  We  empirically demonstrate the superior performance of \texttt{Basil}  compared to UBAR, the state-of-the-art Byzantine-resilient  decentralized learning algorithm over graph, under different Byzantine attacks in the IID setting.    Additionally, we study the performance of \texttt{Basil}  and UBAR with  respect to the   wall-clock time in Appendix H  showing that the training time  for \texttt{Basil} is comparable to UBAR.
 \iffalse In these results, we have shown that the time it takes for UBAR to reach its maximum accuracy is almost the same as the time for \texttt{Basil}  to reach UBAR's maximum achievable accuracy. 
 \fi
 %fast convergence rate while providing up to ${\sim}16 \%$ the high performance of \texttt{Basil}  under different Byzantine attacks compared to UBAR in the IID data distribution setting.   
% \item By using \texttt{Basil}  algorithm,  the issue of imbalanced  data size  at the nodes which has been highlighted  for centralized system in  \cite{portnoy2020realistic}   is no longer there  as each node update its model based  only on one selected update.  
% \item  In our theoretical analysis, we demonstrate provable convergence guarantees for  \texttt{Basil} in the IID setting in the presence of an arbitrary number of Byzantine nodes.
 
 \iffalse
 {\color{red} Under standard convexity assumptions and IID setting, we perform the convergence analysis of  \texttt{Basil}. linear rate of convergence, and remarks in the theoretical results... no need to go into details of theorem, just present the remarks... We present our results in two parts, in Theorems... In Theorem 2, ...Towards this goal, we develop new To achieve this, we first show that performance of monotonicity, performance of models,  sequential, refer to this theorem... then we leverage our theorem to finally bound the performance from optimal...}\fi
\item  For extending the  superior benefits  of \texttt{Basil} to the scenario when data distribution is non-IID across devices, we propose Anonymous Cyclic Data Sharing (ACDS) to be applied on top of \texttt{Basil}. To the best of our knowledge, no prior decentralized Byzantine-robust algorithm has considered the scenario when the data distribution at the nodes is non-IID.  ACDS allows each node to share a random fraction of its local non-sensitive dataset (e.g., landmarks images captured during tours) with all other nodes, while guaranteeing anonymity of the node identity. As highlighted in Section \ref{Related}, there are multiple real-world use cases where anonymous data sharing is sufficient to meet the privacy concerns of the users. 
\item We experimentally  demonstrate that using ACDS  with only $5\%$ data sharing on top of \texttt{Basil}  provides  resiliency to Byzantine behaviors,  unlike UBAR which diverges in the non-IID  setting (Fig. 1(c)).

%\item We experimentally demonstrate that even when each node shares only $5\%$ of its local data  with all other nodes, the test accuracy for the naive sequential training over ring in the non-IID setting can be increased by up to ${\sim}10 \%$ in the absence of Byzantine nodes (Fig. 1(b)). Furthermore, we demonstrate that using ACDS on top of \texttt{Basil}  provides  resiliency to Byzantine behaviors,  unlike UBAR which diverges in the non-IID  setting (Fig. 1(c)). 

\iffalse
\item {\color{blue}Finally, leveraging the performance gains achieved by  \texttt{Basil} through the sequential training over a logical ring,  we provide  \texttt{Basil+} a parallel implementation  of \texttt{Basil}. In particular,  \texttt{Basil+} allows for robust  parallel training across multiple rings, along with sequential training over each ring. This results in decreasing the training time needed for completing one global epoch (visiting all nodes) compared to \texttt{Basil}, which only considers sequential training over one ring. We experimentally demonstrate that \texttt{Basil+} is a scalable Byzantine-robust  algorithm.  }  
\fi
\item As the number of participating nodes in  \texttt{Basil} scales, the increase in the overall latency of sequential training over the logical ring topology may limit the practicality of implementing \texttt{Basil}. Therefore, we propose a parallel extension of \texttt{Basil}, named \texttt{Basil+}, that provides further scalability by enabling Byzantine-robust parallel training across groups of logical rings, while retaining the performance gains of  \texttt{Basil} through sequential training within each group.
\iffalse Furthermore, to demonstrate the scalability gains of \texttt{Basil+}, we carry out different sets of experiments  in Section \ref{sec:V-B}.  
\fi

\end{itemize}
\begin{figure*}[t]
  \centering
   \subfigure[Gaussian Attack  (IID  setting)]{\includegraphics[scale=0.33]{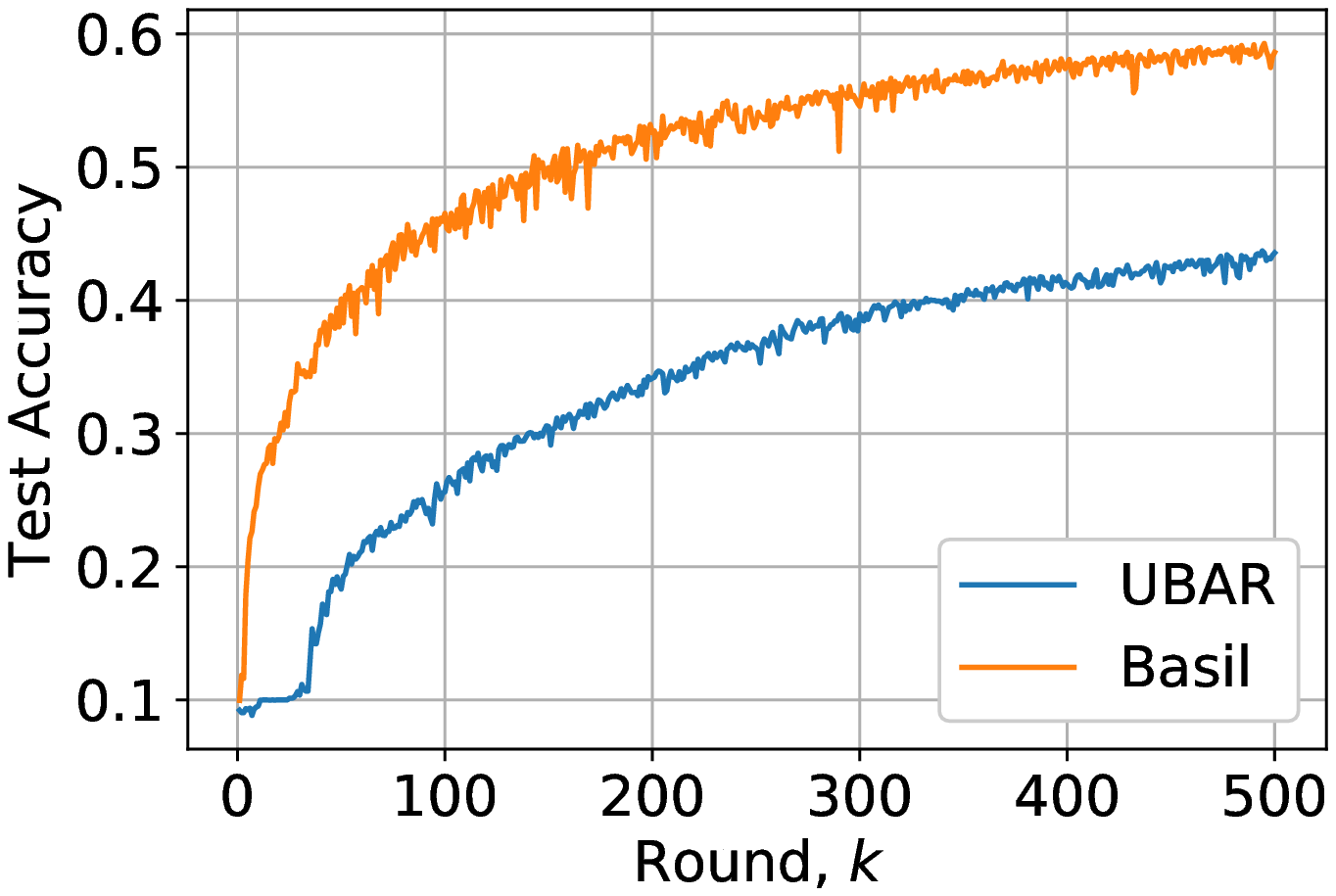}
  }%
  \quad
    \subfigure[No Attack (non-IID setting)]{\includegraphics[scale=0.33]{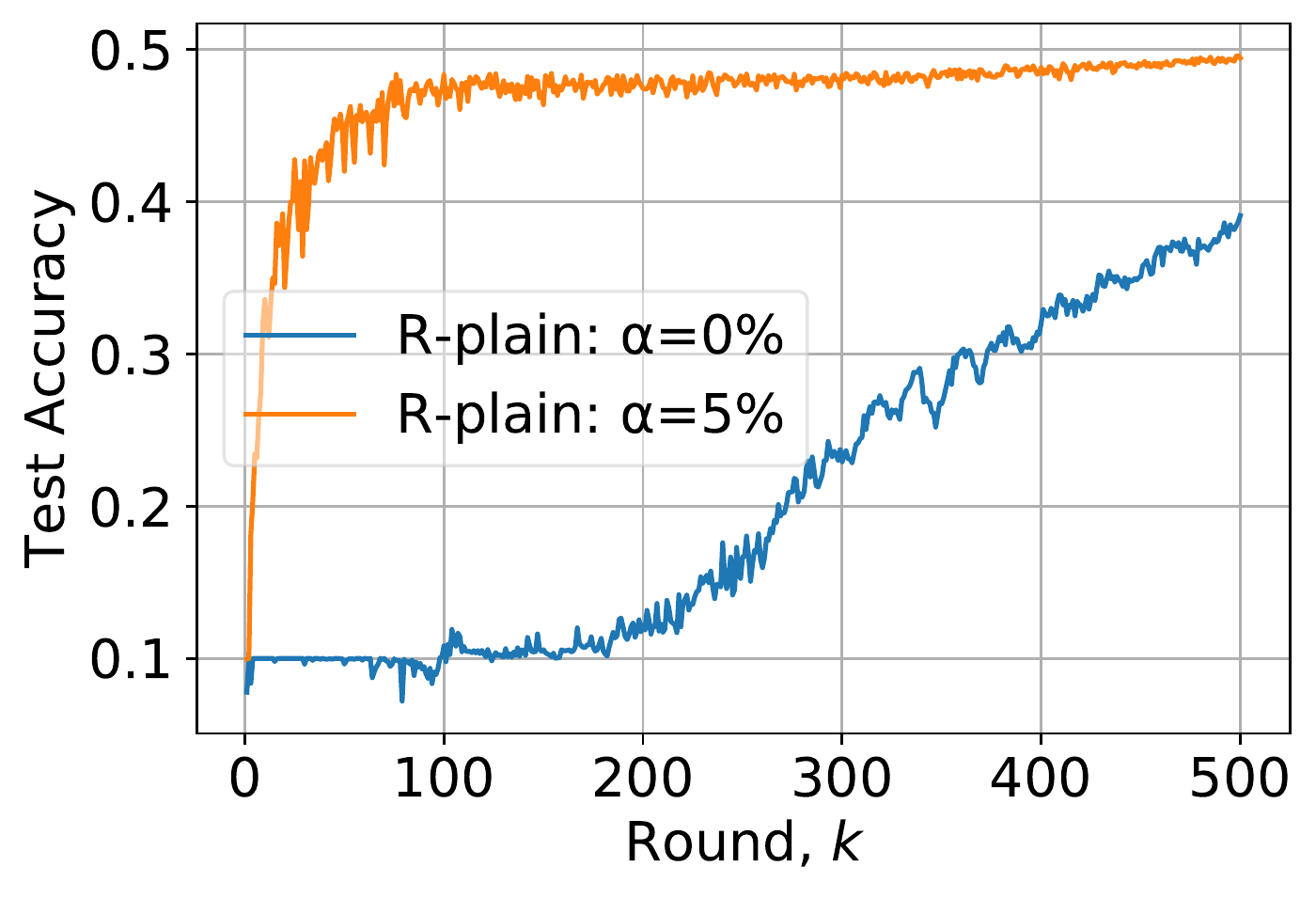}
  }%
  \quad 
  \subfigure[Gaussian Attack ]{\includegraphics[scale=0.33]{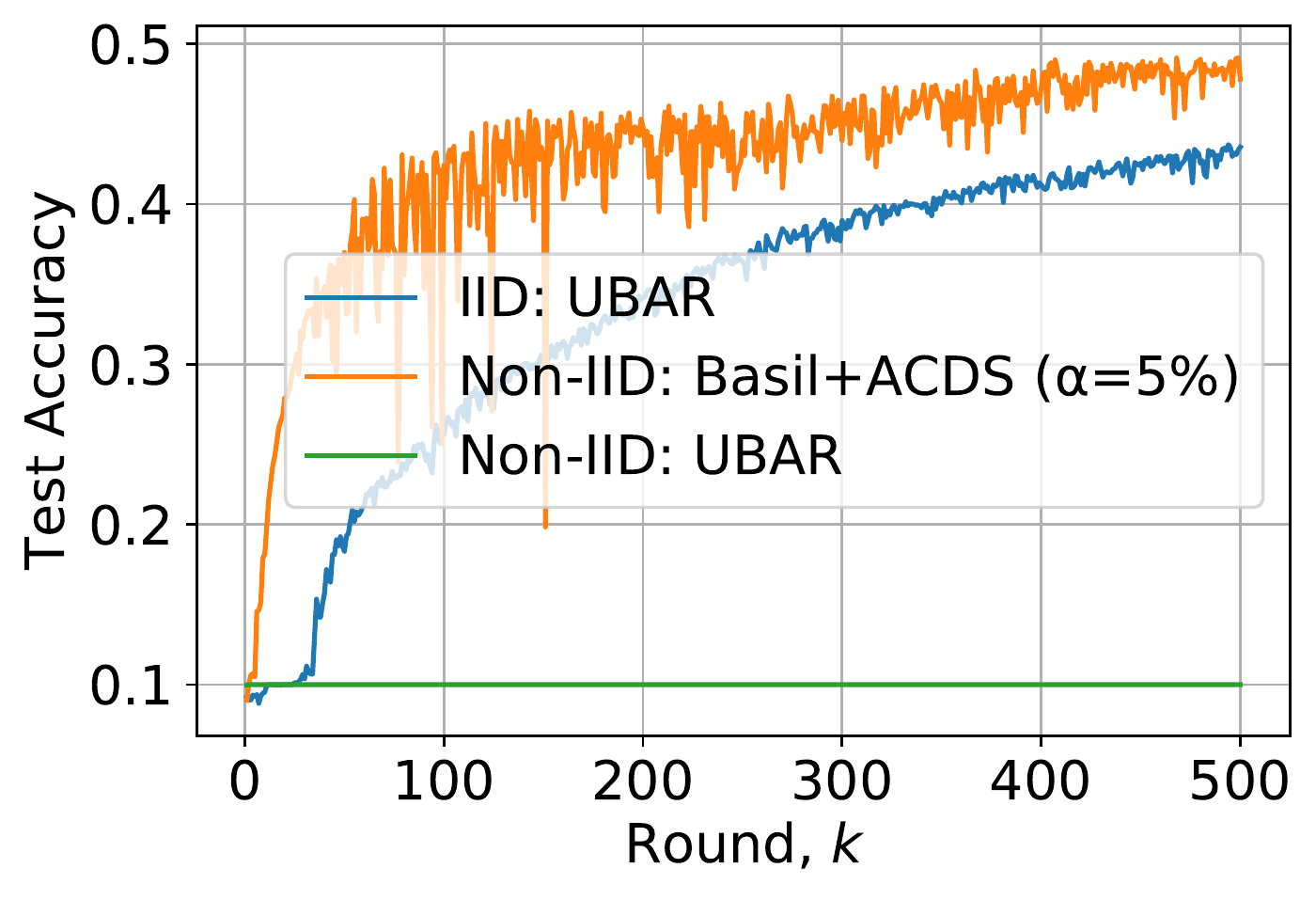}
 }
  \caption{A highlight of the performance benefits of    \texttt{Basil}, compared with state-of-the-art (UBAR) \cite{guo2020byzantineresilient}, for CIFAR10 under different settings: In  Fig.  1(a), we can see the superior performance of \texttt{Basil} over UBAR  with  ${\sim}16 \%$ improvement of the test accuracy under Gaussian attack in    the IID setting. Fig. 1(b) demonstrates that the test accuracy in the non-IID setting by using sequential training over the ring topology can be increased by up to  ${\sim}10 \%$ in the absence of Byzantine nodes, when each node shares only $ 5\%$ of its local data anonymously with other nodes. Fig. 1(c) shows that ACDS on the top of  \texttt{Basil} not only  provides Byzantine robustness to Gaussian attack in the non-IID setting, but also gives higher performance than UBAR in the IID setting. Furthermore, UBAR for the  non-IID setting  completely fails in the presence of this attack. For further details, please refer to Section VI.}
  \label{fig:results_mnist}
 \end{figure*}
\subsection{Related Works}\label{Related}

Many  Byzantine-robust strategies have been proposed recently for  the distributed  training setup (federated learning)  where there is a central server  to orchestrate the training process 
\cite{regatti2020bygars,xie2019zeno,xie2019zenot,krum,By,9712310,so2020byzantine,pillutla2019robust,zhao2020shielding, prakash2020mitigating, Zhang2021FederatedLF}. These Byzantine-robust optimization algorithms combine the gradients received by all workers using robust aggregation rules, to ensure that training is not impacted by malicious nodes. Some of these strategies \cite{krum,By, pillutla2019robust,9712310, so2020byzantine} are based on distance-based approaches, while some others are based on performance-based criteria \cite{regatti2020bygars,xie2019zeno,xie2019zenot,prakash2020mitigating}. The key idea in distance-based defense solutions is to filter the updates that are far from the average of the updates from the benign nodes. It has been shown that distance-based solutions are vulnerable to the sophisticated Hidden  attack proposed in \cite{ pmlr-v80-mhamdi18a}.  In this attack, Byzantine nodes could create gradients that are malicious but indistinguishable from benign gradients in distance.  On the other hand, performance-based filtering strategies   rely on having  some auxiliary dataset   at  the server to  evaluate the model received from each node.
 
Compared to the large number of Byzantine-robust training algorithms for distributed training in the presence of a central server, there have been only a few recent works on Byzantine resiliency in the decentralized training setup with no central coordinator. In particular, to address Byzantine failures in a decentralized training setup over a random graph under the scenario when   the data distribution at the nodes  is IID, the authors in \cite{yang2019bridge} propose using a trimmed mean distance-based approach called BRIDGE to mitigate Byzantine nodes.  However, the authors in \cite{guo2020byzantineresilient} demonstrate that BRIDGE is defeated by   the hidden attack proposed   in \cite{ pmlr-v80-mhamdi18a}. To solve the limitations of the distance-based approaches in the decentralized setup, \cite{ guo2020byzantineresilient} proposes an algorithm called UBAR in which a combination of performance-based and distance-based stages are used to mitigate the Byzantine nodes, where the performance-based stage at a particular node  leverages only its local dataset. As demonstrated numerically in \cite{guo2020byzantineresilient}, the combination of these two  strategies allows UBAR to defeat the Byzantine attack proposed in \cite{pmlr-v80-mhamdi18a}.  However, UBAR is not suitable for the training over resource-constrained edge devices, as the training is carried out in parallel and nodes remain active all the time. In contrast, \texttt{Basil} is a fast and computationally efficient Byzantine-robust algorithm, which leverages a novel sequential, memory-assisted and performance-based criteria for training over a logical ring while filtering the Byzantine users.

Data heterogeneity in the decentralized setting has been studied in some recent works (e.g., \cite{lee2020tornadoaggregate}) in the absence of Byzantine nodes. In particular, the authors of TornadoAggregate \cite{lee2020tornadoaggregate} propose to cluster users into groups based on an algorithm called Group-BY-IID and CLUSTER where both use EMD (earth mover distance) that can approximately model the learning divergences between the models to complete the grouping. However, EMD function relies on having a publicly shared data at each node which can be collected similarly as in \cite{zhao2018federated}. In particular, to improve training on non-IID data in federated learning, \cite{zhao2018federated} proposed sharing of small portions of users' data with the server. The parameter server pools the received subsets of data thus creating a small subset of the data distributed at the clients, which is then globally shared between all the nodes to make the data distribution close to IID. However, the aforementioned data sharing approach is considered insecure in scenarios where users are fine with sharing some of their datasets with each other but want to keep their identities anonymous, i.e., data shares should not reveal who the data owners are. 

There are multiple real-world use cases where anonymous data sharing is sufficient for privacy concerns. For example, mobile users maybe fine with sharing some of their own text data, which does not contain any personal and sensitive information with others, as long as their personal identities remain anonymous. Another example is sharing of some non-private data (such as landmarks images) collected by a person with others. In this scenario, although data itself is not generated at the users, revealing the identity of the users can potentially leak private information such as personal interests, location, or travel history.  Our proposed ACDS strategy is suitable for such scenarios as it guarantees that  the owner  identity of  the  shared data points are kept hidden.

 %Data heterogeneity in the centralized setup has been studied in some recent works \cite{zhao2018federated,9155494, yoshida2020hybridfl,sattler2019robust}. In particular, to improve training on non-IID data, \cite{zhao2018federated} propose sharing of small portions of node  data was introduced for the federated learning setting. The parameter server pools the received subsets of data thus creating a small subset of the data distributed at the clients, which is then globally shared between all the nodes to make the data distribution close to the IID.  Other works consider client sampling, e.g. \cite{9155494}, by selecting client devices to participate in each round of federated learning who can counterbalance the bias introduced by non-IID data.  
 
As a final remark, we point out that for anonymous data sharing, \cite {6389771} proposed an approach which is based on utilizing a secure sum operation along with anonymous ID assignment (AIDA). This involves computational operations at the nodes such as polynomial evaluations and some arithmetic operations such as modular operations. Thus, this algorithm may fail in the presence of Byzantine faults arising during these computations. Particularly, computation errors or software bugs can be present during the AIDA algorithm thus leading to the failure of anonymous ID assignment, or during the secure sum algorithm which can lead to distortion of the shared data.

\section{Problem Statement}
We formally define the decentralized learning system in the presence of Byzantine faults.
\subsection{Decentralized System Model}
We consider a decentralized learning setting in which a set $\mathcal {N} = \{ 1, \dots, N\}$  of $|\mathcal {N}|=N$ nodes collaboratively train a machine learning (ML)  model  $\mathbf{x} \in \mathbb{R}^{d}$, where $d$ is the model size,    based on all the training data samples ${\cup_{n \in \mathcal {N}}}\mathcal{Z}_n$ that are generated and stored at these distributed nodes, where the size of each local dataset is  $|\mathcal{Z}_i| = D_i$ data points.
%The local  dataset   $\mathcal{Z}_n$  at node $n$  consist  of independent and identically distributed (i.i.d.) data samples from a distribution $\mathcal{P}$ while the size of this dataset is   $\mathcal{Z}_n$ samples.  

In this work,   we are motivated by the edge IoT setting, where users want to collaboratively train an ML model, in the absence of a centralized parameter server. The communication in this setting leverages the underlay communication fabric of the internet that connects any pair of nodes  directly via overlay communication protocols. Specifically,  we assume that there is no central parameter server, and consider the setup  where    the training process is carried out in a sequential fashion over a clockwise directed ring. Each node in this ring topology   takes part  in  the  training  process  when  it receives  the  model    from  the  previous counterclockwise  node. In  Section \ref{3.1}, we propose a method in which nodes can consensually agree on a random ordering on a logical ring at the beginning of the training process, so that each node knows the logical ring positions of the other nodes.   Therefore, without loss of generality, we assume for notation simplification that the indices of  nodes in the ring are  arranged in ascending order starting from node $1$.   In this setup,  each node can send its model update to any set of users in the network. 

In this decentralized setting, an unknown $\beta$-proportion of nodes can be Byzantine, where $\beta \in (0,1)$, meaning they can send arbitrary and possibly malicious results to the other nodes. We denote $\mathcal{R}$ (with cardinality $|\mathcal{R}| = r$) and $\mathcal{B}$ (with cardinality $|\mathcal{B}| = b$) as the sets of benign nodes and Byzantine nodes, respectively. Furthermore, Byzantine nodes are uniformly distributed over the ring due to consensus-based random order agreement. Finally, we    assume nodes can authenticate the source of
a message, so no Byzantine node can forge its identity or create multiple fake ones \cite{elmhamdi2020genuinely}.

\subsection{Model Training}
Each node in the set $\mathcal{R}$  of benign   nodes uses its  own dataset to collaboratively train a shared model by solving the following   optimization problem  in the presence of Byzantine nodes: 
\begin{equation}
\label{ma}
\mathbf{x}^* = \arg \min_{\mathbf{x}\in\mathbb{R}^{d}
} \left[ f(\mathbf{x}) := \frac{1}{r} \sum_{i=1}^{r} f_i(\mathbf{x})\right],
\end{equation}
where  $\mathbf{x}$ is the optimization variable,  and $f_i(\mathbf{x})$ is the expected loss function of node $i$ such that $f_i(\mathbf{x}) =  \mathbb{E}_{\zeta_i \sim	 \mathcal{P}_i}[ l_i(\mathbf{x}, \zeta_i) ]$. Here, $l_i(\mathbf{x}, \zeta_i)  \in \mathbb{R}$ denotes the  loss function for  model parameter $\mathbf{x}\in \mathbb{R}^d$ for  a given realization  $\zeta_i$, which is generated from a distribution $\mathcal{P}_i$.

The general update rule in this decentralized setting   is given as follows. At the $k$-th round, the current active node $i$ updates the global model according to:
\begin{align}
\label{Update_main}
\mathbf{x}^{(i)}_{k}=&  \bar {\mathbf{x}}^{(i)}_{k}-\eta_{k}^{(i)}   g_i(\bar {\mathbf{x}}^{(i)}_{k}),
\end{align}
 where $\bar {\mathbf{x}}^{(i)}_{k}{=}\mathcal{A}(\mathbf{x}^{(j)}_{v}, j  \in  \mathcal {N}, v=1, \dots, k)$ is the selected model by node $i$ according to the underlying  aggregation rule $\mathcal {A}$,   $g(\bar {\mathbf{x}}^{(i)}_{k}) $ is  the stochastic gradient computed by node $i$ by using  a random sample  from its local dataset $\mathcal{Z}_i$,  and   $\eta_{k}^{(i)}$ is the learning rate in  round $k$ used by node $i$. 
 
\noindent\textbf{Threat model:} Byzantine node $ i \in \mathcal{B}$ could send faulty or malicious update 
$ \mathbf{x}^{(i)}_{k} = *$, where ``$ *$'' denotes that  $\mathbf{x}^{(i)}_{k} $
can be an arbitrary vector in $\mathbb {R}^d$. Furthermore,  Byzantine nodes cannot forge their identities or create multiple fake ones. This assumption has been used in different prior works (e.g.,  \cite{elmhamdi2020genuinely}). 

Our goal is to design an algorithm for the  decentralized training  setup discussed  earlier, while  mitigating the impact of the Byzantine nodes. Towards achieving this goal, we propose \texttt{Basil} that is described next. 

\section{The Proposed Basil  Algorithm}
Now,  we    describe \texttt{Basil}, our proposed approach  for mitigating both   malicious updates and faulty updates in the IID setting,  where  the local dataset $\mathcal{Z}_i$ at node $i$ consists  of IID data samples from a distribution $\mathcal{P}_i$, where $\mathcal{P}_i =  \mathcal{P} $ for $ i \in \mathcal{N}$, and characterize the complexity of \texttt{Basil}. Note that, in Section \ref{sec-ACDS},  we extend \texttt{Basil}  to the non-IID setting by integrating it to our proposed Anonymous Cyclic Data Sharing scheme. 
\subsection{Basil  for IID Setting} \label{3.1} 
Our proposed \texttt{Basil}
 algorithm  that is given in Algorithm 1 consists of two phases; initialization phase and training phase which are described below. \\
 \underline{\textbf{Phase 1: Order Agreement over the Logical Ring}.}
 Before the training starts,  nodes    consensually agree on their order on the ring by using the    following simple steps. 1) All users first share their IDs with each other, and we assume WLOG that nodes’ IDs can be arranged in ascending order. 2)  Each node     locally generates the order permutation for the users' IDs by using   a pseudo random number generator (PRNG) initialized via a common seed (e.g., $N$). This ensures that all nodes  will generate the same IDs' order for the  ring.
 
 \noindent\underline{\textbf{Phase 2: Robust Training}.}
 As illustrated in Fig.  \ref{Basil}, \texttt{Basil}  leverages sequential training over the logical ring to mitigate the effect of Byzantine nodes. At a high level, in the $k$-th round, the current active node carries out the model update step in \eqref{Update_main}, and then  multicasts its updated model  to the next $S=b+1$ clockwise nodes, where $b$ is the worst case number of Byzantine nodes. We note that multicasting each model to the next $b+1$ neighbors is crucial to make sure that the benign subgraph, which is generated by excluding the Byzantine nodes, is connected. Connectivity of the benign subgraph is important as it ensures that each benign node can still receive information from a few other non-faulty nodes, i.e., the good updates can successfully propagate between the benign nodes. Even in the scenario where all Byzantine nodes come in a row,  multicasting each updated  model to the next $S$ clockwise neighbors  allows the connectivity of benign nodes. 

\begin{figure}
\centering
\includegraphics[width=0.38\textwidth]{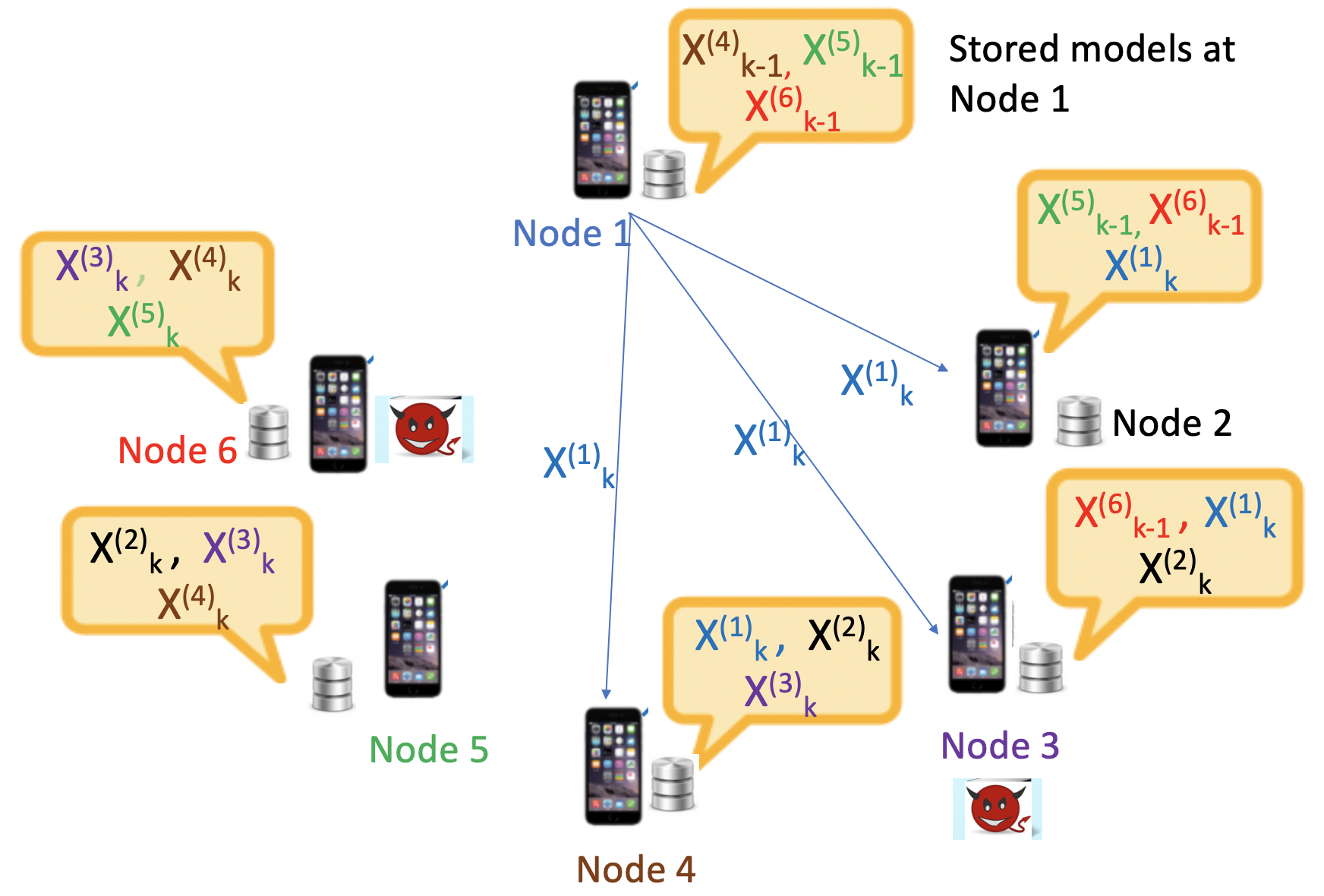}
\caption{ \texttt{Basil}    with  $N=6 $ nodes, where  node $3$ and node $6$ are Byzantine nodes. Node $1$, the current active  benign node in the $k$-th round,   selects  one model out of its stored $3$  models which   gives the lowest  loss when it is evaluated on a mini-batch from its  local dataset $\mathcal{Z}_1$. After that, node $1$ updates the selected model by using the same mini-batch   according to \eqref{Update_main} before multicasting it to the next $3$ clockwise neighbors. }
\label{Basil}
\end{figure}

\iffalse\begin{figure}[]
\vskip 0.2in
\begin{center}
\centerline{\includegraphics[width=.4\columnwidth]{model}}
\caption{ \texttt{Basil}   algorithm with  $N=6 $ nodes where  node $3$ and node $6$ are Byzantine nodes. node $1$, the current active  benign node in the $k$-th round,   selects  one model out of its stored $3$  models which   gives the lowest  loss when it is evaluated on a mini-batch from its  local dataset $\mathcal{Z}_1$. After that, node $1$ updates the selected model by using the same mini-batch   according to \eqref{Update_main} before multicasting it to the next $3$ clockwise neighbors. }
\label{Basil}
\end{center}
\vskip -0.3in
\end{figure}
\fi

We now describe how the aggregation rule $\mathcal{A}_\text{\texttt{Basil} }$ in \texttt{Basil},  that  node $i$ implements for obtaining the model $\bar {\mathbf{x}}^{(i)}_k$ for carrying out the update in \eqref{Update_main}, works. Node $i$ stores the $S$ latest models from its previous $S$ counterclockwise neighbors. %At the beginning of the training, all nodes   store the initial model $\mathbf{x}^0$ as shown in step 4 in Algorithm 1. 
As highlighted above, the reason for storing $S$ models is to make sure that each stored set of models at node $i$ contains at least one good model. When node $i$ is active, it implements our proposed performance-based criteria to pick the best model out of its $S$ stored models. In the following, we formally define our model selection criteria: 
\begin{definition} \label{defi} (\texttt{Basil}  Aggregation Rule) \textit{
 In the $k$-th round over the ring, let  $ \mathcal {N}^{i}_k = \{ \mathbf{y}_1, \dots, \mathbf{y}_{S}\}$   be the  set of    $S$  latest models from the $S$  counterclockwise neighbors of node $i$. We define $\zeta_i$ to be a  random sample from the local dataset $\mathcal{Z}_i$,  and  let $ {l}_i(\mathbf{y}_j) =  {l}_i
({\mathbf{y}_j, \zeta_i
})$ to be the loss function   of node $i$ evaluated on  the model $\mathbf{y}_j\in \mathcal {N}^{i}_k $,  by using  a random sample $\zeta_i$.  The proposed \texttt{Basil}  aggregation rule is defined as
\begin{equation}\label{eq3}
\bar {\mathbf{x}}_k^{(i)} = \mathcal {A}_\text{\texttt{Basil} } ( \mathcal {N}^{i}_k) =  \arg \min_{ \mathbf{y} \in \mathcal {N}^{i}_k }  \mathbb{E} \left[{l}_i({\mathbf{y}}, \zeta_i)\right].
\end{equation} }
\label{def:basil_rule}
\end{definition}
In practice, node $i$ can sample a mini-batch from its dataset and leverage it as validation data to test the performance (i.e., loss function value) of each model of  the neighboring $S$ models, and set $\bar {\mathbf{x}}^{(i)}_k$ to be the model with the lowest loss among the $S$ stored models. As demonstrated in our experiments in Section \ref{expe1}, this practical mini-batch implementation of the \texttt{Basil} criteria in Definition \ref{def:basil_rule} is sufficient to mitigate Byzantine nodes in the network, while achieving superior performance over state-of-the-art.

\RestyleAlgo{ruled}
\SetAlgoNoLine
\begin{algorithm}
\caption{\texttt{Basil}
}\label{alg:fairagg}
{\bf Input:}  $\mathcal{N}$(nodes); $S$(connectivity) ;$\{\mathcal{Z}_i\}_{i \in \mathcal{N}}$(local datasets);  $\mathbf{x}^0$(initial model); $K$(number of rounds) \\

\tikzmk{A}
\textbf{\underline{Initialization}}: \\
 \For {each node  $i \in \mathcal{N}$  }
 { StoredModels [$i$]. insert($\mathbf{x}^0$) \emph{\color{blue}{ //
  queue ``StoredModels [$i$]'' is used by node  $i$ to keep the last inserted  $S$ models, denoted by $\mathcal{N}^i$. It is intialized by inserting $\mathbf{x}^0$  }}} 
 Order $\gets$ RandomOrderAgrement($\mathcal{N}$) \emph{\color{blue}{// users' order generation for the logical ring topology according  to Section III-A.  \\}}
 \tikzmk{B}
\boxit{pink}

% \iffalse
\tikzmk{A}
\textbf{\underline{Robust Training}}:\\
 \For {\text{each round} $k  = 1, \dots, K$  }{
%  \emph{\color{blue}{// Each iteration represents a round over the ring  }}\\
 \For { each node  $ i = 1, \dots, N $ in sequence  }{
 \If{ node  $i \in \text{ benign set } \mathcal{R}$ } {
$\bar {\mathbf{x}}_k^{(i)} \gets   \mathcal {A}_\text{\texttt{Basil} } (\mathcal{N}^i)$  \emph{\color{blue}{//   \texttt{Basil}  performance-based strategy  to select one model from $\mathcal{N}^i$ using \eqref{eq3}}} \\ 
 
 $\mathbf{x}^{(i)}_{k} \gets$ Update$(\bar {\mathbf{x}}_k^{(i)}, \mathcal{Z}_i)$ \emph{\color{blue}{// model update using \eqref{Update_main}}}}
 \Else{$\mathbf{x}^{(i)}_{k} \gets *$ \emph{\color{blue}{// Byzantine node sends faulty model}}}    

 multicast $\mathbf{x}^{(i)}_{k}$ to the next $S$ clockwise neighbors \\
 \For {neighbor   $s  = 1, \dots, S$  } {
 StoredModels [$(i+s) \mod N$]. insert($\mathbf{x}^{(i)}_{k} $)
 {\emph{\color{blue}{//     insert $\mathbf{x}^{(i)}_{k}$ to the queue of the $s$-th neighbor   of node  $i$  }}}
 }
 }
 }
 \tikzmk{B}
\boxit{cyan}
% \fi
{\bf Return $\{\mathbf{x}^{(i)}_{K}\}_{i \in {\mathcal{N}}}$ }  
\end{algorithm}

In the following, we characterize the communication, computation, and storage costs of \texttt{Basil}. 

\noindent \textbf{Proposition 1.}  \textit{ The communication,  computation, and storage  complexities  of \texttt{Basil}  algorithm  are all $\mathcal{O} (Sd)$  for each node at each iteration, where $d$ is the model size. }

The   complexity of  prior graph-based    Byzantine-robust decentralized algorithms UBAR  \cite{guo2020byzantineresilient} and Bridge \cite{yang2019bridge}  is $\mathcal{O} ({H}_id)$, where  ${H}_i$ is the number of neighbours  (e.g., connectivity parameter) of node $i$  on  the graph.   So we conclude that Basil  maintains the same per round complexity as  Bridge and   UBAR, but with higher performance as we will show in  Section VI.

The costs in Proposition 1 can be reduced by relaxing  the connectivity parameter $S$  to   $S <b+1$  while guaranteeing the success of \texttt{Basil} (benign subgraph connectivity)  with high probability, as formally presented in the following.

 \noindent \textbf{Proposition 2.}\textit{ The number of models that each benign node needs to  receive, store and evaluate  from its counterclockwise neighbors for ensuring the  connectivity  and  success of \texttt{Basil} can be relaxed to $S<b+1$ while guaranteeing the success of \texttt{Basil} (benign subgraph connectivity)  with high probability. Additionally, the failure probability of   Basil is given by 
 \begin{align}
\mathbb {P}(\text{Failure}) \leq  \frac{b! (N-S)! }{(b-S)! (N-1)!} \label{Failuremain}, 
\end{align}
where $N$, $b$ are the total number of nodes, and Byzantine nodes, respectively. }
The proofs of Proposition 1-2 are given in Appendix \ref{Appendix A}.

\begin{remark}
In order to further illustrate the impact of choosing $S$ on the probability of failure given in \eqref{Failuremain}, we consider the following numerical examples.  Let the total number of nodes in the system be $N=100$, where $b= 33$ of them are Byzantine, and the storage parameter $S=15$. The failure event probability in  \eqref{Failuremain} turns out to be $ \sim{4 \times 10^{-7}}$, which is negligible. For the case when $S= 10$, the probability of failure becomes $ \sim{5.34 \times 10^{-4}}$, which remains   reasonably  small.
\end{remark}

\subsection{Theoretical Guarantees}\label{Assum}
We derive the convergence  guarantees  of \texttt{Basil} under the following standard assumptions.

\textbf{Assumption 1} (IID data distribution). \textit{ Local dataset $\mathcal{Z}_i$ at node $i$ consists of IID data samples from a distribution $\mathcal{P}_i$, where $\mathcal{P}_i =  \mathcal{P} $ for $ i \in \mathcal{R}$. In other words, $f_i(\mathbf{x}) =  \mathbb{E}_{\zeta_i \sim	 \mathcal{P}_i}[ l(\mathbf{x}, \zeta_i) ] =  \mathbb{E}_{\zeta_j \sim	 \mathcal{P}_j}[ l(\mathbf{x}, \zeta_i) )] = f_j(\mathbf{x})   \,\forall i, j \in \mathcal{R} $. Hence, the global loss function  $f(\mathbf{x}) =   \mathbb{E}_{\zeta_i \sim	 \mathcal{P}_i}[ l(\mathbf{x}, \zeta_i) ]$. }

\textbf{Assumption 2 }(Bounded variance). \textit{ Stochastic gradient $g_i( \mathbf{x})$ is unbiased and variance bounded, i.e., $\mathbb{E}_{\mathcal{P}_i} [ g_i( \mathbf{x})] = \nabla f_i(\mathbf{x}) =    \nabla f(\mathbf{x}) $, and $\mathbb{E}_{\mathcal{P}_i} || g_i( \mathbf{x}) -  \nabla f_i(\mathbf{x})||^2  \leq \sigma^2 $, where $g_i({\mathbf{x}}) $ is  the stochastic gradient computed by node $i$ by using  a  random sample $\zeta_i$  from its local dataset $\mathcal{Z}_i$. }  

\textbf{Assumption 3} (Smoothness\iffalse L-smooth and twice differentiable loss function\fi).  \textit{ The loss functions $f_i's$ are L-smooth and twice differentiable,  i.e., for any $\mathbf{x} \in \mathbb{ R}^d$, we have $||\nabla^2 f_i(\mathbf{x})||_2 \leq  L$.  }

Let $b^i$  be  the number of counterclockwise   Byzantine  neighbors  of node $i$. We divide the set of stored models  $\mathcal{N}_k^{i}$ at   node $i$ in the $k$-th round into two sets.  The first set   $\mathcal{G}^{i}_{k} = \{ \mathbf{y}_1, \dots, \mathbf{y}_{r^{i}}\} $ contains the benign models,  where $ r^i = S - b^i$. We consider scenarios with $S= b+1$, where $b$ is the total number of Byzantine nodes in the network. Without loss of generality, we assume the models in this set are arranged such that the first model is the closest benign node in the neighborhood of node $i$, while the last model is the farthest node. Similarly, we define the second set $\mathcal{B}_k^{i}$ to be the set of models from the counterclockwise   Byzantine neighbors of node $i$ such that $\mathcal{B}_{k}^{i} \cup \mathcal{G}_k^{i} = \mathcal{N}_k^{i}$.

\begin{theorem}\label{lemma1}
When the learning rate  
$\eta_{k}^{(i)}$ for node $i\in \mathcal{R}$ in round $k$ satisfies $\eta_{k}^{(i)} \geq \frac{1}{L}$, the expected loss function $\mathbb{E} \left[ l_i (\cdot) \right] $ of node $i$ evaluated on the set of models in  $\mathcal{N}_k^{i}$  can be arranged as follows: 
\begin{align}\label{222_main}
\mathbb{E} \left[ l_{i} (\mathbf{y}_1) \right] \leq  \mathbb{E} \left[ l_i (\mathbf{y}_2) \right]     \leq  \dots \leq &\mathbb{E} \left[ l_{i} (\mathbf{y}_{r^{i}}) \right] < \mathbb{E}  \left[ l_i (\mathbf{x}) \right]  \nonumber \\  \;  &\forall \mathbf{x} \in \mathcal{B}_{k}^{i}, 
\end{align}
where $\mathcal{G}^{i}_{k} = \{ \mathbf{y}_1, \dots, \mathbf{y}_{r^{i}}\} $ is the set of benign models stored at node $i$.  Hence, the \texttt{Basil} aggregation rule in Definition \ref{defi} is reduced to 
$\bar {\mathbf{x}}_k^{(i)} = \mathcal {A}_\text{\texttt{Basil} } ( \mathcal {N}^{i}_k) =\mathbf{y}_1$. Hence, the model update step in \eqref{Update_main} can be simplified as follows: 
\begin{equation} \label{5333}
\mathbf{x}_k^{(i)} =  \mathbf{y}_1 - \eta_{k}^{(i)}  g_i(\mathbf{y}_1).
\end{equation}
\end{theorem}

\begin{remark}
For the \texttt{Basil} aggregation rule in Definition \ref{defi}, \eqref{222_main} in Theorem \ref{lemma1} implies that for convergence analysis, we can consider only the benign sub-graph which is generated by removing the Byzantine nodes. As described in Section \ref{3.1}, the benign sub-graph is connected. Furthermore, due to \eqref{5333} in Theorem \ref{lemma1}, training via \texttt{Basil} reduces to sequential training over a logical ring with only the set $\mathcal{R}$ of benign nodes and connectivity parameter $S=1$. 
 \end{remark}

Leveraging the results in Theorem 1 and based on the discussion in Remark 1, we prove the linear convergence rate for \texttt{Basil}, under the additional assumption of convexity of the loss functions.
\begin{theorem}
Assume that $f(\mathbf{x})$ is  convex.  Under Assumptions 1-3 stated in this section, \texttt{Basil} with a fixed  learning rate  $\eta = \frac{1}{L}$  at all users  achieves linear convergence with a constant error as follows:
 \begin{align}
 \label{}
 \mathbb{E}\left[ f\left(\frac{1}{T} \sum_{s=1}^{T} \mathbf{x}^{s}\right)\right]-  f(\mathbf{x}^*  )  \leq   \frac{||\mathbf{x}^0- \mathbf{x}^{*}||^2L}{2  T } + \frac{1}{L}\sigma^2,
\end{align}
where $T = Kr$, $K$  is the total number of rounds over the ring and $r$ is the number of benign nodes.  Here $\mathbf{x}^s$ represents the model  after $s$ update steps starting from the  initial model $\mathbf{x}^0$, where $s =r k+i$ with $i = 1, \dots, r$ and $k = 0, \dots, K-1$. Furthermore,  $\mathbf{x}^{*}$ is the optimal solution  in \eqref{ma}  and   $\sigma^2$ is defined in Assumption 2.
\end{theorem}
\begin{remark}
The error bound  for \texttt{Basil} decreases with increasing the total number of  benign nodes $ r = \beta N$, where $\beta \in (0,1)$. 
\end{remark}

 To extend  \texttt{Basil}    to  be robust against software/hardware faults   in  the non-IID setting, i.e., when  the  local dataset $\mathcal{Z}_i$ at node $i$ consists of  data samples from a distribution $\mathcal{P}_i$ with $\mathcal{P}_i \neq   \mathcal{P}_j $ for $  i, j \in \mathcal{N}$ and  $i \neq j$, we present our  Anonymous Cyclic Data Sharing  algorithm (ACDS) in the following section.

\iffalse\begin{figure}[h!]
\vskip 0.1in
\begin{center}
\centerline{\includegraphics[width=.5\columnwidth]{ADSC1}}
\caption{ Two rounds for the  ACDS algorithm within group  $g$ which consists of $n=4$ nodes.}
\label{ACDS1}
\end{center}
\vskip -0.3in
\end{figure}
\fi
\section{Generalizing Basil  to Non-IID Setting  via Anonymous Cyclic Data Sharing  }\label{sec-ACDS}

We propose Anonymous Cyclic Data Sharing (ACDS), an algorithm that can be integrated on the top of \texttt{Basil}   to   guarantee robustness   against software/hardware faults   in the   non-IID  setting.  This algorithm allows   each node  to anonymously  share  a fraction of its local \textit{non-sensitive}  dataset  with  other nodes. In other words, ACDS   guarantees that the identity of the owner of the shared data is kept hidden from all other nodes under no collusion between nodes.
\begin{figure}[h]
\centering
\includegraphics[width=0.4\textwidth]{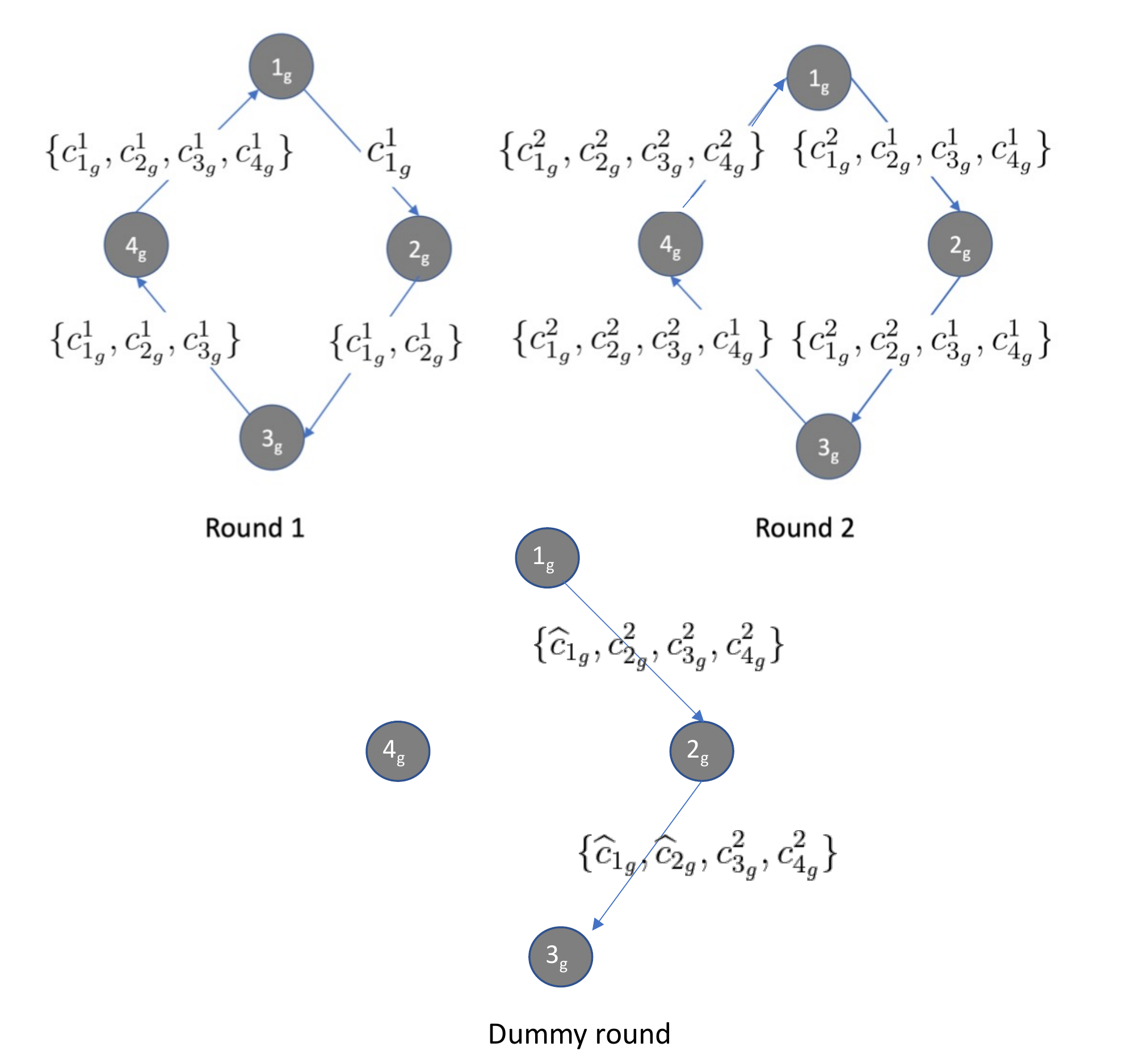}
\caption{ ACDS algorithm within group $g$ with $n =4$ users, where  each node  $i_g \in \mathcal{N}_g$ has two batches $\{c^{1}_{i_g}, c^{2}_{i_g}\}$. Each round starts from node $1_g$ and continues in a clockwise order.  The dummy round is introduced to make sure that node $2_g$ and node $3_g$ get their missing batches $\{c^2_{3_g}, c^2_{4_g}\}$ and $c^2_{4_g}$, respectively. Here, $\widehat c_{i_g}$ represents the dummy batch with the same size as the other batches that are  used by node $i_g \in \mathcal{N}_g$. This dummy batch could be a batch of public data that shares the same features that are used in the learning task.} 

\label{ACDS1}
\end{figure}
\subsection{ACDS Algorithm}
 The ACDS procedure has three phases which are formally described next. The overall algorithm has been summarized in Algorithm 2 and illustrated in Fig. 3.

\underline{\textbf{Phase 1: Initialization}.}
 ACDS  starts by first clustering    the  $N$ nodes into  $G$  groups of rings, where the set of nodes in each group $g \in [G]$ is   denoted by $\mathcal{N}_g =\{1_g, \dots, n_g\}$. Here, node $1_g$ is the starting node in ring $g$, and without loss of generality, we assume all groups have the same size $n = \frac{N}{G}$. Node $ i_g \in \mathcal {N}_g$ for $g \in [G]$ divides its dataset $\mathcal{Z}_{i_g}$ into sensitive  ($\mathcal{Z}_{i_g}^s$) and non-sensitive ($\mathcal{Z}_{i_g}^{ns}$)  portions, which can be done during the data processing phase by each node. Then, for a given hyperparameter $\alpha\in(0,1)$, each node selects $\alpha D$ points from its local non-sensitive dataset at random, where $ |\mathcal{Z}_{i_g}| =D$,  and then partitions these data points into $H$ batches denoted by $\{c_{i_g}^{1}, \dots, c_{i_g}^{H}\}$, where each batch  has $M=\frac{\alpha D}{H}$ data points.  

\RestyleAlgo{ruled}
\SetAlgoNoLine
\begin{algorithm}
\caption{ACDS
}\label{alg:fairagg2}
{\bf Input:}  $\mathcal{N}$ (nodes); $\{\mathcal{Z}_i\}_{i \in \mathcal{N}}$(local datasets); $\alpha$ (data fraction); $H$(number of batches); $G$(number of groups) \\

\tikzmk{A}
\textbf{\underline{Phase 1: Initialization}} \\
  $\{\mathcal{N}_g\}_{g \in [G]} \gets$ Clustering $(\mathcal{N}, G)$ \emph{\color{blue}{// cluster nodes $\mathcal{N}$ into $G$ groups each   of size $n$}} \\
  \For { each node  $i_g \in \mathcal{N}_g$ in parallel   } { $\mathcal{Z}_{{i_g}}^{s} \cup \mathcal{Z}_{{i_g}}^{ns} \gets \text{ Partition}(\mathcal{Z}_{{i_g}})$\emph{\color{blue}{ // partion local data $\mathcal{Z}_{{i_g}}$  into sensitive and non-sensitive parts}}.\\
  $\{ c^1_{i_g}, \dots,  c^H_{i_g}\} \gets $ RandomSelection$( \mathcal{Z}_{{i_g}}^{ns}, \alpha, H)$\emph{\color{blue}{ // random selection of  $H$ batches, each of  size  $\frac{\alpha D}{H}$,  from  $\mathcal{Z}_{{i_g}}^{ns}$}}.\\

   DStored$[i_g]$ =list()  \emph{\color{blue}{ // a list used by node  $i_g$  to store the shared data from other nodes}}
  }
 DShared[$g$]= list(),   $ \forall g \in [G]$ \emph{\color{blue}{ // a list that is  used to circulate the data within group $g$ \\
 }
}
\tikzmk{B}
\boxit{pink}
 
 \tikzmk{A}\textbf{\underline{Phase 2: Within Group Anonymous Data Sharing}}\\
 \For {each group $g  = 1, \dots, G$ in parallel }{
 \For {each batch $ h = 1, \dots, H $}{
 \For {each node  $i_g \in \mathcal{N}_g$ in sequence}
{
DShared$[g]$; DStored$[i_g] \gets$ \texttt{RobustShare}(DShared$[g]$; DStored$[i_g]$, $g$, $i_g$, $c^{h-1}_{i_g}$, $c^{h}_{i_g}$)\\
Send DShared$[g]$ to the next clockwise neighbor  
 } }
   \emph{\color{blue}{ // start the dummy round}}\\
 \For { each node  $i_g \in \mathcal{N}_g\backslash\{1_g,n_g\}$ in parallel   } {DShared$[g]$; DStored$[i_g] \gets$ \texttt{RobustShare}(DShared$[g]$; DStored$[i_g]$, $g$, $i_g$, $H$, $c^{H}_{i_g}$, $\widehat{c_{i_g}}$)
 }
 }
 \DontPrintSemicolon
  \SetKwFunction{FMain}{RobustShare}
  \SetKwProg{Fn}{Function}{:}{}
  \Fn{\FMain{DShared$[g]$; DStored$[i_g]$, $g$, $i_g$, $h$, $c^{h-1}_{i_g}$, $c^{h}_{i_g}$}}{
        \If { $h > 1$} {DShared$[g]$.remove($c^{h-1}_{i_g}$) \emph{\color{blue}{//    remove the batch $c^{h-1}_{i_g}$  from ``DataShared$[g]$''}}\\}
  DStored$[i_g].$add(DShared$[g]$)  \emph{\color{blue}{//  copy  the data in ``DShared$[g]$''  to   ``DStored$[i_g]$'' }} \\
  DShared$[g]$.add($c^{h}_{i_g}$)  {\emph{\color{blue}{// add the $h$-th batch  $c^{h}_{i_g}$ that will be shared  with  other nodes  to ``DShared$[g]$}}}\\
  DShared$[g]$.shuffle() {\emph{\color{blue}{//shuffle  the data in the list ``DShared$[g]$''  }}} \;
        \KwRet DShared$[g]$; DStored$[i_g]$\;
  }

\tikzmk{B}
 \boxit{cyan}
 
   \tikzmk{A}\textbf{\underline{Phase 3: Global Sharing}}\\
 \For {$g  = 1, \dots, G$ in parallel  }{
 node $1_g$ multicasts  DStored$[1_g] \bigcup  \{ c^1_{1_g}, \dots,  c^H_{1_g}\}$  with all nodes in $ \bigcup_{g'  \in [N]  \backslash \{g\}}\mathcal{N}_{g'}$} \tikzmk{B}
 \boxit{orange}
\end{algorithm}

\underline{\textbf{Phase 2: Within Group Anonymous Data Sharing}.} In this phase, for $g \in [G]$, the set of nodes $\mathcal{N}_g$ in  group $g$ anonymously share their data batches $\{ c^{j}_{1_g}, \dots, c^{j}_{n_g} \}_{j \in [H]}$   with each other. The anonymous data sharing within group $g$  takes $H+1$ rounds\iffalse, where the last round is the dummy round\fi. In particular,  
as shown in Fig.  \ref{ACDS1}, within group  $g$ and during the first round $h =1$,    node $1_g$ sends the first batch $c^{1}_{1_g}$ to its clockwise neighbor, node $2_g$. Node $2_g$ then stores the received batch. After that, node $2_g$ augments the received batch with its own first batch $c^{1}_{2_g}$ and shuffles them together before sending them to node $3_g$. More generally, in the intra-group cyclic sharing over the ring, node $i_g$ stores the received set of shuffled data points from batches $\{ c^{1}_{1_g}, \dots, c^{1}_{(i-1)_g} \}$ from its counterclockwise neighbor node  $(i-1)_g$. Then, it adds its own batch $c^{1}_{i_g}$ to the received set of data points, and shuffles them together, before sending the combined and shuffled dataset to  node $(i+1)_g$.

For round $1 <h \leq H$, as shown in Fig. \ref{ACDS1}-(round 2), node $1_g$ has the data points from the set of batches $\{ c^{(h-1)}_{1_g}, \dots,  c^{(h-1)}_{n_g} \}$ which were received from node $n_g$ at the end of round $(h-1)$. It first removes its old batch of data points  $c^{(h-1)}_{1_g}$ and then stores the remaining set of data points. After that, it adds its $h$-th  batch, $c^{(h)}_{1_g}$ to this remaining set, and then shuffles the entire set of data points in the new set of batches $\{ c^{h}_{1_g}, c^{{h-1}}_{2_g},  \dots, c^{h-1}_{n_g} \}$, before sending them to node $2_g$. More generally, in the $h$-th round for $1<h\leq H$, node $i_g$ first removes its batch $c^{h-1}_{i_g}$ from the received set of batches and then stores the set of remaining data points. Thereafter, node $i_g$ adds its $c^{h}_{i_g}$ to the set  $\{ c^{h}_{1_g}, \ldots, c^{h}_{(i-1)_g},c^{h-1}_{(i)_g}, \dots, c^{h-1}_{n_g} \}\backslash \{ c^{h-1}_{i_g}\}$, and then  shuffles the    set of data points in the new set  of batches  $\{ c^{h}_{1_g},  \dots, c^{h}_{i_g}, c^{h-1}_{(i+1)_g}, \dots, c^{h-1}_{n_g}  \}$ before sending them to node $(i+1)_g$.   

After $H$ intra-group communication iterations within  each group as described above, all nodes within each group have completely shared their $H-1$ batches with each other. However, due to the sequential nature of unicast communications, some nodes are still missing the batches shared by some clients in the $H^{th}$ round. For instance, in Fig. 3, after the completion of the second round, node  $2_g$ is missing the last batches $c^2_{3_g}$ and $c^2_{4_g}$. Therefore, we propose a final \textit{dummy} round, in which we repeat the same procedure adopted in  rounds $ 1<h \leq H$, but with the following slight modification: node $i_g$ replaces  its batch  $c^{H}_{i_g}$   with a dummy batch  $  c^{\text{dummy}}_{i_g}$. This dummy batch could be a batch of public data points that share the same feature space that is used in the learning task. This completes the anonymous cyclic data sharing within  group $g \in [G]$. 

\underline{\textbf{Phase 3: Global Sharing}.} 
For each $g \in [G]$, node $1_g$ shares the dataset $\{ c^{j}_{1_g}, \dots, c^{j}_{n_g} \}_{j \in [H]}$, which it has gathered in phase 2,  with all other nodes in the other groups.    

We note that implementation of ACDS only needs to be done once before training. As we demonstrate later in Section \ref{expe1}, the one-time overhead of the ACDS algorithm dramatically improves convergence performance when data is non-IID. In the following proposition, we describe the  communication cost/time of ACDS. 

\noindent \textbf{Proposition 3 }(Communication cost/time of ACDS). \textit{ Consider   ACDS algorithm with a set of   $N$ nodes divided equally into $G$ groups with $n$ nodes in each group. Each node  $i\in [N]$  has $H$ batches each of size $M = \frac{\alpha D}{H}$ data points, where $\alpha D$ is the fraction of shared local data, such that  $\alpha \in (0,1)$ and $D$ is the local data size. By letting $I$ to be the size of each data point in bits, we have the following: \\
\underline{ (1) Worst case communication cost per node  ($C_{\text{ACDS}}$)} 
\begin{equation}\label{eq-cost-remark}
    C_{\text{ACDS}} = \alpha D I (\frac{1}{H} +n (G+1)).
\end{equation}
%(3n+1)HI 
\underline{ (2)  Total communication time for completing ACDS ($T_{\text{ACDS}}$) }
When the  upload/download bandwidth of each node is $R$ b/s, we have the following 
\begin{align}
T_{\text{ACDS}} = \frac{\alpha D I}{HR} \left[n^2 (H+0.5) + n \left(H (G-1)  - 1.5\right) \right].
\end{align}
}

\begin{remark}
The worst case communication cost in \eqref{eq-cost-remark} is with respect to  the first node $1_g$, for $g \in [G]$,   that has more communication cost than the other nodes in group $g$ for its  participation in the global sharing phase  of ACDS.
\end{remark}
\begin{remark} To illustrate the  communication overhead resulting from using ACDS, we consider the following numerical example. Let the total number of nodes in the system   be $N = 100$ and  each node  has $ D = 500$ images from the CIFAR10 dataset, where  each image of size $I = 24.5$ Kbits\footnote{Each image in CIFAR10 dataset has $3$ channels each  of size $32 \times 32$ pixels, and each pixel takes value from $0-255$.}. When the communication  bandwidth  at each node  is $R= 100$ Mb/s (e.g., 4G speed), and each node  shares only $\alpha = 5 \%$ of its dataset in the form of $H = 5$ batches each with size $M = 5 $ images, the latency, and communication cost  of ACDS with $G = 4$ groups are  $11$ seconds and $~75$ Mbits, respectively. We note that the communication  cost for ACDS and  completion time of the algorithm   are  small with respect to the training process that requires sharing large model for large number of iteration as demonstrated in Section \ref{expe1}.
\end{remark}
The proof of Proposition 3 is presented in Appendix \ref{ACDS-time}.

In the following, we discuss the anonymity guarantees of ACDS.
\subsection{Anonymity Guarantees of  ACDS}
In the first round of the scheme, node $2_g$ will know that the source of the received batch   $c^1_{1_g}$ is node $1_g$.  Similarly and more general, node $i_g$ will know that each data point in the received set of batches $\{ c^1_{1_g},  \dots, c^{1}_{(i-1)_g} \}$ is generated by  one of the previous $i-1$ counterclockwise neighbors.  However, in the next $H-1$ rounds, each received data point   by any node will be equally likely generated from any one of the remaining $n-1$ nodes in this group. Hence, the   size of the candidate pool from which each node could take a guess for the owner of each data point is small specially for the first set of nodes in the ring. In order to provide   anonymity for the entire data and decrease the risk in the first round of the ACDS scheme, the size of the batch can be reduced to just one data point. Therefore, in the first round node $2_g$ will only know one data point from node $1_g$. This comes on the expense of increasing the number of rounds. Another key consideration is that the number of nodes in each group trades the level of anonymity with the communication cost. In particular, the communication cost per node in the ACDS algorithm is $\mathcal{O}(n)$, while the anonymity level, which we measure by the number of possible candidates  for a given data point, is $(n-1)$. Therefore, increasing $n$, i.e., decreasing the number of groups  $G$, will decrease the communication cost but increase the anonymity level.

\section{\texttt{Basil}+: Parallelization of \texttt{Basil}} \label{sec7}
 In this section, we describe our modified protocol   \texttt{Basil}+ which     allows for  parallel training across multiple rings, along with sequential training over each ring.  This results in decreasing the training  time needed for completing one global epoch  (visiting all nodes) compared to \texttt{Basil} which only considers sequential training over one ring.
 
 \subsection{Basil+ Algorithm }
 At a high level,   \texttt{Basil}+  divides nodes into $G$  groups   such that each   group in parallel  performs sequential training over a logical ring using  \texttt{Basil} algorithm. After   $\tau$ rounds   of sequential training within  each group,    a robust circular     aggregation  strategy is implemented to have a robust average model from all the groups. Following the robust circular     aggregation  stage,   a final multicasting step is implemented such that  each group can use   the resulting  robust average model.  This entire process is repeated for $K$  global rounds.  
 
 We now formalize the execution of  \texttt{Basil}+ through the following four stages.  \\
 
\noindent \underline{\textbf{Stage 1:  \texttt{Basil}+ Initialization}.} The protocol     starts by  clustering   the set of  $N$ nodes equally  into  $G$  groups of rings with $n = \frac{N}{G}$ nodes in each group.   The set of nodes in group $g$ is    denoted by $\mathcal{N}_g =\{u^g_1, \dots, u_n^g\}$, where node $u_1^g$ is the starting node in  ring $g$, where  $g= 1, \dots, G$.  The clustering of nodes follows a random splitting agreement protocol similar to the one in Section \ref{3.1} (details are presented in Section \ref{sec-7.3}).  	The connectivity parameter within each group is set to be $S = \min (n-1, b+1)$, where b is the worst-case number of Byzantine nodes.   This choice of $S$ guarantees that each benign subgraph within each group is  connected with high probability, as described in Proposition 3.  \\
 
 \noindent\underline{\textbf{Stage 2: Within  Group Parallel Implementation of \texttt{Basil}}.}  Each group $g \in [G]$ in parallel  performs the training  across   its nodes for $\tau$ rounds  using  \texttt{Basil} algorithm.    \\
 
 \noindent \underline{\textbf{Stage 3: Robust Circular Aggregation Strategy}.} We denote
  \begin{equation}\label{eq-s}
      \mathcal{S}_g = \{u^g_{n-1}, u^g_{n-2},  \dots, u_{n-S+1}^g\}, 
  \end{equation}
  to be the set of  $S$ counterclockwise neighbors of node $u^g_1$.   The robust circular    aggregation    strategy consists of $G-1$ steps performed
sequentially over the $G$ groups, where the $G$ groups form a global ring.  At step $g$, where   $g \in \{ 1, \dots, G-1\}$, the set of nodes  $\mathcal{S}_g$ send their  aggregated models  to  each node  in the set $\mathcal{S}_{g+1}$. The reason for sending $S$ models from one group to another  is to ensure the connectivity of the global ring when removing  the Byzantine nodes. The average aggregated   model at node $u^{g+1}_i \in \mathcal{S}_{g+1} $ is  given as follows:
\begin{align}
\label{Update1}
\mathbf{z}_{i}^{g+1}=& \frac{1}{g+1} \left( \mathbf{x}^{(i,g+1)}_{\tau}+ g \bar {\mathbf{z}}_i^{g+1}\right),
\end{align}
where $\mathbf{x}^{(i,g+1)}_{\tau}$ is the local updated model at node $u^{g+1}_i$ in  ring $g+1$ after $\tau$ rounds of  updates according to   \texttt{Basil} algorithm. Here,    $\bar{\mathbf{z}}_i^{g+1}$ is  the selected model by node $u^{g+1}_i $  from the set of  received models from the set $\mathcal{S}_{g}$. More specifically, by letting
\begin{equation}\label{eq-L}
 \mathcal{L}_{g} = \{ \mathbf{z}_{n-1}^{g}, \mathbf{z}_{n-2}^{g}, \dots, \mathbf{z}_{n-S+1}^{g} \}
\end{equation}   be the set of  average  aggregated  models  sent  from the set of nodes $\mathcal{S}_{g}$ to each node  in the set $\mathcal{S}_{g+1}$,  we define  $\bar{\mathbf{z}}_i^{g+1}$ to  be the model selected from $\mathcal{L}_{g}$ by node $u^{g+1}_i \in \mathcal{S}_{g+1} $ using the Basil aggregation rule as follows: 
 \begin{equation}\label{eq12--}
 \bar {\mathbf{z}}_i^{g+1} = \mathcal {A}_\text{\texttt{Basil} } (  \mathcal{L}_{g}) =  \arg \min_{ \mathbf{y} \in \mathcal{L}_{g} }  \mathbb{E} \left[{l}^{g+1}_i({\mathbf{y}}, \zeta^{g+1}_i)\right].
\end{equation} 
 \underline{\textbf{Stage 4: Robust Multicasting}.}
The final stage is the multicasting stage.  The set of nodes in $\mathcal{S}_G$ send the final set of robust aggregated models $\mathcal{L}_{G}$ to  $\mathcal{S}_1$. Each node  in the  set   $\mathcal{S}_1$     applies the aggregation rule in \eqref{eq12--} on the set   of received models $\mathcal{L}_{G}$. Finally,  each  benign node in the set $\mathcal{S}_1$  sends the filtered model ${\mathbf{z}}_i^{1}$   to all  nodes in this set $\cup_{g=1}^G \mathcal{U}_g$, where  $\mathcal{U}_g$ is defined as follows 
\begin{equation}\label{eq-U}
    \mathcal{U}_g = \{u^g_1, u^g_{2},  \dots, u_{S}^g\}.
\end{equation}Finally, all nodes in this set $\cup_{g=1}^G \mathcal{U}_g$  use the aggregation rule in \eqref{eq12--} to get the best model  out of this set $\mathcal{L}_{1}$ before updating it according to \eqref{Update_main}.  These four stages are repeated for $K$ rounds. 

\RestyleAlgo{ruled}
\SetAlgoNoLine

\begin{algorithm}
\caption{\texttt{Basil}+}
\label{alg:fairagg3}
 
 {\bf Input:} $\mathcal{N}$; $S$; $\{\mathcal{Z}_i\}_{i \in \mathcal{N}}$; $\mathbf{x}^0$; $\tau$, $K$
 
 \tikzmk{A}\underline{\textbf{Stage 1: Initialization}}: 

 \For {each node  $i \in \mathcal{N}$  }
{
  $\{\mathcal{N}_g\}_{g \in [G]} \gets$ RandomClusteringAggrement$(\mathcal{N},G)$ \emph{\color{blue}{//cluster ther nodes into $G$ groups each of size $n$  according to Section IV-A}}
} 
$\mathbf{x}^{(i,g)} \gets \mathbf{x}^0, \quad \forall i \in \mathcal{N}_g, g \in [G]$ 
  \tikzmk{B}
 \boxit{pink} 
 
 \For{ each global round $k = 1, \dots, K$}
 { 
  \tikzmk{A}
\textbf{\underline{Stage 2: Within Groups  Robust Training}}:

 \For { each group $g  = 1, \dots, G$ in parallel  }
 {
  \texttt{Basil}{(}$\mathcal{N}_g, S, \{\mathcal{Z}_i\}_{i \in {\mathcal{N}_g}},   \{\mathbf{x}^{(i,g)}\}_{i \in {\mathcal{N}_g}}, \tau ${)}$\rightarrow\{\mathbf{x}^{(i,g)}_{\tau}\}_{i \in {\mathcal{N}_g}}$  
  \emph{\color{blue}{//apply \texttt{Basil} algorithm within each group for $\tau$ rounds}}
  
 $\mathbf{z}_i^g \gets  \mathbf{x}^{(i,g)}_{\tau}, \quad \forall i \in \mathcal{N}_g$ 
 }
 \tikzmk{B}
 \boxit{cyan}
 
 \tikzmk{A}\textbf{\underline{Stage 3: Robust Circular Aggregation}}:
 \For { each group $g  = 1, \dots, G-1$ in sequence  }
 {
 
 \For {each node $u_i^{g+1} \in \mathcal{S}_{g+1}$ in parallel}
 {  
 
 \emph{\color{blue}{//  the set $\mathcal{S}_{g+1}$ is defined in \eqref{eq-s}}}
 
 $\mathcal{L}_g \gets \{ \mathbf{z}_i^{g}\}_{u_i^{g} \in \mathcal{S}_g}$ \emph{\color{blue}{//   each  node $u_i^{g+1} \in \mathcal{S}_{g+1}$  receives the set of models $\mathcal{L}_g$ (defined in \eqref{eq-L})  from the nodes in   $ \mathcal{S}_{g}$. }}

  \If{ node  $u_i^{g+1}  \in \text{benign set } \mathcal{R}$ } 
  {
  
 $\bar{\mathbf{z}}_i^{g+1} \gets \mathcal {A}_\text{\texttt{Basil} } (\mathcal{L}_g)$ \emph{\color{blue}{//    \texttt{Basil}  performance based strategy  to select one model from  $\mathcal{L}_g$ using  \eqref{eq12--}}}  \\
 $ \mathbf{z}_{i}^{g+1} \gets \frac{1}{g+1} \left( \mathbf{x}^{(i,g+1)}_{\tau}+ g \bar {\mathbf{z}}_i^{g+1}\right)$ {\emph{\color{blue}{//get proper average model from the first $g+1$ groups }}}
  
  }
 \Else{ $\mathbf{z}_{i}^{g+1}  \gets *$ \emph{\color{blue}{// Byzantine node sends faulty model} }}

 }

 }
 \tikzmk{B}
 \boxit{orange}

 \tikzmk{A}
 \textbf{\underline{Stage 4: Robust Multicasting}}:
 
 \For {each node  $u_i^{1} \in \mathcal{S}_{1}$ in parallel}
 {
  $\mathcal{L}_G \gets \{ \mathbf{z}_i^{(G)}\}_{u_i^{G} \in \mathcal{S}_G}$.\\
 
 $\bar{\mathbf{z}}_i^{1} \gets \mathcal {A}_\text{\texttt{Basil} } (\mathcal{L}_G)$\\
  
  $ \mathbf{z}_{i}^{1} \gets \bar{\mathbf{z}}_i^{1}$ 
  
  }
%   \iffalse
 \For {each node  $u_i^g  \in \cup_{g=1}^G \mathcal{U}_g$ in parallel  }
 {
 \emph{\color{blue}{//  the set $\mathcal{U}_g$ is defined in \eqref{eq-U}}}
 $\mathcal{L}_1 \gets \{ \mathbf{z}_i^{1}\}_{u_i^{1} \in \mathcal{S}_1}$ \emph{\color{blue}{//   each  node $u_i^{g}$  receives the set of models $\mathcal{L}_1$  from the nodes in   $ \mathcal{S}_{1}$. }}
 
 $\bar{\mathbf{z}}_i^{g} \gets \mathcal {A}_\text{\texttt{Basil} } (\mathcal{L}_1)$

    $\mathbf{x}^{(i,g)} \gets  \bar{\mathbf{z}}_i^{g}$
}
% \fi
\tikzmk{B}
\boxit{yellow}

}

 {\bf Return $\{\mathbf{x}^{(i,g)}_{K}\}_{{i \in \mathcal{N}_g}, g \in [G]}$ } 

 \end{algorithm}

We compare between the training time of \texttt{Basil} and \texttt{Basil}+ in the following proposition.

  \noindent \textbf{Proposition 4.}  \textit{ Let $T_{\text{comm}}$, $T_{\text{perf-based}}$, and  $T_{\text{SGD}}$ respectively denote the time needed to multicast/receive  one model, the time to evaluate $S$ models according to \texttt{Basil} performance-based criterion, and the time to take one step of model update. The total training time  for completing   one  global round  when using   \texttt{Basil} algorithm, where one global round  consists  of $\tau$  sequential rounds over the ring,  is 
  \begin{equation} \label{trainig_time}
  T_\text
  {\texttt{Basil}} \leq (\tau nG) T_{\text{perf-based}} + (\tau nG) T_{\text{comm}} + (\tau nG)T_{\text{SGD}},
  \end{equation}
compared to the  training time  for  \texttt{Basil}+ algorithm, which is given as follows
\begin{align} \label{trainig_time+}
  T_\text
  {\texttt{Basil}+} &\leq (\tau n +G + 1 ) T_{\text{perf-based}} + (SG+\tau n-1) T_{\text{comm}} \nonumber\\
  &\quad\quad\quad\quad\quad\quad\quad\quad\quad\quad\quad\quad\quad+ (\tau n)T_{\text{SGD}}.
  \end{align}
  }
  \begin{remark}
   According to this proposition, we can see the training time  of  \texttt{Basil} is  polynomial  in $nG$, while  in \texttt{Basil}+, the training time is linear in both $n$ and $G$.  
The proof of Proposition 4 is given in Appendix \ref{prop3}. 
  \end{remark}

In the following section, we discuss the random clustering method used in the Stage 1 of \texttt{Basil}+. 
 \subsection{Random  Clustering Agreement } \label{sec-7.3}
 In practice, nodes can  agree on   a random clustering   by using similar approach as in Section \ref{3.1} by  the    following simple steps. 1) All nodes first share their IDs with each other, and we assume WLOG that nodes’ IDs can be arranged in ascending order,  and  Byzantine nodes cannot forge their identities or create multiple fake ones \cite{elmhamdi2020genuinely}. 2)  Each node     locally   random splits the nodes into $G$ subsets by using  a pseudo random number generator (PRNG) initialized via a common seed (e.g., N). This ensures that all  nodes  will generate the set of nodes. To know the nodes order within each local group, the method in Section \ref{3.1} can be used.

\subsection{The Success of  \texttt{Basil}+ }\label{sec-7.2}
We will consider different scenarios for the connectivity parameter $S$ while evaluating the success of \texttt{Basil}+.
\underline{\textbf{Case 1: $S = \min (n-1, b+1)$}}
We set the connectivity parameter for \texttt{Basil}+ to   $S = \min (n-1, b+1)$.  By setting  $S =  b+1$ when  $n>b+1$, this ensures the connectivity  of each ring (after removing the Byzantine nodes) along with the global ring. On the other hand, by setting $S =  n-1$ if $n \leq b+1$,  \texttt{Basil}+ would only fail if  at least one group has    a number of Byzantine nodes  of   $n$ or $n-1$.  We define    $B_j$ to be the failure event in which  the number of Byzantine nodes in a  given group  is $n$ or $n-1$. The failure event $B_j$   follows  a Hypergeometric distribution with parameters $(N,b,n) $, where $N$, $b$, and $n$ are the total number of nodes, total number of Byzantine nodes,  number of nodes in each group, respectively. The probability of failure is given as follows  
\begin{align}
\mathbb {P}(\text{Failure}) {=}\mathbb {P}( \bigcup_{j=1}^{G}B_j) 
 \overset {(a)}\leq \sum_{j=1}^{G}\mathbb {P}(B_j)
{=}  \frac{\binom{b}{n} + \binom{b}{n-1} \binom{N-b}{1}}{\binom{N}{n}} G \label{Failure1},
\end{align}
where (a) follows from the union bound.  

In order to further illustrate the impact of choosing the group size $n$  when setting  $S = n-1 $ on the probability of failure given in \eqref{Failure1}, we consider the following numerical examples.  Let the total number of nodes in the system be $N=100$, where $b= 33$ of them are Byzantine. By setting    $n = 20$ nodes in each group,  the  probability in  \eqref{Failure1} turns out to be $ \sim{5 \times 10^{-10}}$, which is negligible. For the case when $n= 10$, the probability of failure becomes $ \sim{1.2 \times 10^{-4}}$, which remains   reasonably  small.

\underline{\textbf{Case 2: $S<n-1$}}
Similar to the failure analysis of  \texttt{Basil} given in Proposition 2, we relax the connectivity parameter $S$ as stated in the following proposition. 

\noindent \textbf{Proposition 5.}\textit{ The connectivity parameter $S$ in \texttt{Basil+}  can be relaxed to $S<n-1$ while guaranteeing the success of the algorithm (benign local/global subgraph connectivity)  with high probability. The failure probability of \texttt{Basil+} is given by  
\begin{equation} \label{Prop4}
\mathbb {P}(F) \leq  G \sum_{i = 0}^{\min{ (b,n)}}   \left(  \prod_{s=0 }^{S-1} \frac{ \max \left(i-s , 0 \right)}{(N-s) } n  \right) \frac{\binom{b}{i}\binom{N-b}{n-i}}{\binom{N}{n}}, 
\end{equation}
where $N$, $n$, $G$, $S$ and $b$ are the number of total nodes, number of nodes in each group, number of groups, the connectivity parameter, and the number of Byzantine nodes.  }

The proof of Proposition 5 is presented in Appendix \ref{prop4}. 

\begin{remark}In order to further illustrate the impact of choosing $S$ on the probability of failure given in \eqref{Prop4}, we consider the following numerical examples.  Let the total number of nodes in the system be $N=400$, where $b= 60$ of them are Byzantine and $n=100$,  and the connectivity  parameter $S=10$. The probability of failure event  in  \eqref{Prop4} turns out to be $ \sim{ 10^{-6}}$, which is negligible. For the case when $S= 7$, the probability of failure becomes $ \sim{ 10^{-4}}$, which remains   reasonably  small.
 \end{remark}

\section{Numerical Experiments}\label{expe1}
We start by evaluating the performance gains of \texttt{Basil}.  After that, we give the set of experiments of \texttt{Basil+}. We note that in Appendix H, we have included additional experiments for Basil including the wall-clock time performance compared to UBAR, performance of \texttt{Basil} and ACDS for CIFAR100 dataset, and performance comparison between \texttt{Basil} and \texttt{Basil}+.  
\subsection{Numerical Experiments for Basil}

\iffalse
{\color{red} this approach can also be fine
https://arxiv.org/pdf/2011.06223.pdf}
{\color{red} follow UBAR description}
{\color{red} dump all the material, will restructure later}
\fi

\noindent\textbf{Schemes}:
We consider  four  schemes as described next.
\begin{itemize}
\item {G-plain}: This is for graph based topology. At the start of each round, nodes exchange their models with their neighbors. Each node then finds the average of its model with the received neighboring models and uses it to carry out an SGD step over its local dataset.
\item {R-plain}:  This is for ring based topology with $S = 1$. The current active node carries out an SGD step over its local dataset by using the model received from its previous counterclockwise neighbor. 
    \item {UBAR}: This is the prior state-of-the-art for mitigating Byzantine nodes in decentralized training over graph, and is described in Appendix \ref{UBAR1}. 
    
   \item {\texttt{Basil}}: This is our proposal. 
    %In order to have roughly same number of \textit{neighbors} as UBAR scheme, we set $S=p N$.
\end{itemize}

%{\color{red} We note that we choose to only compare with UBAR as it gives higher performance than the prior works in decentralized training such BRIDGE \cite{yang2019bridge} and   }

\noindent\textbf{Datasets and Hyperparameters}:
There are a total of $100$ nodes, in which $67$ are benign. For decentralized network setting for simulating UBAR and G-plain schemes, we follow a similar approach as described in the experiments in \cite{ guo2020byzantineresilient} (we provide the details in Appendix \ref{impUBAR}). For \texttt{Basil}  and R-plain, the nodes are arranged in a logical ring, and $33$ of them are randomly set as Byzantine nodes. Furthermore, we set $S = 10$  for \texttt{Basil}   which gives us the connectivity  guarantees discussed  in  Proposition 2. We use a decreasing learning rate of $0.03/(1+0.03\,k)$. We consider   CIFAR10 \cite{Krizhevsky2009LearningML} and use a neural network with $2$ convolutional layers and $3$ fully connected layers. The details are included in Appendix \ref{model}. The training dataset is partitioned equally among all nodes. Furthermore, we report the worst test accuracy among the benign clients in our results.  We also conduct similar evaluations on the MNIST dataset. The experimental results lead to the same conclusion and can be found in Appendix \ref{expe}.  Additionally, we emphasize that \texttt{Basil} is based on sequential training over a logical ring, while UBAR is based on  parallel training over a graph topology. Hence, for consistency of experimental evaluations, we consider the following common definition for training round: 

\begin{definition} [Training round]
 With respect to the total number of SGD computations, we define a round over a logical ring to be   equivalent to one parallel iteration over a graph. This definition aligns with our motivation of training with resource constrained edge devices, where user's computation power is limited.
\end{definition}

\begin{figure*}[t]
  \centering
  \subfigure[No Attack]{\includegraphics[scale=0.28]{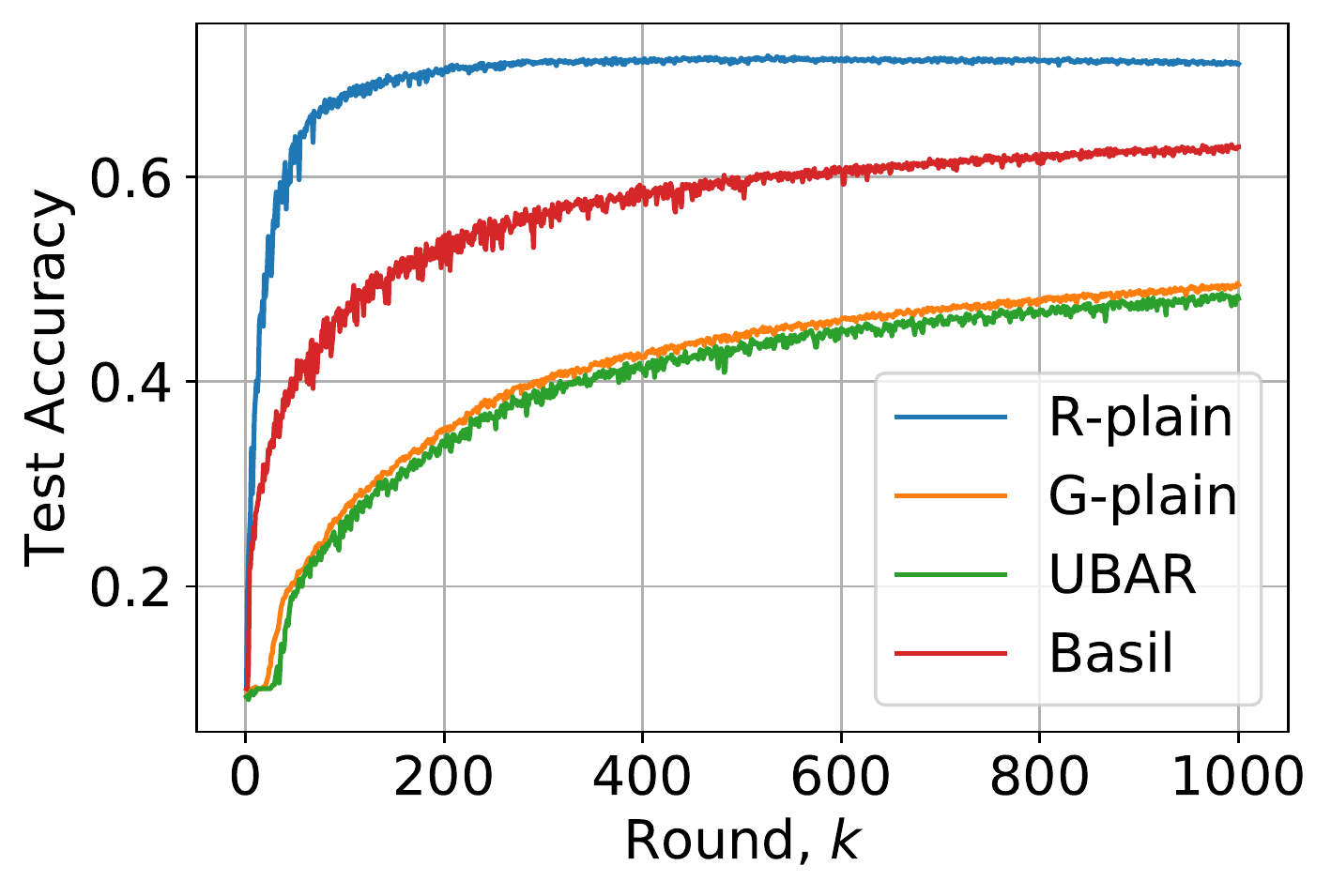}
  }%\quad
  \subfigure[Gaussian Attack]{\includegraphics[scale=0.28]{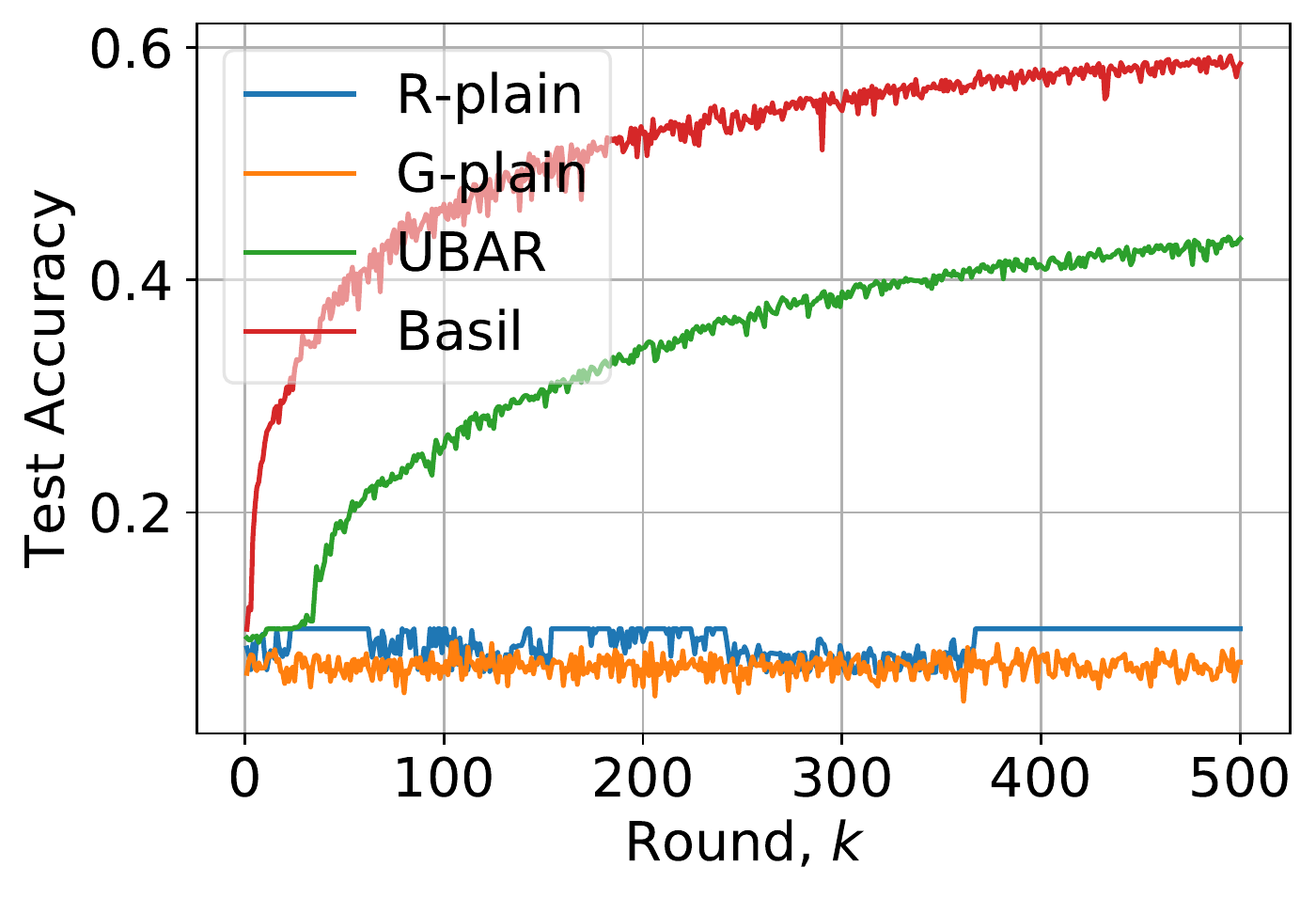}
  }%\quad
  \subfigure[Random Sign Flip]{\includegraphics[scale=0.28]{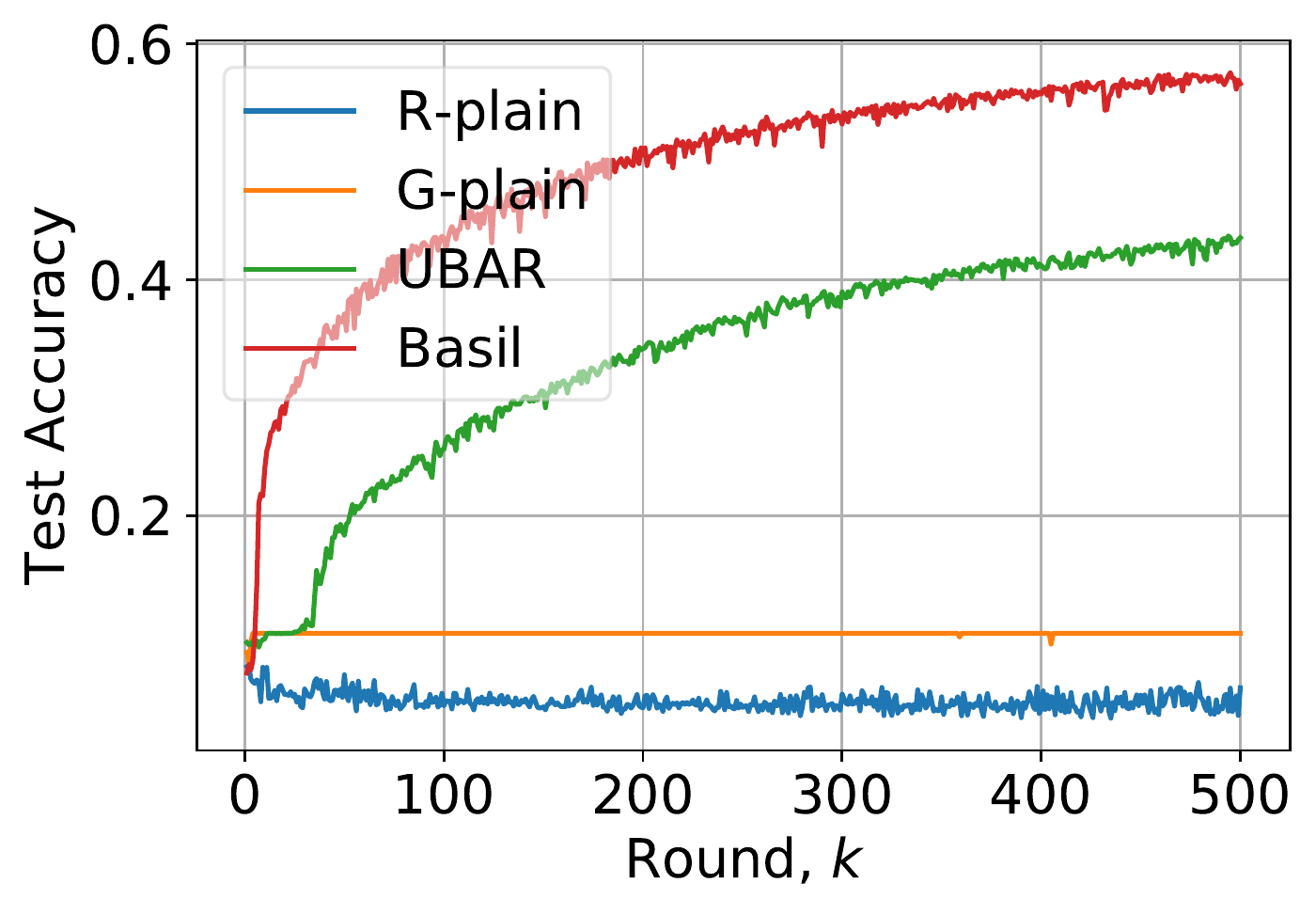}
  }
  \subfigure[Hidden Attack]{\includegraphics[scale=0.28]{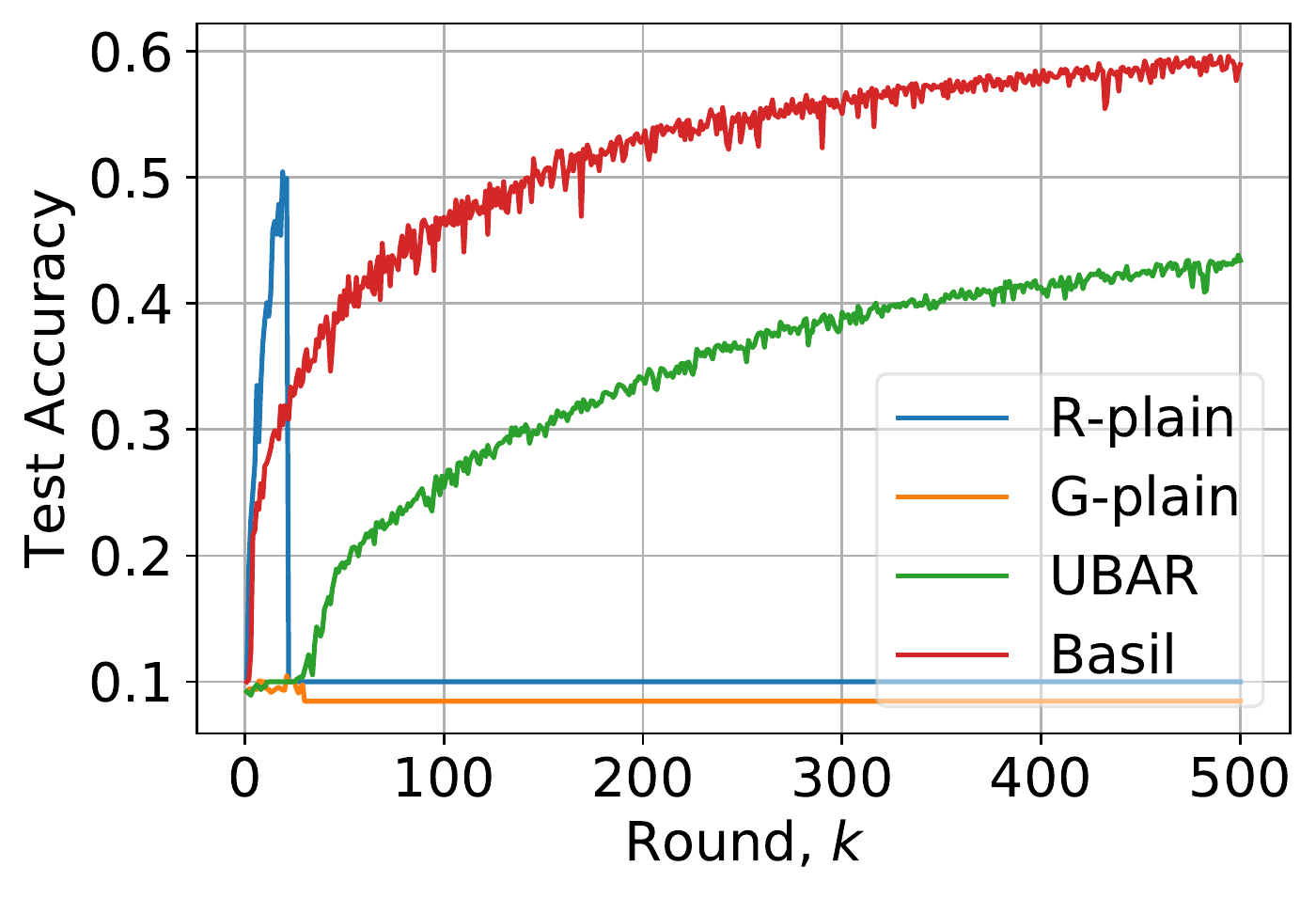}
  }%\quad
  \caption{Illustrating the performance of \texttt{Basil}   using  CIFAR10 dataset under IID data distribution setting.}
  \label{fig:results_cifar10}
 \end{figure*}

 \begin{figure*}[]
  \centering
  \subfigure[No Attack]{\includegraphics[scale=0.28]{cifar10_no_byzant_plot_1.pdf}
  }%\quad
  \subfigure[No Attack]{\includegraphics[scale=0.28]{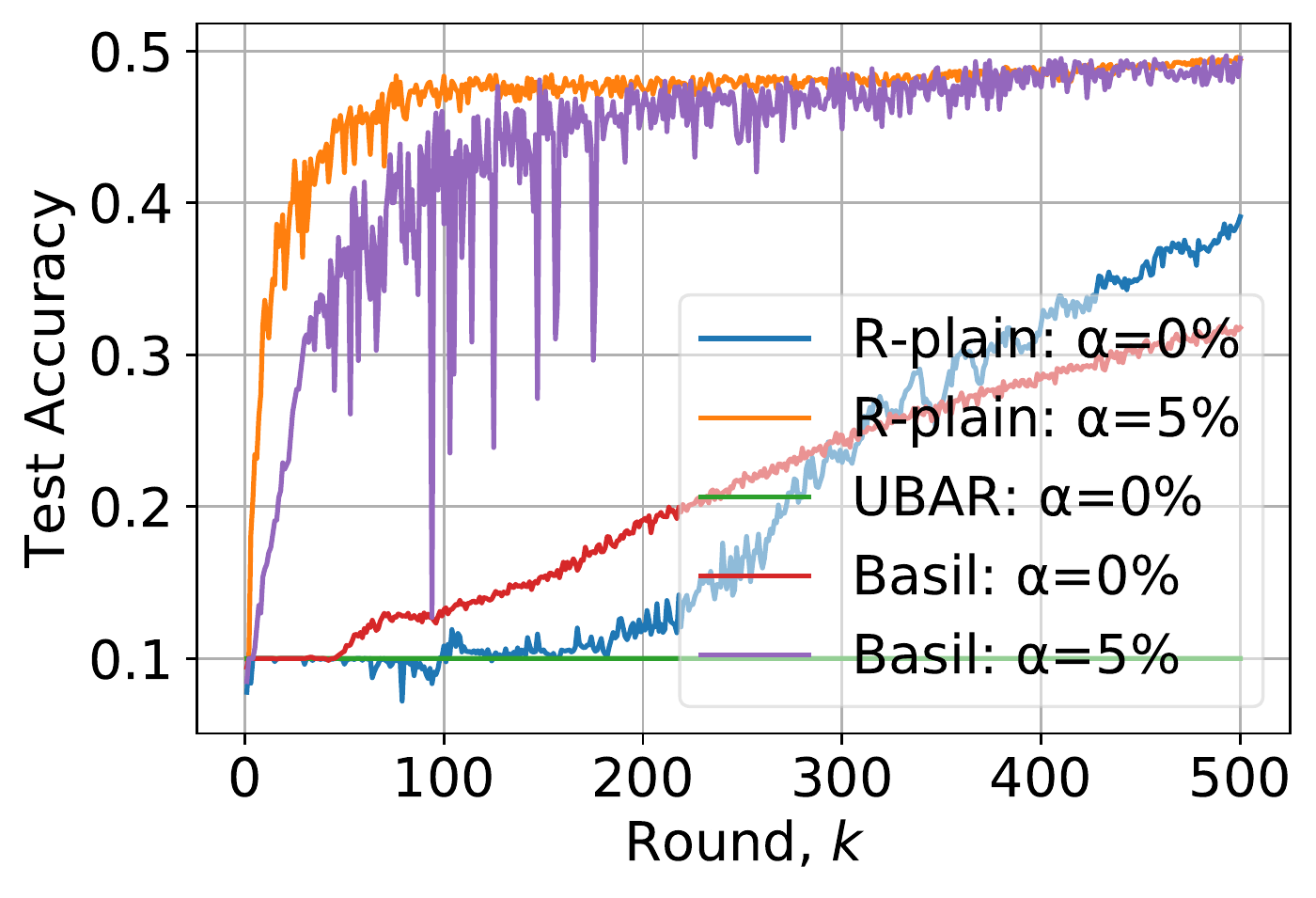}
  }%\quad
  \subfigure[Gaussian Attack]{\includegraphics[scale=0.28]{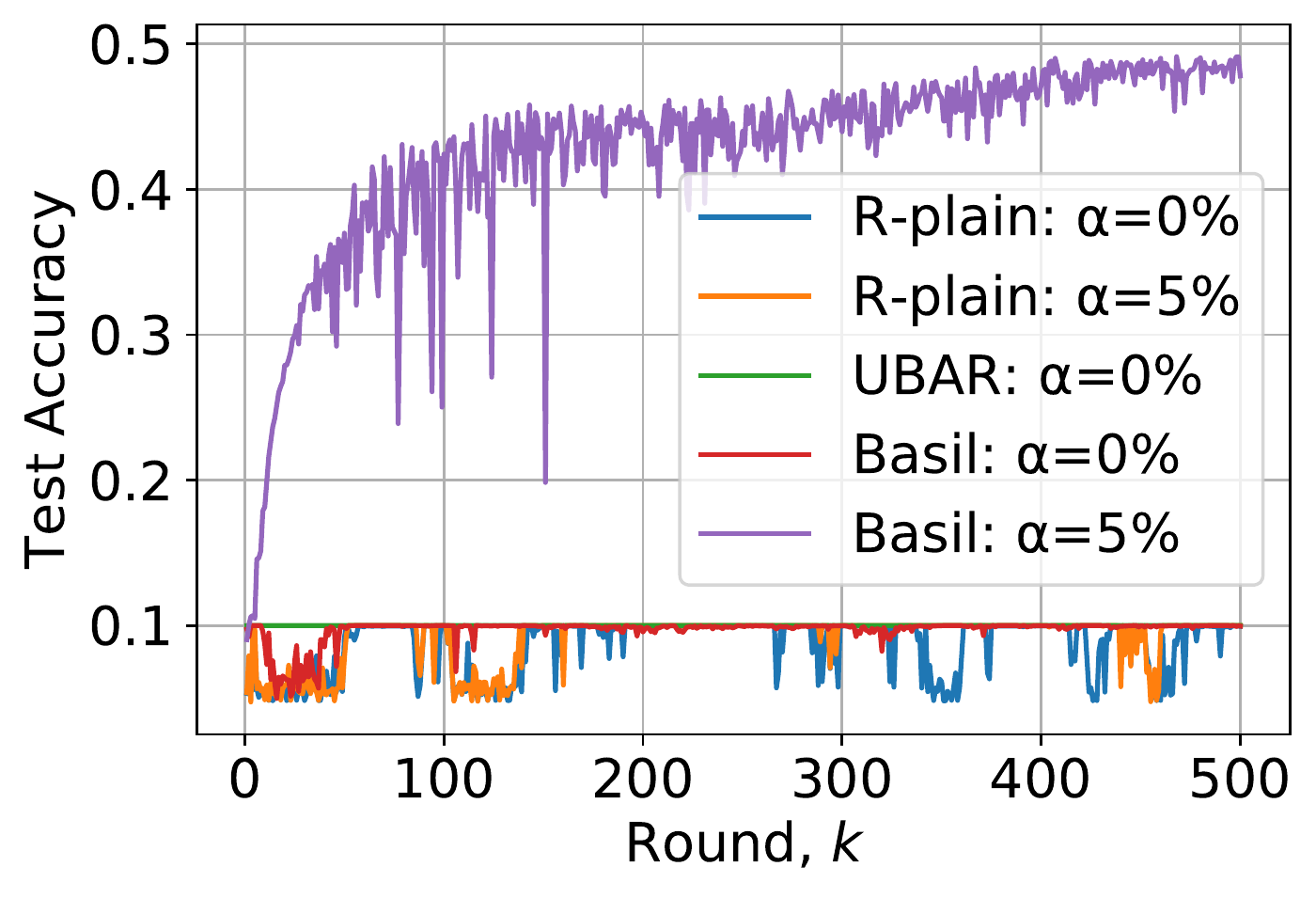}
  }
  \subfigure[Random Sign Flip]{\includegraphics[scale=0.28]{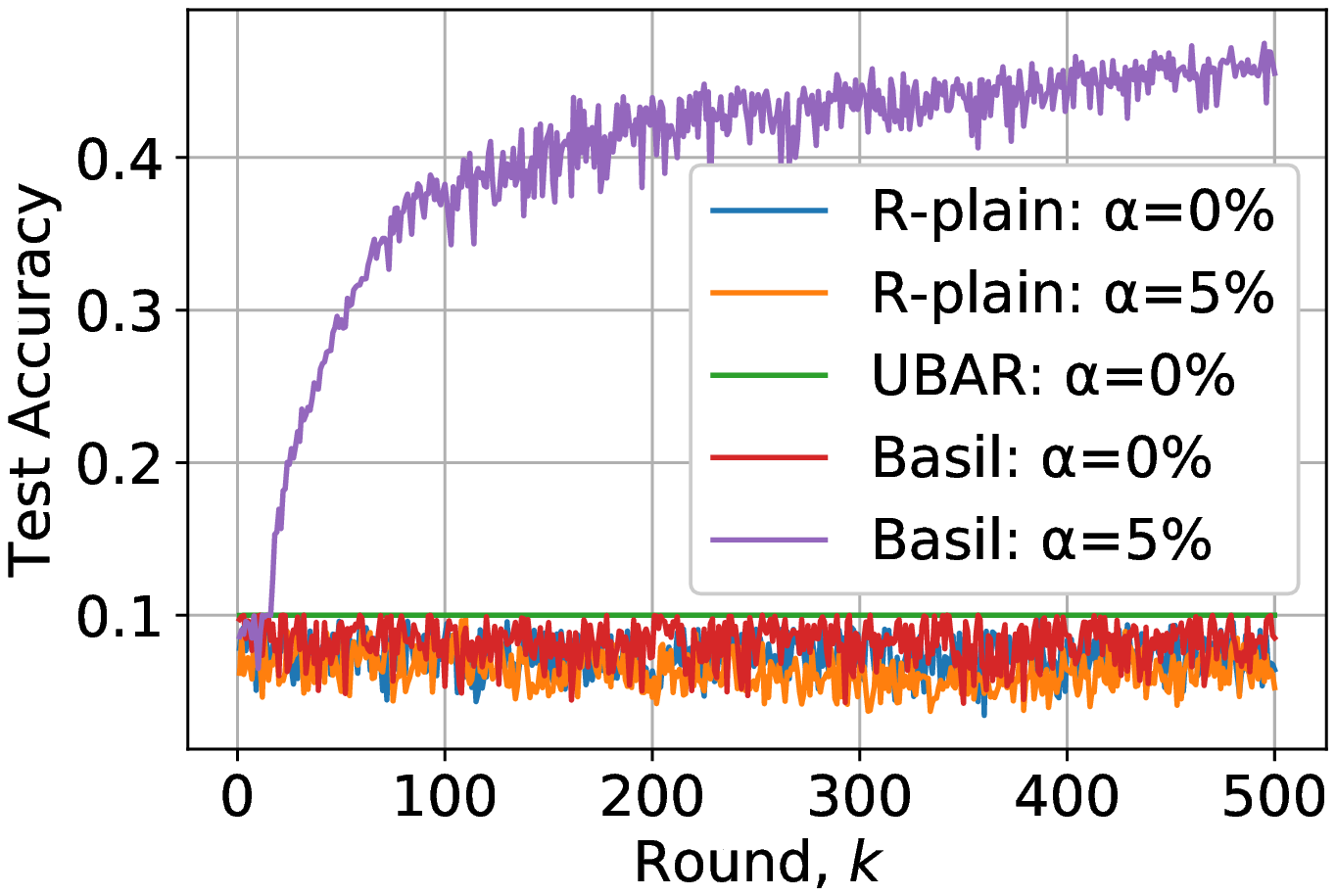}
  }%\quad
  \caption{Illustrating the performance of \texttt{Basil}   using  CIFAR10 dataset under non-IID data distribution setting.}
  \label{fig:cifar10_sign_flip}
 \end{figure*}

\noindent\textbf{Byzantine Attacks}:
We consider a variety of attacks, that are described as follows. \textit{Gaussian Attack}: Each Byzantine node replaces its model parameters with entries drawn from a Gaussian distribution with mean $0$ and standard distribution $\sigma=1$. \textit{ Random Sign  Flip}: We observed in our experiments that the naive sign flip attack, in which Byzantine nodes flip the sign of each parameter before exchanging their models with their neighbors, is not strong in the R-plain scheme. To make the sign-flip attack stronger, we propose a layer-wise sign flip, in which Byzantine nodes randomly choose to flip or keep the sign of the entire elements in each neural network layer.   \textit{Hidden Attack}: This is the attack that degrades the performance of distance-based defense approaches, as proposed in \cite{pmlr-v80-mhamdi18a}.  Essentially, the Byzantine nodes are assumed to be omniscient, i.e., they can collect the models uploaded by all the benign clients. Byzantine nodes then design their models such that they are undetectable from the benign ones in terms of the distance metric, while still degrading the training process. For hidden attack, as the key idea is to exploit similarity of models from benign nodes, thus, to make it more effective, the Byzantine nodes launch this attack after $20$ rounds of training.

\noindent\textbf{Results (IID Setting)}:
We first present the results for the IID data setting. The training dataset is first shuffled randomly and then partitioned among the nodes. As can be seen from  Fig. 4(a), \texttt{Basil}  converges much faster than both UBAR and G-plain even in the absence of any Byzantine attacks, illustrating the benefits of ring topology based learning over graph based topology. We note that the total number of gradient updates after $k$ rounds in the two setups are almost the same. We can also see that R-plain gives higher performance than \texttt{Basil}. This is because in \texttt{Basil}, a small mini-batch is used for performance evaluation, hence in contrast to R-plain, the latest neighborhood model may not be chosen in each round resulting in the loss of some update steps. Nevertheless, Figs.  4(b), (c) and 4(d) illustrate that \texttt{Basil} is not only Byzantine-resilient, it maintains its superior performance over UBAR with ${\sim}{16\%}$ improvement in test accuracy, as opposed to R-plain that suffers significantly. 
Furthermore, we would like to highlight that as \texttt{Basil}  uses a performance-based criterion for mitigating Byzantine nodes, it is robust to the Hidden attack as well. Finally, by  considering the poor convergence of R-plain under  different Byzantine attacks, we conclude that \texttt{Basil}  is a good solution with  fast convergence, strong Byzantine resiliency and acceptable computation overhead.

\noindent\textbf{Results (non-IID Setting)}:
 For simulating the  non-IID  setting, we sort the training data as per class, partition the sorted data into $N$ subsets, and assign each node  $1$ partition.  By applying ACDS   in the absence of Byzantine nodes while  trying  different  values for $\alpha$, we found that $\alpha = 5 \%$  gives a good performance while a small   amount of shared data from each node. Fig. 5(a) illustrates that test accuracy for R-plain in the non-IID setting can be increased by up to ${\sim}10 \%$  when each node shares only $\alpha = 5\%$ of its local data with other nodes.  Fig. 5(c), and   Fig. 5(d)   illustrate that \texttt{Basil}  on the top of ACDS with $\alpha = 5 \%$ is robust to software/hardware faults represented in Gaussian model and random sign flip. Furthermore,  both \texttt{Basil}   without ACDS  and UBAR completely fail in the presence of these faults. This is because the two defenses are using performance-based criterion which  is not meaningful in the non-IID  setting. In other words, each node  has only data from one class,  hence it becomes  unclear whether a high  loss value  for a given model can be attributed to  the Byzantine nodes, or to the very heterogeneous nature of the data.   Additionally,   R-plain with $\alpha = 0\% , 5\%$ completely fail in the presence of these faults.

Furthermore,   we can observe  in Fig. 5(b) that \texttt{Basil}  with $\alpha =0$ gives low performance. This confirms that non-IID data distribution degraded the convergence behavior. For UBAR, the performance is completely degraded, since in   UBAR   each node selects the  set of  models  which gives a lower loss than its own local  model, before using them in the update rule.  Since performance-based is not meaningful in this setting,  each node might   end up  only  with its own model. Hence, the model of each node does not completely propagate over the graph, as also demonstrated in Fig. 5(b), where UBAR fails completely. This is different from the ring setting, where the model is propagated over the ring.

 \begin{figure*}[t]
  \centering
  \subfigure[No Attack]{\includegraphics[scale=0.33]{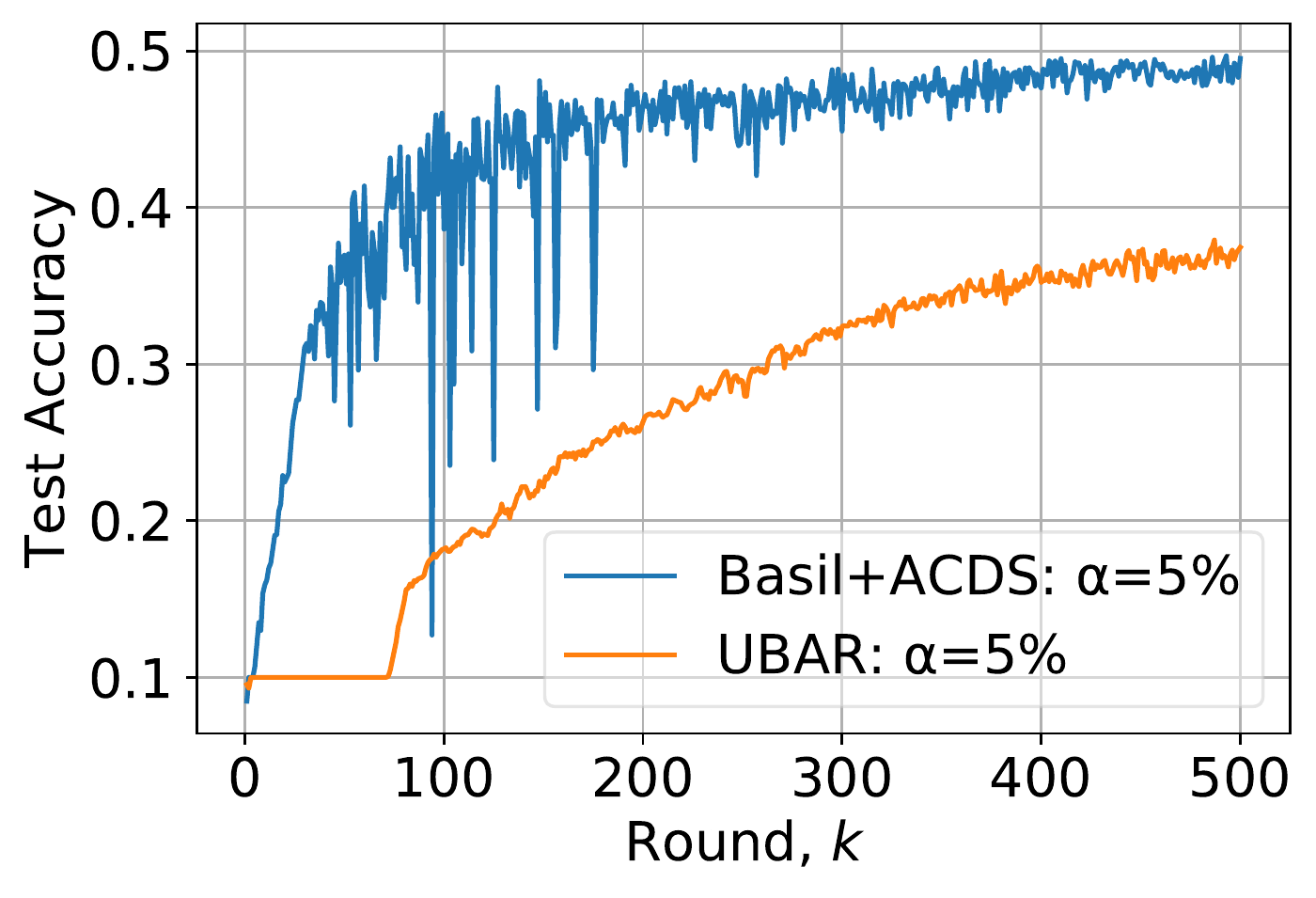}
  }
  \subfigure[Gaussian Attack]{\includegraphics[scale=0.33]{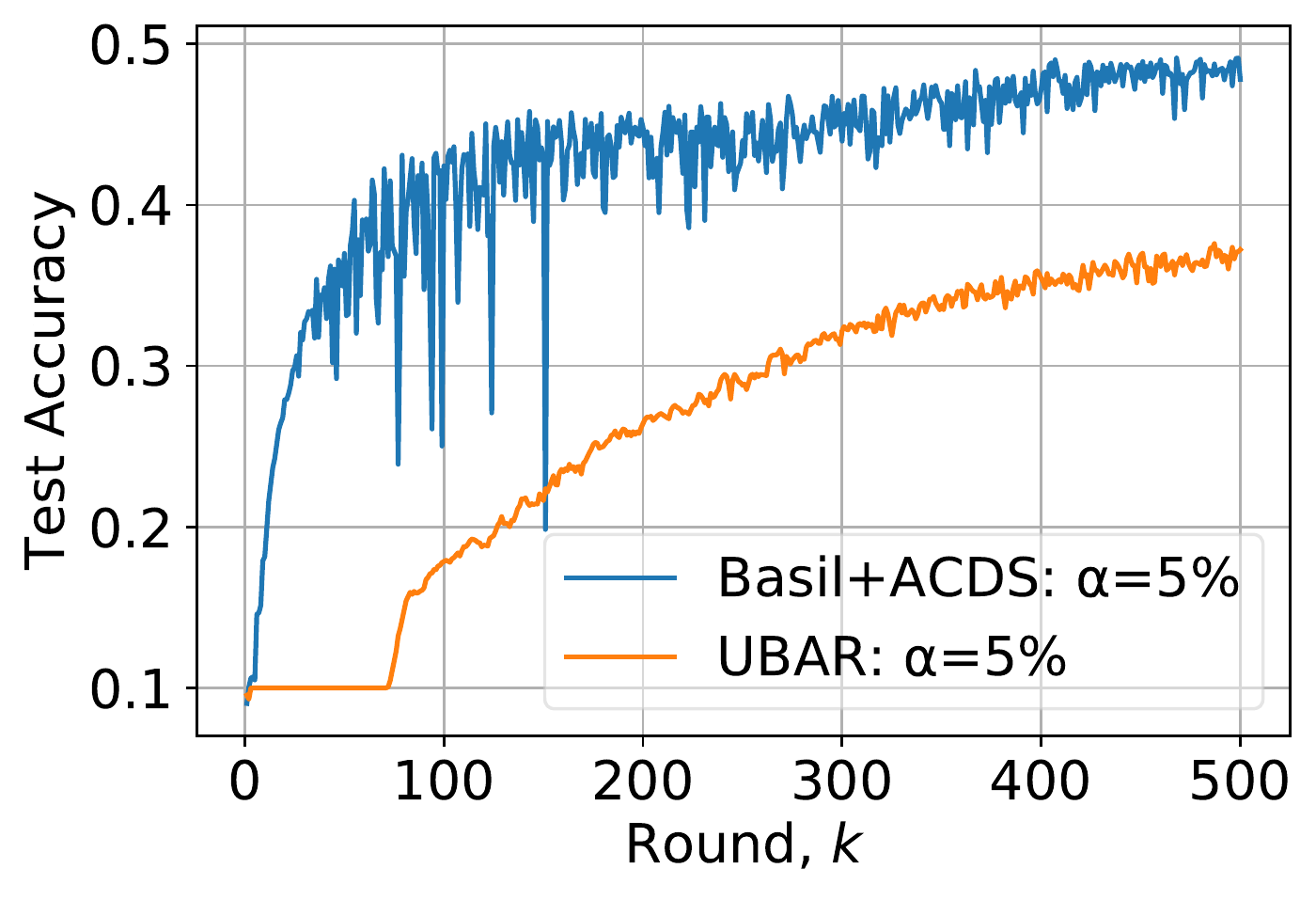}
 }%\quad
  \subfigure[Random Sign Flip Attack]{\includegraphics[scale=0.33]{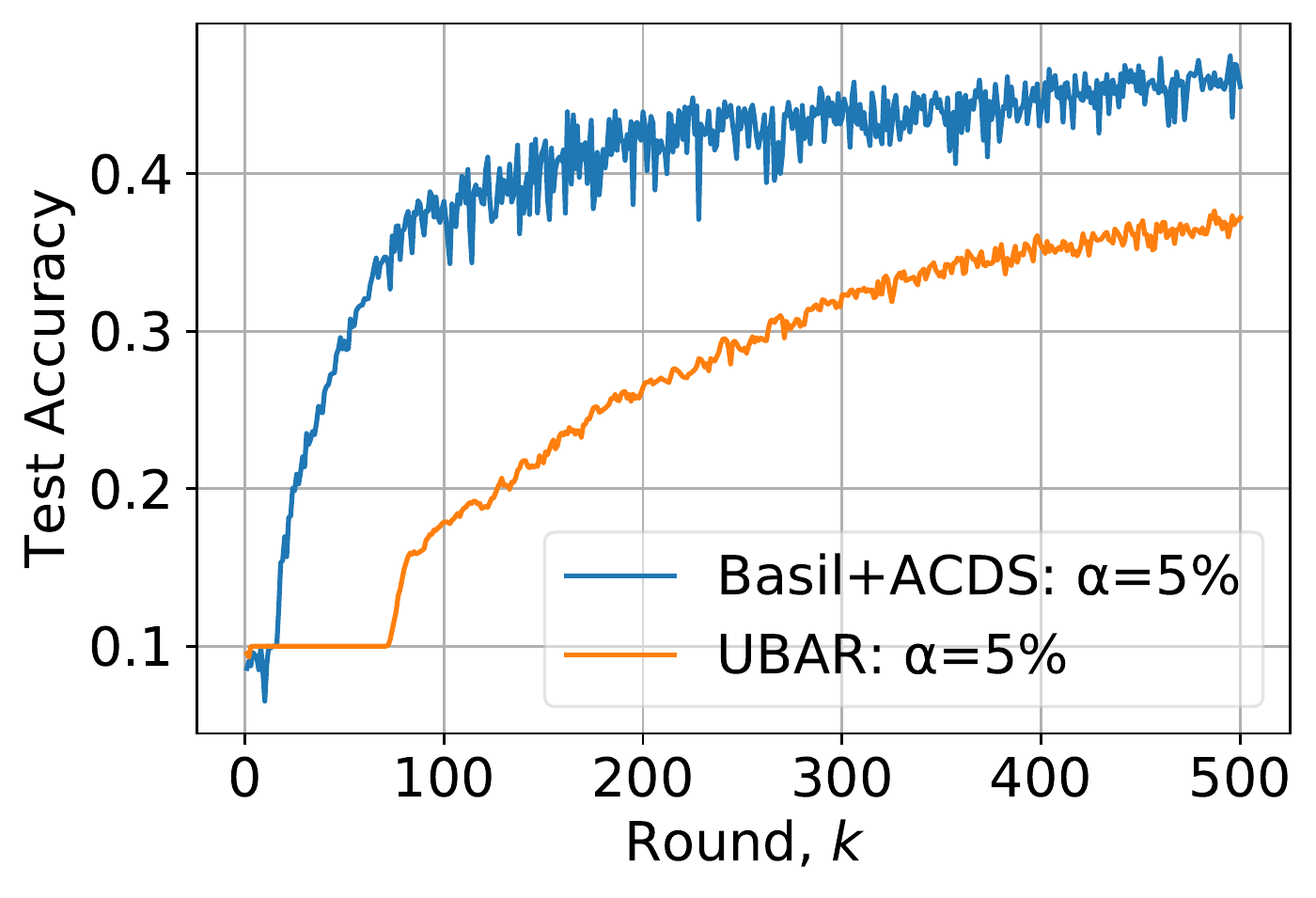}
  }
  \caption{Illustrating the performance of \texttt{Basil}   compared with UBAR  for CIFAR10 under non-IID data distribution setting  with $ \alpha = 5\%$ data sharing.}
  \label{fig:Basil VS Mozi with data sharing}
 \end{figure*}

 \begin{figure*}[t]
  \centering
  \subfigure[Varying N, (b=0.33×N, S =10)]{\includegraphics[scale=0.28]{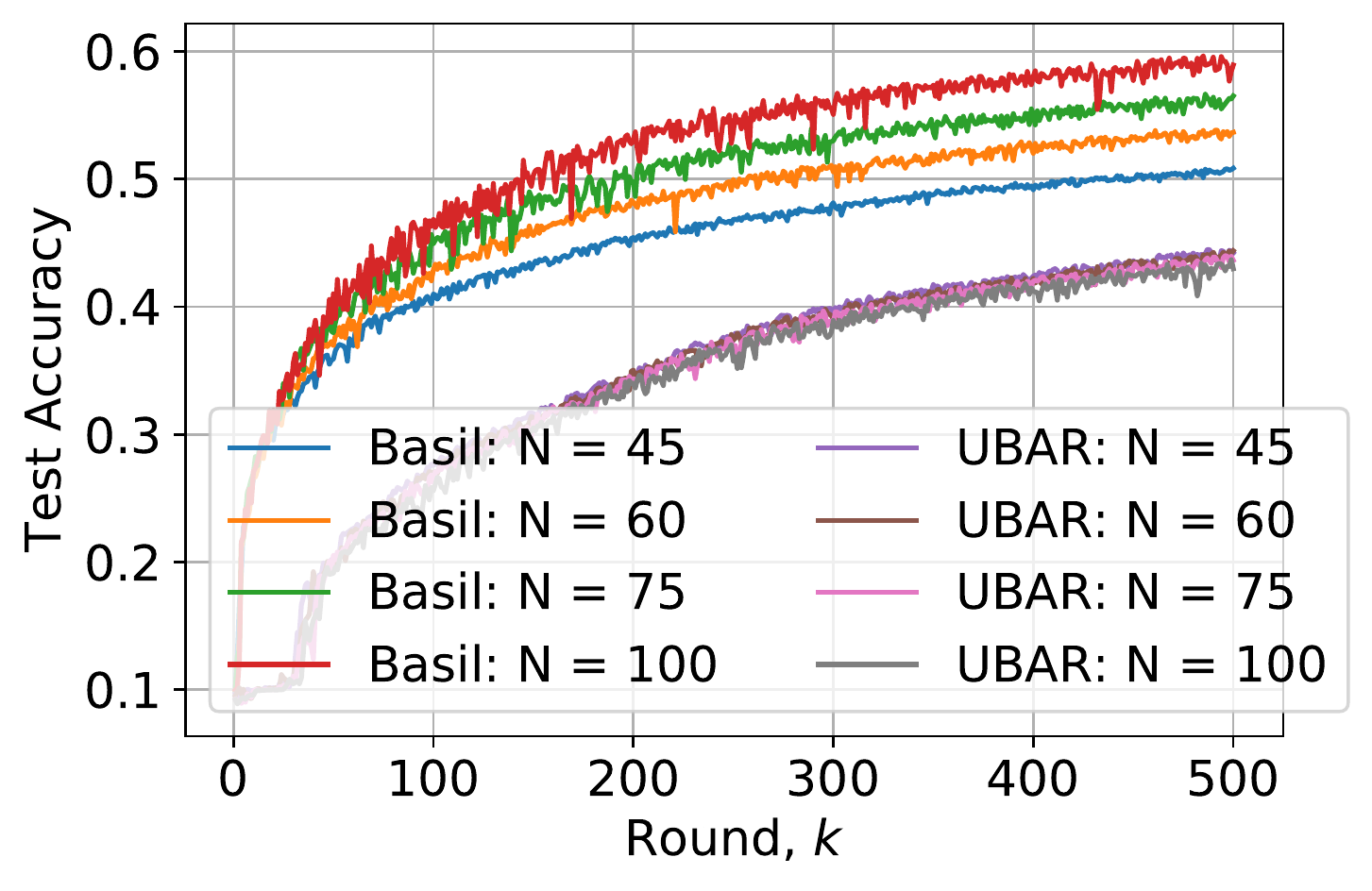}
  }
  \subfigure[Varying b, (N=100, S=10) ]{\includegraphics[scale=0.28]{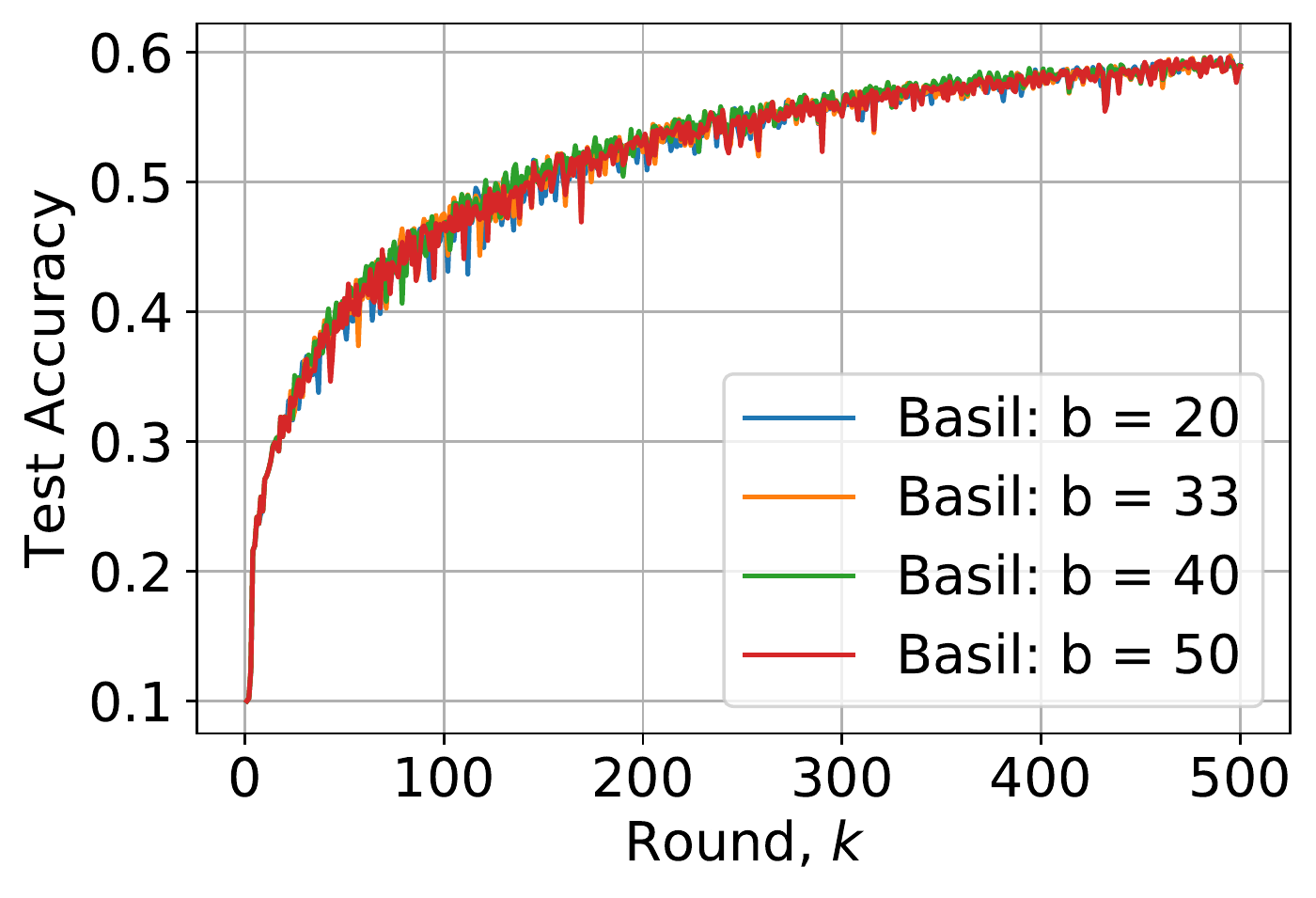}
 }%\quad
  \subfigure[Varying S, (N=100, b=33) ]{\includegraphics[scale=0.28]{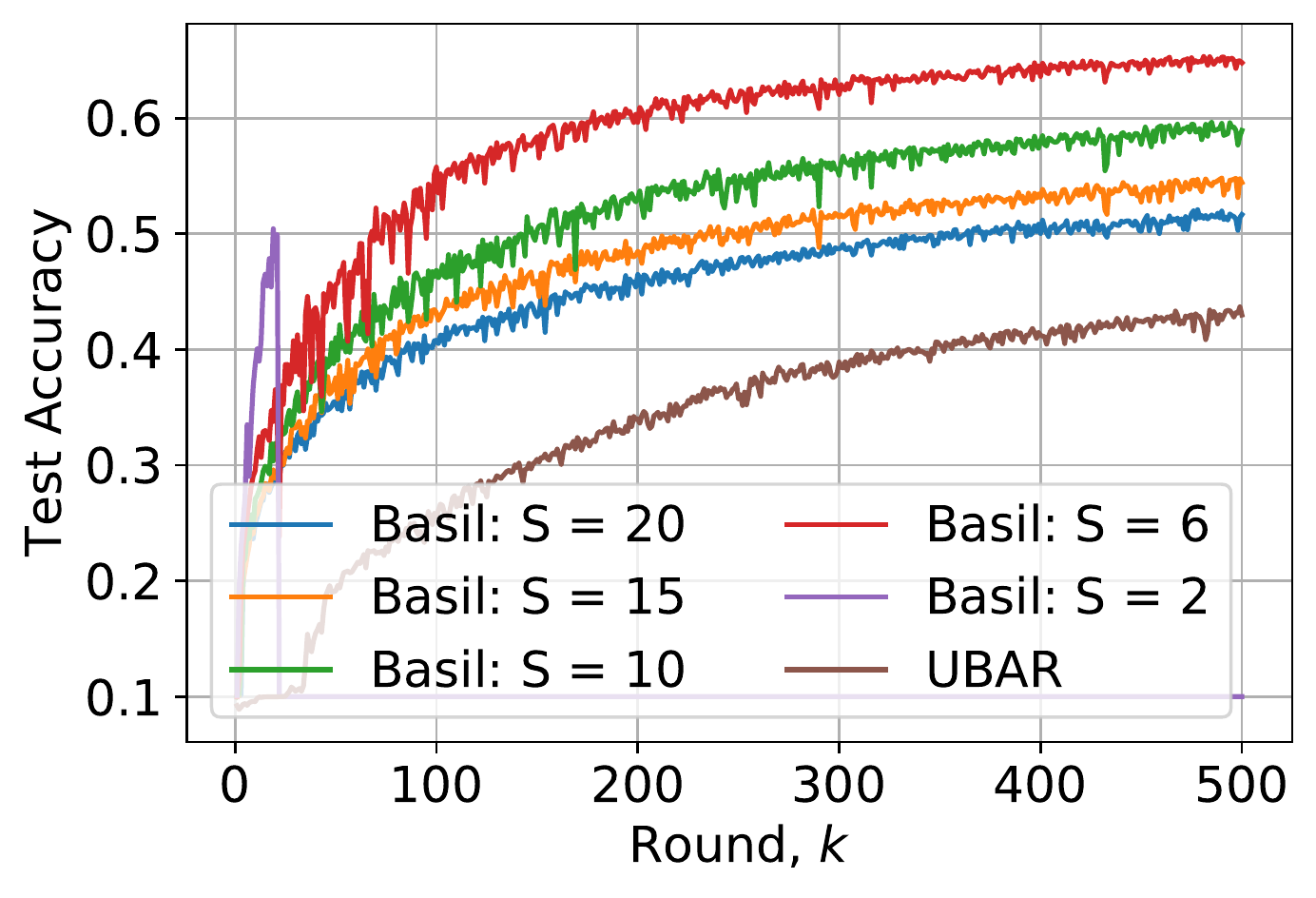}
}
  \subfigure[\texttt{Basil}  with ACDS in the non-iid setting (N=100, b=33, S=10)]  {\includegraphics[scale=0.28]{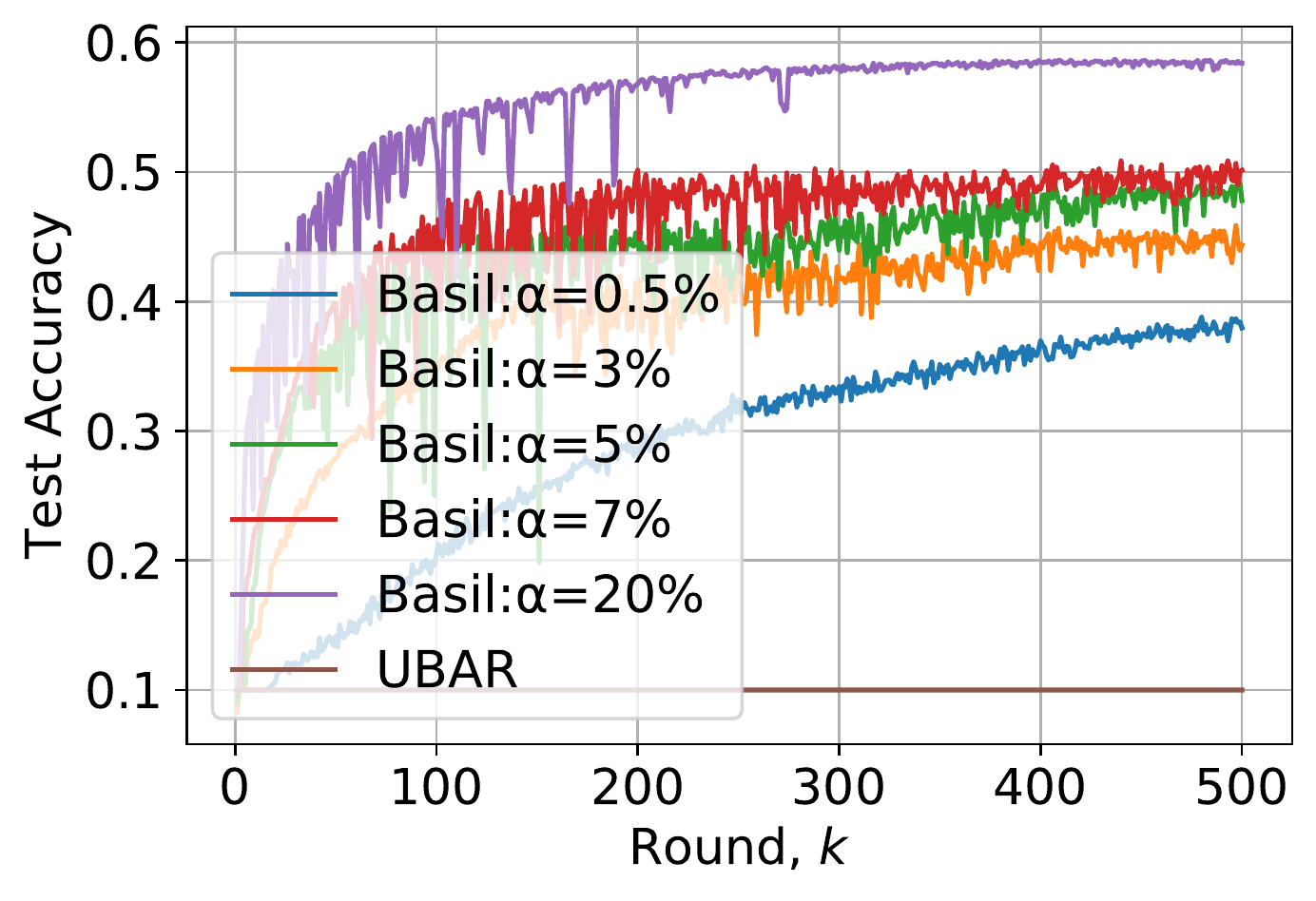}
}

  \caption{ Ablation studies. Here $N$, $b$, $S$, $\alpha$  are the total nodes, Byzantine nodes, connectivity, and fraction of shared data.  For non-IID data, Gaussian attack is considered,  while for others plots with IID  hidden attack
 is used. Using the same NNs in Section \ref{model}.}
  \label{ab}
 \end{figure*}

In Fig. 5, we showed that UBAR performs quite poorly for non-IID data setting, when no data is shared among the clients. 
We note that achieving anonymity in data sharing in  graph based decentralized learning in general and UBAR in particular  is an open problem. Nevertheless, in Fig. 6, we further show that even $5\%$ data sharing is done in UBAR, performance remains quite low in comparison to \texttt{Basil}+ACDS.

Now, we compute the communication cost overhead  due to leveraging ACDS for the experiments associated with Fig. 5. By considering the setting discussed in Remark 5  for ACDS with $G =4 $ groups for data sharing and each node sharing $\alpha =5 \% $ fraction of its local dataset, we can see from Fig. 5  that \texttt{Basil}  takes $500$ rounds to get $\sim 50 \%$ test accuracy. Hence, given that the model used in this section is of size $3.6$ Mbits ($117706$ trainable parameters each represented by $32$ bits), the communication cost overhead resulting from using ACDS for data sharing is only $4\%$.

 \iffalse{\color{blue}
In Appendix, we have performed some ablation studies to show the effect of different parameters such as number of nodes $N$, number of Byzantine nodes $B$, connectivity parameter $S$ and the fraction of data sharing $\alpha$, on the performance of Basil. }\fi

\noindent \textbf{Further ablation studies}:
 We perform ablation studies to show the effect of different parameters on the performance of \texttt{Basil}: number of nodes $N$, number of Byzantine nodes $b$, connectivity parameter $S$, and the fraction of data sharing $\alpha$. For the ablation studies corresponding to $N$, $b$, $S$, we consider the IID setting described previously, while for the $\alpha$, we consider the non-IID setting. 

Fig. 7(a) demonstrates that, unlike UBAR, \texttt{Basil} performance scales with  the number of nodes $N$. This is because in any given round, the sequential training over the logical ring topology accumulates SGD updates of clients along the logical ring, as opposed to parallel training over the graph topology in which an update from any given node only propagates to its neighbors. Hence, \texttt{Basil} has better accuracy than UBAR. Additionally, as described in Section \ref{sec7}, one can also leverage \texttt{Basil+} to achieve further scalability for large $N$ by parallelizing \texttt{Basil}. We provide the experimental evaluations corresponding to \texttt{Basil+} in Section \ref{sec:V-B}. 
  
  To study the effect of different number of Byzantine nodes in the system, we conduct experiments with different $b$. Fig. 7(b) demonstrates that \texttt{Basil} is quite robust to different number of Byzantine nodes. 
  
  Fig. 7(c) demonstrates the impact of the connectivity parameter $S$. Interestingly, the convergence rate decreases as $S$ increases. We posit that due to the noisy SGD based training process, the closest benign model is not always selected, resulting in loss of some intermediate update steps. However, decreasing $S$ too much results in larger increase in the connectivity failure probability of \texttt{Basil}. For example, the upper bound on the failure probability when $S = 6$ is less than $0.09$. However, for an extremely low value of $S=2$, we observed consistent failure  across all simulation trials, as also illustrated in Fig. 7(c).  Hence,  a careful choice of $S$ is important. 
  
  Finally, to study the relationship between privacy and accuracy when \texttt{Basil} is used alongside ACDS, we carry out numerical analysis by varying $\alpha$ in the non-IID setting described previously. Fig. 7(d) demonstrates that as $\alpha$ is increased, i.e., as the amount of shared data is increased, the convergence rate increases as well. Furthermore, we emphasize that even having $\alpha = 0.5 \%$ gives reasonable performance when data is non-IID, unlike UBAR which fails completely.

\subsection{Numerical Experiments for \texttt{Basil+}}\label{sec:V-B}

In this section, we demonstrate the achievable gains of \texttt{Basil+} in terms of  its  scalability, Byzantine robustness, and superior performance over UBAR. 

\noindent \textbf{Schemes}:
We consider  three  schemes, as described next. 
\begin{itemize}
\item  \textit{Basil+}: Our proposed scheme. 
\item \textit{R-plain+}: We implement a parallel extension of R-plain. In particular, nodes are divided into $G$ groups. Within each group, a sequential R-plain training process is carried out, wherein the current active node carries out local training using the model received from its previous counterclockwise neighbor. After $\tau$ rounds of sequential R-plain training within each group, a circular aggregation is carried out along the $G$ groups. Specifically, the model from the last node in each group gets averaged. The average model from each group is then used by the first node in each group in the next global round. This entire process is repeated for $K$ global rounds.
\end{itemize}

    %In order to have roughly same number of \textit{neighbors} as UBAR scheme, we set $S=p N$.

\noindent
\textbf{Setting}:
We use CIFAR10 dataset \cite{Krizhevsky2009LearningML} and use the neural network with $2$ convolutional layers and $3$ fully connected layers described in Section \ref{expe1}. The training dataset is partitioned uniformly among the set $\mathcal{N}$ of all nodes, where $|\mathcal{N}|=400$. We set the batch size to $80$ for local training and performance evaluation in \texttt{Basil+} as well as UBAR. Furthermore, we consider epoch based local training, where we set the number of epochs to $3$. We use a decreasing learning rate of $0.03/(1+0.03\,k)$, where $k$ denotes the global round. For all schemes, we report the average test accuracy among the benign clients. For \texttt{Basil+}, we set the connectivity parameter to $S=6$ and the number of intra-group rounds to $\tau = 1$.  The implementation of UBAR is given in Section \ref{impUBAR}. 
   \begin{figure}[h]
  \centering
  \includegraphics[scale=0.36]{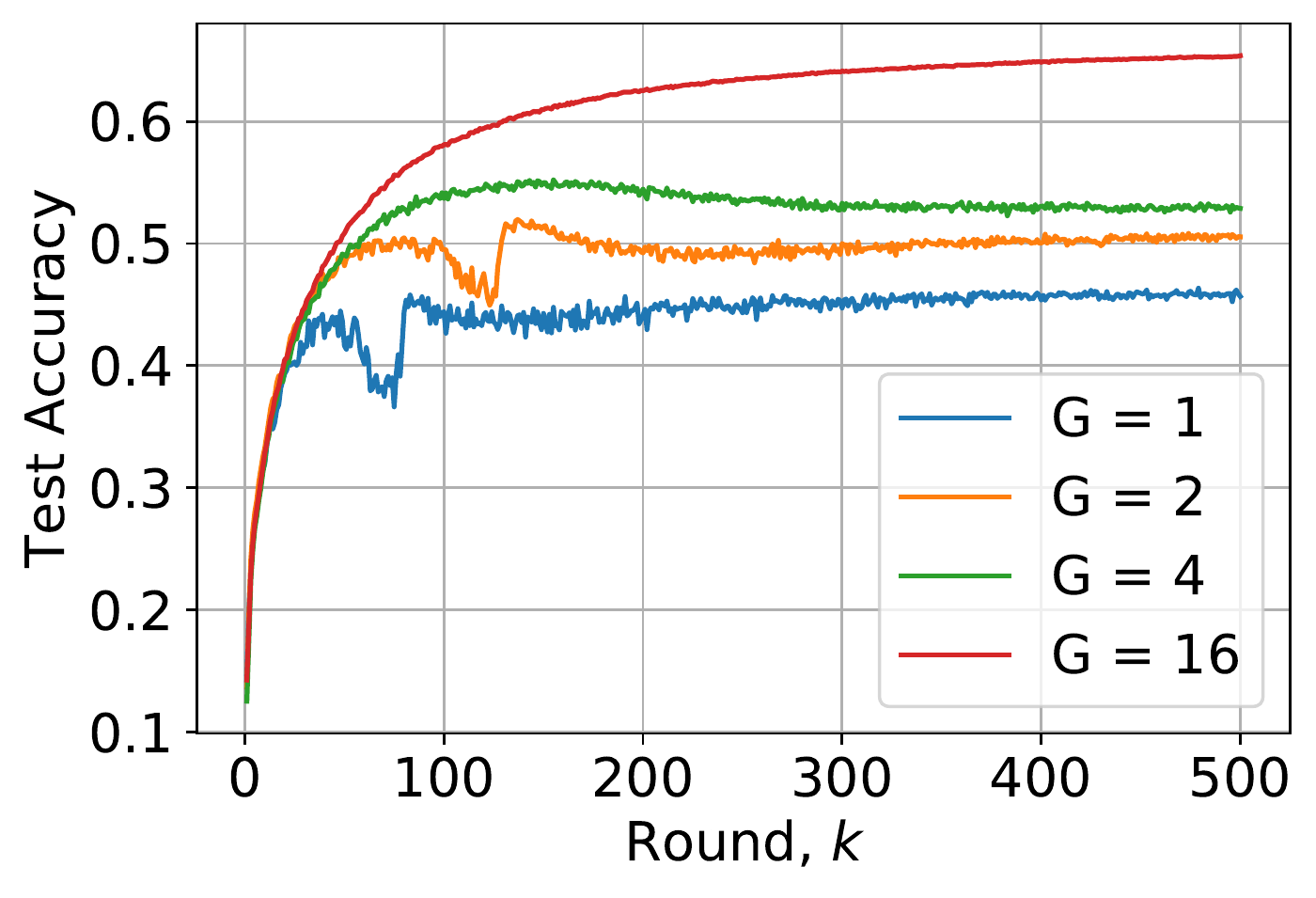}
 %\quad
  %\subfigure[Gaussian Attack]{\includegraphics[scale=0.5]{Figures_supplementary/MNIST_Mozi_5_percent_gaussian_attack}
  %\label{fig:mnist_no_byzant21}}
  \caption{The scalability gain of \texttt{Basil}+ in the presence of Byzantine nodes as number of nodes increases. Here $G$ is the number of groups, where each group has $n=25$ nodes.}
   \label{scalbility_Basil+}
 \end{figure}
 
 \begin{figure*}[t]
  \centering
  \subfigure[Number of active nodes $N_a  = 25$]{\includegraphics[scale=0.33]{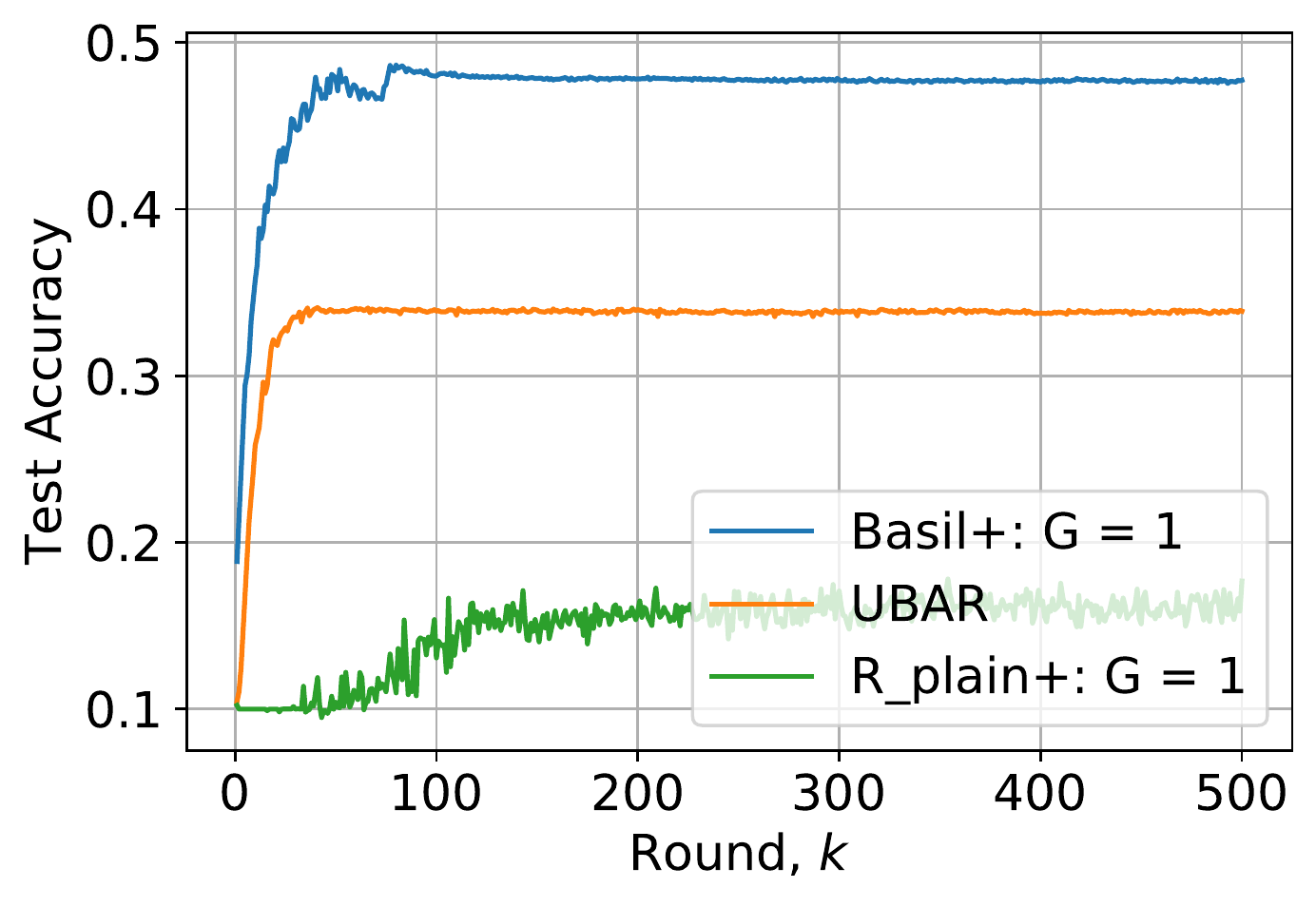}
  } \quad \quad
  \subfigure[Number of active nodes $N_a  = 50$]{\includegraphics[scale=0.33]{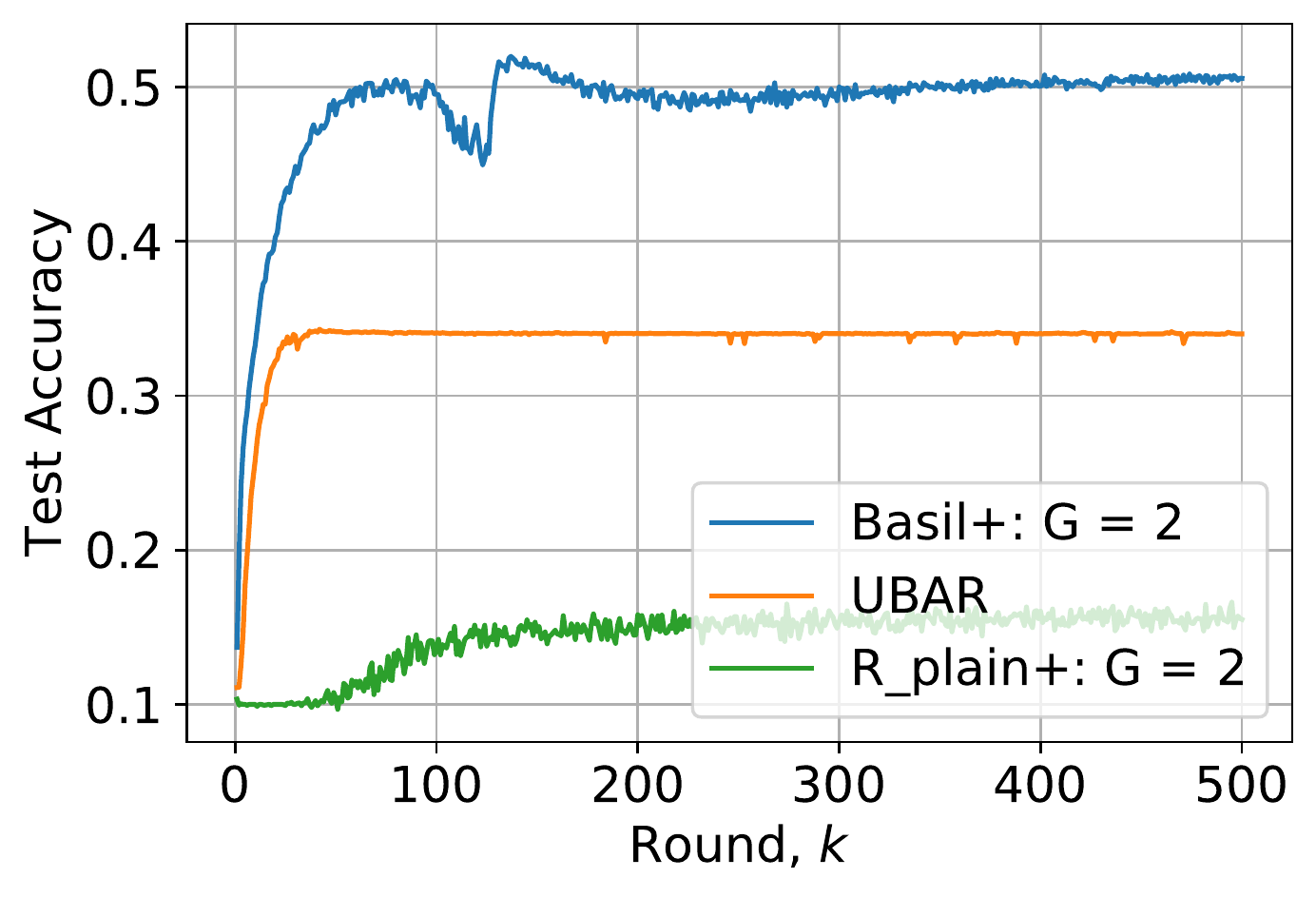}
  }\quad \quad 
  \\
  \subfigure[Number of active nodes $N_a  = 100$]{\includegraphics[scale=0.33]{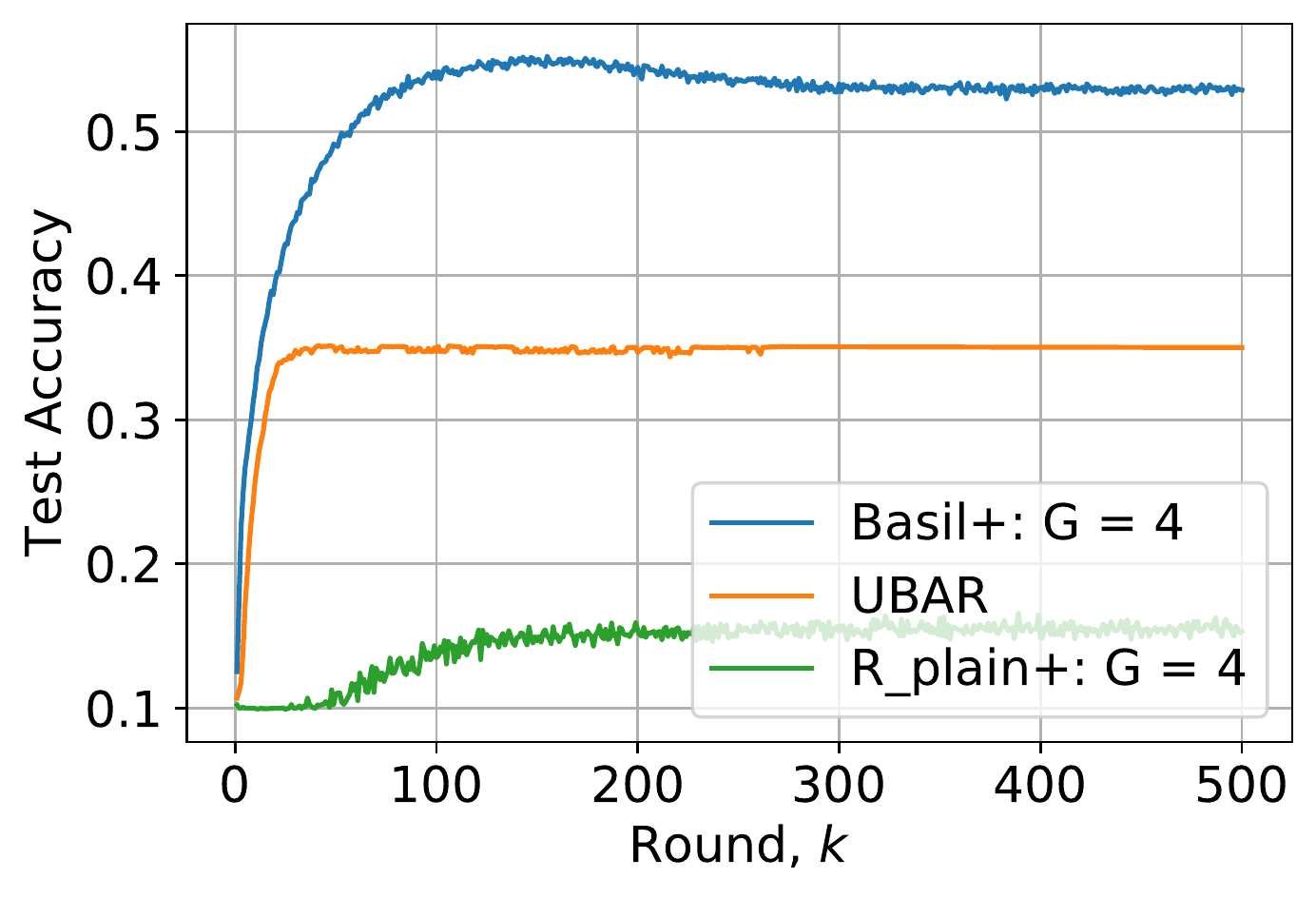}
  }\quad \quad 
  \subfigure[Number of active nodes $N_a  = 400$]{\includegraphics[scale=0.33]{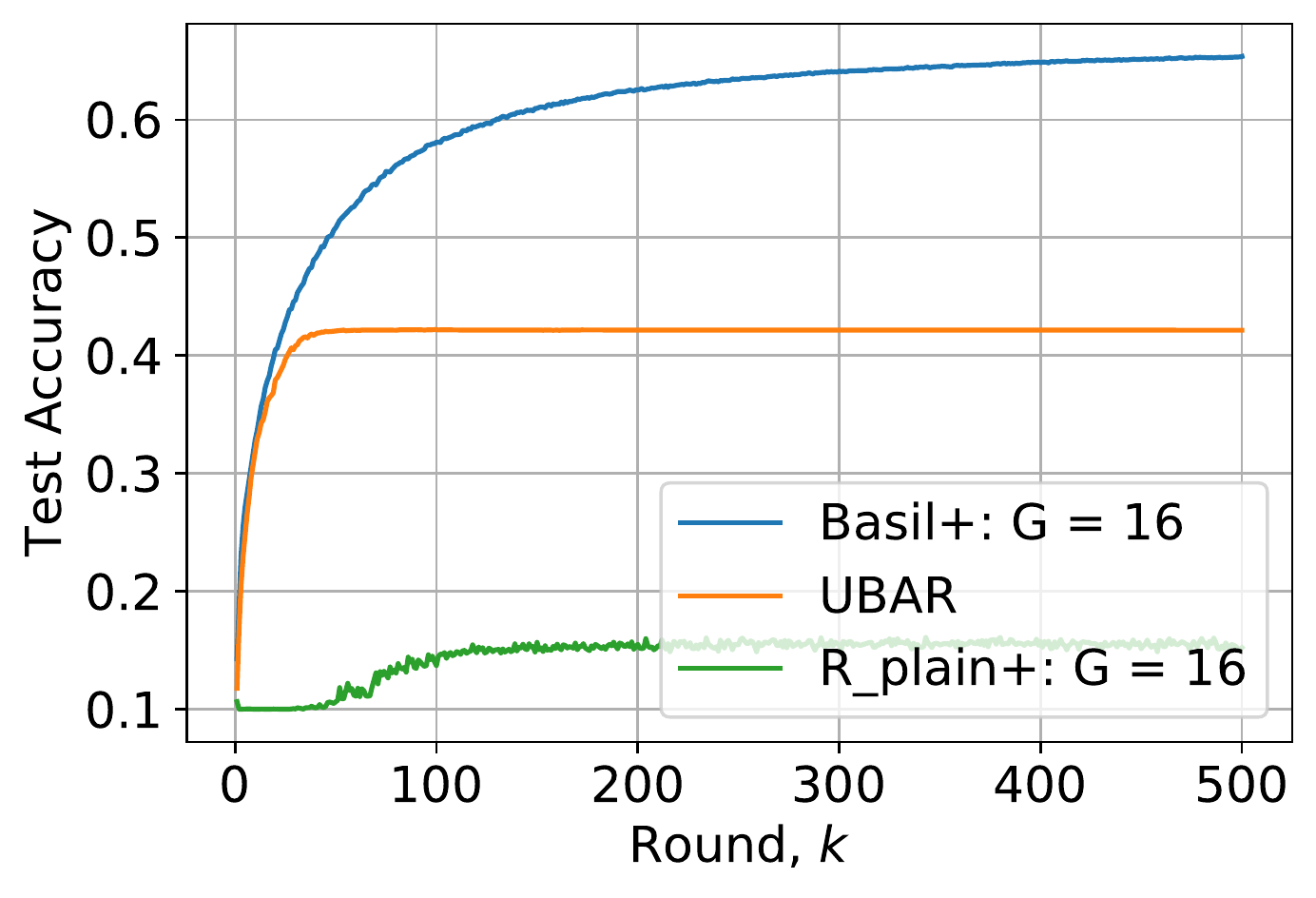}
  }\quad \quad 
  \caption{Illustrating the performance gains  of \texttt{Basil+} over  UBAR and R-plain+  for CIFAR10 dataset under different number of nodes $N_a$.}
  \label{fig:Basil VS Mozi+}
 \end{figure*}

\noindent \textbf{Results}:
For studying how the three schemes perform when more nodes participate in the training process, we consider different cases for the number of participating nodes $\mathcal{N}_a \subseteq \mathcal{N}$, where $|\mathcal{N}_a| = N_a$. Furthermore, for the three schemes, we set the total number of Byzantine nodes to be $\floor{\beta N_a}$ with $\beta = 0.2$. \iffalse Starting from having only a set of $ N_a =  25$ nodes   in the system out of the $400$ nodes, we will see how the three different schemes behaves as more nodes join the system in the presence of Byzantine nodes.    
\fi
 \iffalse
  \begin{figure*}[!htb]
  \centering
  \subfigure[]{\includegraphics[scale=0.5]{Basil+.pdf}
  \label{The scalbility gain of Basil+}}%\quad
  %\subfigure[Gaussian Attack]{\includegraphics[scale=0.5]{Figures_supplementary/MNIST_Mozi_5_percent_gaussian_attack}
  %\label{fig:mnist_no_byzant21}}
  \caption{The scalability gain of \texttt{Basil}+ in the presence of Byzantine nodes}
 \end{figure*}
\fi

Fig.  \ref{scalbility_Basil+} shows  the performance of \texttt{Basil+} in the presence of   Gaussian  attack  when the number of groups increases, i.e., when  the number of participating nodes  $N_a$ increases. Here, for all the four scenarios in Fig.  \ref{scalbility_Basil+},  we  fix the number of nodes in each group to $n=25$ nodes. As we can see  \texttt{Basil+} is able to mitigate the Byzantine behavior while achieving scalable performance as the number of nodes increases. In particular, \texttt{Basil+} with $G= 16$ groups  (i.e., $N_a = 400$ nodes) achieves a test accuracy which is higher by an absolute margin of $20\%$ compared to the case of $G = 1$ (i.e., $N_a = 25$ nodes). Additionally, while \texttt{Basil+} provides scalable model performance when the number of groups increases, the overall increase in training time scales gracefully due to parallelization of training within groups. In particular, the key difference in the training time between the two cases of $G=1$ and $G=16$ is in the stages of robust aggregation and global multicast. In order to get further insight, we set $T_{\text{comm}}$, $T_{\text{perf-based}}$ and $T_{\text{SGD}}$ in equation (29) to one unit of time. Hence, using (29), one can show that the ratio of the training time for $G=16$ with $N_a = 400$ nodes to the training time for $G=1$ with $N_a = 25$ nodes is just $1.5$. This further demonstrates the scalability of \texttt{Basil+}.

In Fig. \ref{fig:Basil VS Mozi+}, we compare the performance of the three schemes for different numbers of participating nodes $N_a$. In particular, in both \texttt{Basil+} and  {R-plain+}, we have $N_a=25\,G$, where $G$ denotes the number of groups, while for UBAR, we consider a parallel graph with $N_a$ nodes. As can be observed from Fig.  \ref{fig:Basil VS Mozi+}, \texttt{Basil+} is not only robust to the Byzantine nodes, but also gives superior performance over UBAR in all the 4 cases shown in Fig.  \ref{fig:Basil VS Mozi+}. The key reason of \texttt{Basil+} having higher performance than UBAR is that training in \texttt{Basil+} includes sequential training over logical rings within groups, which has better performance than graph based decentralized training.

 \section{Conclusion and Future Directions}
We propose \texttt{Basil}, a fast and computationally efficient Byzantine-robust  algorithm for decentralized training over a logical ring. We provide the theoretical convergence guarantees of \texttt{Basil} demonstrating its    linear convergence rate. Our experimental results in the IID setting show the superiority of \texttt{Basil}  over the state-of-the-art algorithm for decentralized training.  Furthermore, we generalize \texttt{Basil}  to the non-IID setting by integrating it with   our proposed Anonymous Cyclic Data Sharing (ACDS) scheme. Finally, we propose \texttt{Basil+} that enables a parallel implementation of \texttt{Basil}  achieving further scalability.   

One interesting future direction is to explore some techniques such as data compression or data placement and coded shuffling to reduce the communication cost resulting from using ACDS. Additionally, it is interesting to see how  some  differential privacy (DP) methods can be adopted by  adding  noise to the shared data in ACDS to provide further privacy while studying the impact of the added noise in the overall training performance.

\iffalse
Note that, when the nodes  even  do only one epoch under the assumption that the data size is different among the nodes, in this case, how to design the algorithm based on that given that the nodes may declare different data size. 
refer to model averaging.

These include crashes and computation errors,
stalled processes, biases in the way the data samples are distributed among the processes,
Due to the inherent unpredictability of this abnormal (sometimes adversarial) behavior, it is typically
modeled as Byzantine failure (Lamport et al., 1982), meaning that some worker machines may behave
completely arbitrarily and can send any message to the master machine that maintains and updates
an estimate of the parameter vector to be learned.
\fi

\bibliographystyle{IEEEtran}
\bibliography{references}

\begin{appendices}
\section*{Overview}
In the following, we summarize the content of the appendix.
\begin{itemize}
 \item In Appendix \ref{Appendix A}, we   prove Propositions 1 and 2.
 \item In Appendix \ref{Convergence}, we prove  the convergence guarantees of \texttt{Basil}.
 \item In Appendix \ref{drop}, we describe how \texttt{Basil} can be robust to nodes dropout.
 \item In Appendix D, Appendix  \ref{prop3} and Appendix \ref{prop4}, the proofs of Propositions 3,  4 and 5 are presented.  
\item  In Appendix \ref{UBAR1}, we    present UBAR \cite{guo2020byzantineresilient}, the recent  Byzantine-robust decentralized algorithm.
\item  In Appendix \ref{expe}, we  provide  additional experiments.
\end{itemize}

 \section{Proof of Proposition 1 and Proposition 2 }\label{Appendix A}
 
  \noindent \textbf{Proposition 1.}  \textit{ The communication,  computation, and storage  complexities  of \texttt{Basil}  algorithm  are all $\mathcal{O} (Sd)$  for each node in each iteration, where $d$ is the model size. }

\textit{Proof.} Each node receives and stores the latest $S$ models, calculates the loss by using each model out of the $S$ stored models, and multicasts its updated model to the next $S$ clockwise neighbors. Thus, this results in $\mathcal{O} (Sd)$ communication, computation, and storage costs.  $\hfill \square$

  \noindent \textbf{Proposition 2.}\textit{ The number of models that each benign node needs to receive, store and evaluate from its counterclockwise neighbors for ensuring the  connectivity  and  success of \texttt{Basil} can be relaxed to $S<b+1$ while guaranteeing the success of \texttt{Basil} (benign subgraph connectivity)  with high probability.}

\textit{Proof.}
This can be proven by showing 
 that the benign subgraph, which is generated by removing the Byzantine nodes, is connected with high probability when each node multicasts its updated model to the next $S<b+1$ clockwise neighbors instead of   $b+1$ neighbors. Connectivity of the benign subgraph is important as it ensures that each benign node can still receive information from a few other non-faulty nodes. Hence, by letting each node store and evaluate the latest $S$  model updates,   this  ensures that each  benign node has the chance to select one of the benign updates. 

%can relax the number of models that each node stores and evaluates in each round to be $S$ instead of being $b+1$, where $S<b+1$, while guaranteeing that  each benign node has at least one benign within its  set of $S$ clockwise neighbors.
%\color{blue}{ where the second equality }follows from the definition of factorial
More formally, when each node multicasts its  model to the next $S$ clockwise neighbors, we define    $A_j$ to be the failure event in which $S$ Byzantine nodes come in a row where $j$ is the starting node of these $S$ nodes. When $A_j$ occurs, there is at least one pair of benign nodes that have no link between them. The probability of $A_j$ is given as follows: 
\begin{equation}
\mathbb {P}(A_j) =  \prod_{i=0 }^{S-1} \frac{ (b-i) }{(N-i) } = \frac{b! (N-S)! }{(b-S)! N!} \label{Prob}, 
\end{equation}
 where the second equality follows from the definition of factorial, while  $b$,  and $N$ are the      number of Byzantine nodes and the total number of nodes  in the system, respectively. Thus, the probability of having a disconnected benign subgraph in \texttt{Basil}, i.e., $S$ Byzantine nodes coming in a row, is given as follows: 
\begin{align}
\mathbb {P}(\text{Failure}) = \mathbb {P}( \bigcup_{j=1}^{N}(A_j)) 
 \overset {(a)}\leq \sum_{j=1}^{N}\mathbb {P}(A_j)
\overset {(b)}= \frac{b! (N-S)! }{(b-S)! (N-1)!} \label{Failure}, 
\end{align}
where (a) follows from the union bound and (b) follows from \eqref{Prob}.  $\hfill \square$

 \section{Convergence Analysis}\label{Convergence}
 In this section, we prove the two theorems  presented in Section III.B in the main paper. We  start the proofs by stating  the  main assumptions and  the update rule  of \texttt{Basil}.

\textbf{Assumption 1} (IID data distribution). \textit{ Local dataset $\mathcal{Z}_i$ at node $i$ consists of IID data samples from a distribution $\mathcal{P}_i$, where $\mathcal{P}_i =  \mathcal{P} $ for $ i \in \mathcal{R}$. In other words, $f_i(\mathbf{x}) =  \mathbb{E}_{\zeta_i \sim	 \mathcal{P}_i}[ l(\mathbf{x}, \zeta_i) ] =  \mathbb{E}_{\zeta_j \sim	 \mathcal{P}_j}[ l(\mathbf{x}, \zeta_i) ] = f_j(\mathbf{x})   \,\forall i, j \in \mathcal{R} $. Hence,  the global loss function  $f(\mathbf{x}) =   \mathbb{E}_{\zeta_i \sim	 \mathcal{P}_i}[ l(\mathbf{x}, \zeta_i )]$. }

\textbf{Assumption 2 }(Bounded variance). \textit{ Stochastic gradient $g_i( \mathbf{x})$ is unbiased and variance bounded, i.e., $\mathbb{E}_{\mathcal{P}_i} [ g_i( \mathbf{x})] = \nabla f_i(\mathbf{x}) =    \nabla f(\mathbf{x}) $, and $\mathbb{E}_{\mathcal{P}_i} || g_i( \mathbf{x}) -  \nabla f_i(\mathbf{x})||^2  \leq \sigma^2 $, where $g_i({\mathbf{x}}) $ is  the stochastic gradient computed by node $i$ by using  a  random sample $\zeta_i$  from its local dataset $\mathcal{Z}_i$. }  

\textbf{Assumption 3} (Smoothness\iffalse L-smooth and twice differentiable loss function\fi).  \textit{ The loss functions $f_i's$ are L-smooth and twice differentiable,  i.e., for any $\mathbf{x} \in \mathbb{ R}^d$, we have $||\nabla^2 f_i(\mathbf{x})||_2 \leq  L$.  }

Let $b^i$  be the number of   Byzantine nodes out of the $S$ counterclockwise neighbors   of node $i$. We divide the set of stored models  $\mathcal{N}_k^{i}$ at   node $i$ in the $k$-th round into two sets.  The first set   $\mathcal{G}^{i}_{k} = \{ \mathbf{y}_1, \dots, \mathbf{y}_{r^{i}}\} $ contains the benign models,  where $ r^i = S - b^i$. We consider scenarios with $S= b+1$, where $b$ is the total number of Byzantine nodes in the network. Without loss of generality, we assume the models in this set are arranged such that the first model is from the closest benign node in the neighborhood of node $i$, while the last model is from the farthest node. Similarly, we define the second set $\mathcal{B}^i_{k}$ to be the set of models from the counterclockwise Byzantine neighbors of node $i$  such that $\mathcal{B}_{k}^{i} \cup \mathcal{G}_k^{i} = \mathcal{N}_k^{i}$.
 
 The general update rule in \texttt{Basil}   is given as follows.  At the $k$-th  round, the current active node $i$ updates the global model according to the following rule:
\begin{align}
\label{Update}
\mathbf{x}^{(i)}_{k}=&  \bar {\mathbf{x}}^{(i)}_{k}-\eta_{k}^{(i)}   g_i(\bar {\mathbf{x}}^{(i)}_{k}),
\end{align} 
where $\bar {\mathbf{x}}^{(i)}_{k}$ is given as follows
\begin{equation}\label{AA}
\bar {\mathbf{x}}_k^{(i)} = \arg \min_{ \mathbf{y} \in \mathcal {N}_{k}^i }  \mathbb{E} \left[{l}_i({\mathbf{y}}, \zeta_i)\right].
\end{equation}
\subsection{Proof of Theorem 1}
We first  show  that if 
  node $i$ completed the performance-based criteria in \eqref{AA} and selected   the model $\mathbf{y}_1 \in \mathcal{G}^{i}_{k}$,  and updated  its model  as follows:
\begin{equation}\label{5555}
\mathbf{x}_k^{(i)} =  \mathbf{y}_1 - \eta_{k}^{(i)}  g_i(\mathbf{y}_1), 
\end{equation}
we will have 
\begin{equation}\label{133}
\mathbb{E} \left[ {\ell}_{i+1} (\mathbf{x}_{k}^{(i)}) \right]    \leq \mathbb{E} \left[ {\ell}_{i+1} (\mathbf{y}_1) \right],  
\end{equation}
where ${\ell}_{i+1} (\mathbf{y}_1) = {\ell}_{i+1}(\mathbf{y}_1,\zeta_{i+1}) $ is the loss function of node $i+1$ evaluated on a random sample $\zeta_{i+1}$ by using the model $\mathbf{y}_1$. 

The proof of \eqref{133} is as follows: 
By using   Taylor’s theorem, there exists a  $\gamma$ such that
\begin{align}
{\ell}_{i+1}({\mathbf{x}_k^{(i)}} )  =&  {\ell}_{i+1}\left( \mathbf{y}_1 - \eta_{k}^{(i)}  g_i(\mathbf{y}_1) \right)\\ =& {\ell}_{i+1} (\mathbf{y}_1) -  \eta_{k}^{(i)}  g_i(\mathbf{y}_1) ^T   g_{i+1}({\mathbf{y}_1}) \nonumber \\&+\frac{1}{2}  \eta_{k}^{(i)}  g_i(\mathbf{y}_1) ^T  \nabla^2 {\ell}_{i+1}({\gamma} ) \eta_{k}^{(i)}  g_i(\mathbf{y}_1),
\end{align}
%\\ \leq& {\ell}_i^{(k+1)} ( \mathbf{x}^{k-1}) - \eta_{k}^{(i)}  g_k(\mathbf{x}^{k-1}) g_{(k+1)}(\mathbf{x}^{k-1}) +\frac{ \eta_{k}^{(i)}^2 L}{2} ||g_k(\mathbf{x}^{k-1})||^2
where   $\nabla^2 {\ell}_{i+1} $ is the stochastic Hessian matrix. By using  the following  assumption 
\begin{equation}
|| \nabla^2 {\ell}_{i+1}({\mathbf{x}_k^{(i)}} )  ||_2 \leq L ,  
\end{equation} \text{ for all random  samples $\zeta_{i+1}$ and  any  model $ \mathbf{x} \in \mathbb{R}^d$},  where  $L$ is the Lipschitz constant, we get 
\begin{align}
{\ell}_{i+1}({\mathbf{x}_k^{(i)}} ) \leq &  {\ell}_{i+1} (\mathbf{y}_1) -  \eta_{k}^{(i)}   g_i(\mathbf{y}_1) ^T  g_{i+1}({\mathbf{y}_1}) \nonumber \\ +&\frac{ (\eta_{k}^{(i)})^2 L}{2} ||g_i(\mathbf{y}_1)||^2. 
\end{align}
By taking the expected value of both sides of this expression (where
the expectation is taken over the randomness in the sample selection), we get 
\begin{align}
&\mathbb{E} \left[{\ell}_{i+1}({\mathbf{x}_k^{(i)}} )  \right] \leq  \mathbb{E} \left[ {\ell}_{i+1} (\mathbf{y}_1) \right] - \eta_{k}^{(i)}  \mathbb{E} \left[g_i(\mathbf{y}_1)^T    g_{i+1}({\mathbf{y}_1})\right] \nonumber \\ +&\frac{ (\eta_{k}^{(i)})^2 L}{2} \mathbb{E}||g_i(\mathbf{y}_1)||^2 \nonumber \\   \overset{a}=  & \mathbb{E} \left[ {\ell}_{i+1} (\mathbf{y}_1) \right]  - \eta_{k}^{(i)}   \mathbb{E}\left[g_i(\mathbf{y}_1)^T\right]  \mathbb{E} \left[g_{i+1}({\mathbf{y}_1})\right]  \nonumber \\ +&\frac{ (\eta_{k}^{(i)})^2L}{2} \mathbb{E}||g_i(\mathbf{y}_1)||^2 \nonumber \\ \leq &  \mathbb{E} \left[ {\ell}_{i+1} (\mathbf{y}_1) \right]  - \eta_{k}^{(i)}   \mathbb{E}\left[g_i(\mathbf{y}_1)^T\right]  \mathbb{E} \left[g_{i+1}({\mathbf{y}_1})\right]   \nonumber \\+& (\eta_{k}^{(i)})^2L \mathbb{E}||g_i(\mathbf{y}_1)||^2 \nonumber \\
\overset{b} \leq &  \mathbb{E} \left[ {\ell}_{i+1} (\mathbf{y}_1) \right] - \eta_{k}^{(i)}  ||\nabla f(\mathbf{y}_1)||^2 +(\eta_{k}^{(i)})^2 L ||\nabla f(\mathbf{y}_1)||^2   \nonumber \\+&  (\eta_{k}^{(i)})^2 L\sigma^2 \nonumber \\ 
  =  &  \mathbb{E} \left[ {\ell}_{i+1} (\mathbf{y}_1) \right]   {-}  ||\nabla f(\mathbf{y}_1)||^2 \left( \eta_{k}^{(i)} {-}  (\eta_{k}^{(i)})^2 L \right)  + (\eta_{k}^{(i)})^2 L \sigma^2,
\end{align}
where (a) follows from that the samples are drawn from independent data distribution, while (b) from    Assumption 1 along with    
\begin{align}\label{e13}
\mathbb{E}||g_i(\mathbf{y}_1)||^2 = & \mathbb{E}||g_i(\mathbf{y}_1) - \mathbb{E} \left[g_i(\mathbf{y}_1)\right]||^2  +  ||  \mathbb{E} [g_i(\mathbf{y}_1)]||^2 \nonumber \\  \leq &  \sigma^2 + ||  \nabla f(\mathbf{y}_1)||^2. 
\end{align}
 Let $ C_k^i = \frac{|| \nabla f(\mathbf{y}_1)||^2 }{|| \nabla f(\mathbf{y}_1)||^2 + \sigma^2} = \frac{||   \mathbb{E} [ g_{i} (\mathbf{y}_1 ] ||^2 }{||  \mathbb{E} [g_{i} (\mathbf{y}_1 ]||^2  + \sigma^2} $,   which implies that   $C_k^i \in [0,1]$.    By selecting the learning rate as $\eta_{k}^{(i)} \geq \frac{1}{L} C_k^i$,  we get 
\begin{equation}\label{13}
\mathbb{E} \left[ l_{i+1} (\mathbf{y}_1) \right]    \leq \mathbb{E} \left[ l_{i+1}(\mathbf{y}_1) \right]. 
\end{equation}

Note that, nodes can just use a learning rate $\eta \geq \frac{1}{L}$, since    $C_k^i\in [0,1]$, while still achieving \eqref{13}. This completes the first part of the proof.

By using \eqref{13}, it can be easily seen that the update rule in equation \eqref{Update}  can be  reduced to the case where  each node   updates its model based on the model received  from the closest benign node \eqref{5555} in its neighborhood, where this follows from using  induction.  

Let's consider this example. Consider a ring with $N$ nodes and by using $S=3$ while ignoring the Byzantine nodes for a while (assume all nodes are benign nodes).  We   consider the first   round $k=1$.   With a little abuse of notations, we can get the following, the updated model by  node 1 (the first node in the ring) $\mathbf{x}_1 = h (\mathbf{x}_0)$ is a function of the initial model $\mathbf{x}_0$ (updated by using the model $\mathbf{x}_0$). Now, node 2 has to select one model from the set of two models  $\mathcal{N}_k^2 = \{\mathbf{x}_1= h (\mathbf{x}_0), \mathbf{x}_0\}$. The selection is performed   by evaluating the expected loss function of node 2 by using the criteria given in \eqref{AA} on the models on the set $\mathcal{N}_k^2$. According to \eqref{13}, node 2 will select  the model $\mathbf{x}_1$ which results in lower expected loss. Now, node 2 updates its model based on the model $\mathbf{x}_1$, i.e.,   $\mathbf{x}_2= h (\mathbf{x_1})$.  After that, node 3 applies the aggregation rule in \eqref{AA} to selects one model from this set of models  $\mathcal{N}_k^3 =  \{\mathbf{x}_2= h (\mathbf{x_1}), \mathbf{x}_1= h(\mathbf{x_0}),  \mathbf{x}_0\}$.  By using \eqref{13} and Assumption 1, we get 
\begin{align}
\mathbb{E} [ l_{3} (\mathbf{x}_2) ] \leq  \mathbb{E} [ l_{3} (\mathbf{x}_1) ]     \leq \mathbb{E} [ l_{3} (\mathbf{x}_{0}) ],  \label{155}
\end{align}
and node 3 model will be updated according to the model  $\mathbf{x}_2$, i.e.,   $\mathbf{x}_3= h (\mathbf{x}_2)$. 

More generally, the set of stored benign  models  at node $i$   is given by $\mathcal{N}_k^{i} =  \{\mathbf{y}_1= h (\mathbf{y}_2), \mathbf{y}_2= h (\mathbf{y}_3), \dots,   \mathbf{y}_{r^{i}} = h( \mathbf{y}_{r^{i}-1})\}$, where $r^{i}$ is the number of  benign  models in the set $\mathcal{N}_k^{i}$. According to   \eqref{13},  we will have the following \begin{align}\label{222}
\mathbb{E} \left[ l_{i} (\mathbf{y}_1) \right] {\leq}  \mathbb{E} \left[ l_{i} (\mathbf{y}_2) \right]     {\leq}  \dots \leq &\mathbb{E} \left[ l_{i} (\mathbf{y}_{r^{i}}) \right] {\leq} \mathbb{E} \left[ l_{i} (\mathbf{x}) \right] \;  \forall \mathbf{x} \in \mathcal{B}_{k}^{i}, 
\end{align}
where the last inequality in \eqref{222} follows from the fact that the Byzantine nodes are sending faulty models and their expected loss is supposed to be  higher than the expected loss of the benign nodes. 

According  to this discussion  and  by removing the Byzantine nodes thanks to \eqref{222}, we can only consider the benign subgraph  which is generated by removing the Byzantine nodes according to the discussion  in Section III-A in the main paper. Note that by letting each active  node send its updated model to the next $b+1$ nodes, where $b$ is the total number of Byzantine nodes,  the  benign subgraph  can always be connected. By considering the benign subgraph (the logical rings without Byzantine nodes), we   assume without  loss of generality that  the indices of benign  nodes in the ring are arranged in ascending order starting from node 1 to node $r$.  In this benign subgraph, the update rule will be given as follows  
\begin{equation}\label{177}
\mathbf{x}_k^{(i)} =  \mathbf{x}_k^{(i-1)}- \eta_{k}^{(i)}  g_i(\mathbf{x}_k^{(i-1)}). 
\end{equation}

\subsection{Proof of Theorem 2}
By using   Taylor’s theorem,  there  exists a  $\gamma$ such that 
\begin{align}
&{f}({\mathbf{x}_k^{(i+1)}} )  \overset{a}=  {f}\left( \mathbf{x}_{k}^{(i)} - \eta_{k}^{(i+1)}  g_{i+1}({\mathbf{x}}^{(i)}_{k}) \right) \nonumber \\ &= {f} (\mathbf{x}_{k}^{(i)}) -  \eta_{k}^{(i+1)} \left( g_{i+1}({\mathbf{x}}^{(i)}_{k})\right)^T   \nabla{f}({\mathbf{x}_k^{(i)}} )  \nonumber \\ &+\frac{1}{2}  \eta_{k}^{(i+1)} \left( g_{i+1}({\mathbf{x}}^{(i)}_{k})\right)^T   \nabla^2 {f}({\gamma} ) \eta_{k}^{(i+1)}  g_{i+1}({\mathbf{x}}^{(i)}_{k}) \nonumber  \\  &\overset{b}\leq  {f} (\mathbf{x}_{k}^{(i)}) -  \eta_{k}^{(i+1)} \left( g_{i+1}({\mathbf{x}}^{(i)}_{k})\right)^T   \nabla{f}({\mathbf{x}_k^{(i)}} )    \nonumber \\ &+\frac{L}{2}  \eta_{k}^{(i+1)}  ||g_{i+1}({\mathbf{x}}^{(i)}_{k}))||^2,
\end{align}
where (a) follows from the update rule in \eqref{177}, while $f$ is the  global loss function in equation (1) in the main paper, and (b) from  Assumption 3 where  $|| \nabla^2 {f}({\gamma} )|| \leq L $. Given  the model $\mathbf{x}_k^{(i)}$, we take expectation over the randomness in selection of sample $\mathcal{\zeta}_{i+1}$ (the random sample used to get the model $\mathbf{x}_k^{(i+1)}$). We recall that $\mathcal{\zeta}_{i+1}$ is drawn according to the distribution $\mathcal{P}_{i+1}$ and is independent of the  model  $\mathbf{x}_k^i$). Therefore, we get the following set of equations:
\begin{align}
&\mathbb{E} [{f}({\mathbf{x}_k^{(i+1)}} ) ]  \leq    f (\mathbf{x}_{k}^{(i)})  - \eta_{k}^{(i)}  \mathbb{E} \left[(g_{i+1}({\mathbf{x}}^{(i)}_{k}))\right]^T \nabla{f}({\mathbf{x}_k^{(i)}} )   \nonumber \\ & +\frac{ (\eta_{k}^{(i)})^2 L}{2} \mathbb{E}||g_{i+1}({\mathbf{x}}^{(i)}_{k})||^2 \nonumber  \\
\overset{a} \leq & f (\mathbf{x}_{k}^{(i)}) - \eta_{k}^{(i)}  ||\nabla f(\mathbf{x}_k^{(i)})||^2 + \frac{(\eta_{k}^{(i)})^2 L}{2} ||\nabla f(\mathbf{x}_k^{(i)})||^2   \nonumber \\ &+  \frac{(\eta_{k}^{(i)})^2 L}{2}\sigma^2 \nonumber \\ 
  =  &  f (\mathbf{x}_{k}^{(i)})    -  ||\nabla f(\mathbf{x}_k^{(i)})||^2 \left( \eta_{k}^{(i)} -  \frac{(\eta_{k}^{(i)})^2 L}{2} \right)  \nonumber \\ & + \frac{ (\eta_{k}^{(i)})^2 L}{2} \sigma^2 \nonumber \\ \overset {b} \leq&
 f(\mathbf{x}_{k}^{(i)})   - \frac{\eta_{k}^{(i)}}{2} ||\nabla f(\mathbf{x}_k^{(i)})||^2  + \frac{ \eta_{k}^{(i)}}{2} \sigma^2, \label{e18}
\end{align}
where (a) follows from \eqref{e13}, and (b)  by selecting $\eta_{k}^{(i)} \leq \frac{1}{L} $. Furthermore, in  the  proof of Theorem 1, we choose the learning to be  $\eta_{k}^{(i)} \geq \frac{1}{L}$. Therefore,  the learning rate will be given by  $\eta_{k}^{(i)} = \frac{1}{L}$.   By the convexity of the loss function $f$, we  get the next inequality from the inequality in \eqref{e18}
\begin{align}\label{1888}
\mathbb{E}[{f}({\mathbf{x}_k^{(i+1)}} )]  \leq &  f (\mathbf{x}^*)  +\langle\,  \nabla f(\mathbf{x}_k^{(i)}), \mathbf{x}_k^{(i)} -  \mathbf{x}^* \rangle    \nonumber \\ & -  \frac{\eta_{k}^{(i)}}{2} ||\nabla f(\mathbf{x}_k^{(i)})||^2  + \frac{ \eta_{k}^{(i)}}{2} \sigma^2.
\end{align}
We now back-substitute $g_{i}({\mathbf{x}}^{(i)}_{k})$
into \eqref{1888} by using $\mathbb {E} [g_{i+1}({\mathbf{x}}^{(i)}_{k})] = \nabla f(\mathbf{x}_k^{(i)}) $ and $||\nabla f(\mathbf{x}_k^{(i)})||^2  \geq  \mathbb{E} || g_{i+1}({\mathbf{x}}^{(i)}_{k})||^2 -\sigma^2   $: 
\begin{align}
&\mathbb {E}[{f}({\mathbf{x}_k^{(i+1)}})]  \leq   f (\mathbf{x}^*)  + \langle\,  \mathbb {E}[ g_{i+1}({\mathbf{x}}^{(i)}_{k})], \mathbf{x}_k^{(i)}  -  \mathbf{x}^* \rangle   \nonumber \\ &- \frac{\eta_{k}^{(i)}}{2} \mathbb {E} ||g_{i+1}({\mathbf{x}}^{(i)}_{k})||^2  + \eta_{k}^{(i)} \sigma^2 \nonumber \\ &=    f (\mathbf{x}^*)  + \mathbb {E} [ \langle\,   [g_{i+1}({\mathbf{x}}^{(i)}_{k})], \mathbf{x}_k^{(i)} -  \mathbf{x}^* \rangle  \nonumber \\&   - \frac{\eta_{k}^{(i)}}{2} ||g_{i+1}({\mathbf{x}}^{(i)}_{k})||^2  ]  + \eta_{k}^{(i)} \sigma^2. 
\end{align}
By completing the square of the middle two terms to get:  
\begin{align}
&\mathbb {E}[{f}({\mathbf{x}_k^{(i+1)}} )]  \leq    f (\mathbf{x}^*)   \nonumber \\+& \mathbb {E} \left[ \frac{1}{2\eta_{k}^{(i)}} \left( ||   \mathbf{x}_k^{(i)} {-}  \mathbf{x}^* ||^2  {-}  || \mathbf{x}_k^{(i)} {-} \mathbf{x}^*  {-}  \eta_{k}^{(i)}  g_{i+1}({\mathbf{x}}^{(i)}_{k}) ||^2 \right)\right]    \nonumber \\+& \eta_{k}^{(i)} \sigma^2. \nonumber \\    = &   f (\mathbf{x}^*)  {+} \mathbb {E} \left[ \frac{1}{2\eta_{k}^{(i)}} \left( ||   \mathbf{x}_k^{(i)} {-}  \mathbf{x}^* ||^2   {-} || \mathbf{x}_k^{(i+1)} {-} \mathbf{x}^*  ||^2\right)\right] \nonumber \\&  {+} \eta_{k}^{(i)} \sigma^2.
\end{align}
For $K$ rounds and $r$ benign nodes, we note that the total number of SGD steps are $T = Kr$. We let $s = kr+i$ represent the number of updates happen to the initial model $\mathbf{x}^0$, where   $i = 1, \dots, r$ and $k = 0, \dots, K-1$. Therefore,  ${\mathbf{x}_k^{(i)}} $ can be written as   $\mathbf{x}^s$. With the modified notation, we can now take the expectation in the above expression over the entire sampling process during training and then by summing the above equations for $s = 0, \dots, T-1$, while taking $\eta = \frac{1}{L}$, we have the following: 
\begin{align}\label{we}
 &\sum_{s=0}^{T-1}   \left( \mathbb{E}[ {f}({\mathbf{x}^{s+1}} )] -  f(\mathbf{x}^{*}) \right) \nonumber \\ &\leq   \frac{L}{2} \left( ||   \mathbf{x}_0 -  \mathbf{x}^* ||^2   - \mathbb{E} [  || \mathbf{x}_k^{(T)} - \mathbf{x}^*  ||^2]\right) +\frac{1}{L}T \sigma^2.
%\epsilon \leq \frac{(r_2-r_1)^2}{4N^2} \frac{m}{G} n  \sum_{g=0}^{G-1} \frac{2(G-g )-1}{(K_g-1)^2 }.
\end{align}

 By using the convexity of $f(.)$, we get
 \begin{align}
 \mathbb{E}\left[ f\left(\frac{1}{T} \sum_{s=1}^{T} \mathbf{x}^{s}\right)\right]-  f(\mathbf{x}^*  ) \leq & \frac{1}{T} \sum_{s=0}^{T-1}  \mathbb{E}\left[ f(\mathbf{x}^{s+1})\right]-  f(\mathbf{x}^* )  \nonumber \\& \leq   \frac{||\mathbf{x}^0- \mathbf{x}^{*}||^2 L}{2  T } + \frac{1}{L} \sigma^2.
\end{align}

%F(\bm{\theta}^{*}) + < \mathbb{E} [\bar{\mathbf{p}}^{(t)}], \bm{\theta}^{(t)} - \bm{\theta}^{*}> - \frac{\eta}{2}\mathbb{E}||  \bar{\mathbf{p}}^{(t)}||^2 + \eta \sigma_{\text{HeteroSAg}}^2 \nonumber \\ 
%Rearranging the terms, summing up over $T=Kb$ updates, where $k$ is number of complete rounds over the ring and $b$ is the number of nodes we are considering over the ring, 
\section{Joining and Leaving of Nodes}\label{drop}
\texttt{Basil} can handle the scenario of 1) node dropouts out of the $N$ available nodes  2) nodes rejoining the system. 
 
\subsection{Nodes Dropout }  For handling node dropouts, we allow for extra communication between nodes. In particular, each active node can multicast its model to the $S{=}b{+}d{+}1$  clockwise neighbors, where $b$ and $d$ are respectively the number of Byzantine nodes and  the worst case number of dropped nodes, and each node  can store only the latest $b{+}1$ model updates. By doing that, each benign node will have at least 1 benign update even in the worst case where all Byzantine nodes appear in a row and $d$ (out of $S$) counterclockwise nodes drop out. 
 
\subsection{Nodes Rejoining}
To address a node \textit{rejoining} the system, this rejoined node can re-multicast its ID  to all other nodes. Since benign nodes know the correct order of the nodes (IDs) in  the ring according to Section \ref{3.1}, each active node out of the $L{=} b{+}d{+}1$ counterclockwise neighbors of  the  rejoined node sends its model to it, and this rejoined node stores the latest  $b{+}1$ models.
We note that handling participation of new fresh nodes during training is out of scope of our paper, as we consider mitigating Byzantine nodes in decentralized training with a \textit{fixed} number of $N$ nodes

\section{Proof of Proposition 3}\label{ACDS-time}
We first prove the communication cost given in Proposition 3, which corresponds to node $1_g$, for $g \in [G]$. We recall from Section   IV  that in ACDS, each node  $i \in \mathcal{N}$ has $H$ batches each of size $\frac{\alpha D}{H}$ data points. Furthermore, for each   group $g \in [G]$, the anonymous cyclic data sharing phase (phase 2) consists of   $H+1$   rounds.  The communication cost of node $1_g$ in the first round is $\frac{\alpha D}{H} I$ bits, where $\frac{\alpha D}{H} $ is the size of one batch and $I$ is the size of one data point in bits.  The cost of each round $h \in [2,H+1]$ is $ n\frac{\alpha D}{H} I$, where node $1_g$ sends  the set of shuffled data from the $n$ batches  $\{c^h_{1_g}, c^{h-1}_{2_g}, \dots,  c^{h-1}_{n_g}\}$ to node $2_g$. Hence, the total communication cost for node $1_g$ in this phase is given by $C_{\text{ACDS}}^{\text{phase-2}} = \alpha D I(\frac{1}{H} +n)$.  In phase 3, node $1_g$ multicasts its set of shuffled data from batches $\{c^h_{1_g}, c^{h}_{2_g}, \dots,  c^{h}_{n_g}\}_{h\in[H]}$ to all nodes in the other groups at a cost of $n \alpha D I $ bits.  Finally, node $1_g$ receives $(G-1)$ set of batches $\{c^h_{1_{g'}}, c^{h}_{2_{g'}}, \dots,  c^{h}_{n_{g'}}\}_{h\in[H], g' \in [G]\backslash \{g\}}$ at a cost of $(G-1)n \alpha D I$. Hence, the communication cost of the third phase of ACDS is given by $C_{\text{ACDS}}^{\text{phase-3}} = \alpha D nG I$. By adding the cost of Phase 2 and Phase 3, we get the first result in Proposition 3.

Now, we prove the communication time of ACDS by first computing the time needed to complete the anonymous data sharing phase (phase-2), and then compute the time for the multicasting phase.  The second phase of ACDS consists of  $H+1$ rounds.  The communication time of the first round   is given by $T_{R_1}= \sum_{i =1}^n  iT$, where $n$ is the number of nodes in each group. Here,  $T = \frac{\alpha D I}{HR}$  is  the time needed  to send one batch of size $\frac{\alpha D I}{H}$ data points with $R$ being the communication bandwidth in b/s, and $I$ is the size of one data points in bits. On the other hand, the time for  each round  $h \in [2,H]$, is given by  $T_{R_h}= n^2 T  $, where each node sends $n$ batches.  Finally, the time for completing the dummy round, the $(H+1)$-th round,  is given by  $T_{R_{H+1}}= n (n-2) T  $ where only the first $n-2$ nodes in the ring participate in this round as discussed in Section IV. Therefore, the total time for completing the anonymous cyclic data sharing phase (phase 2 of ACDS) is given by $T_{\text{phase-2}} = T_{R_1} +  (H-1) T_{R_h} + T_{R_{H+1}} =  T(n^2(H+0.5) - 1.5n)$ as this phase happens in parallel for all the $G$ groups. The time for completing the multicasting phase is $T_{\text{phase-3}} = (G-1)n HT $, where each node  in  group $g$  receives $n H$ batches from each node $1_{g'}$ in group $g' \in [G]\backslash \{g\}$. By adding $T_{\text{phase-2}}$ and $T_{\text{phase-3}}$, we get the communication  time of ACDS  given in Proposition 3.

\section{Proof of Proposition 4}\label{prop3}

We recall  from Section III that the per round   training time of    \texttt{Basil}  is divided into four parts. In particular, each active node (1) receives the model  from  the  $S$ counterclockwise neighbors; (2) evaluates the $S$ models using the \texttt{Basil}  aggregation rule; (3) updates the model by taking one step of SGD; and (4) multicasts the model to the next $S$ clockwise neighbors. 

%To avoid the double count of the communication time when evaluating the per round training time of Basil, we consider the communication time for receiving $S$  models instead of the time for multicasting  of one model.  

Assuming training begins at time $0$, we define  $E_i^{(k)}$ to be the wall-clock time at which node $i$ finishes the training in round $k$. We also  define   $T_{\text{com}} = \frac{32d}{R}$ to be the time  to receive one  model of size $d$  elements each of size $32$ bits, where each node receives only one  model at each step  in the ring as the training is sequential. Furthermore,  we let  $T_{\text{comp}} =  T_{\text{perf-based}} + T_{\text{SGD}} $ to be the time needed to evaluate $S$ models  and  perform one step of SGD  update.

We assume that each node $i \in \mathcal{N}$ becomes active and starts evaluating the $S$ models (using \eqref{eq3}) and taking the SGD model update step (using \eqref{Update_main}) only when it receives the  model  from  its counter clock-wise neighbor.  Therefore, for the first round, we have the following time recursion:
\begin{align} 
    E_1^{(1)} &=   T_{\text{SGD}} \\
    E_n^{(1)} &{=} E_{n-1}^{(1)}  + T_{\text{com}}{+}  (n-1)T_{\text{perf-based}}+ T_{\text{SGD}} \text{ for }  n \in [2,S]\\
    E_n^{(1)} &=  E_{n-1}^{(1)}+ T_{\text{com}}+ T_{\text{comp}}  \text{ for }  n \in [S+1,N], 
\end{align}
where (42) follows from the fact that node $1$ just takes one step of model update using the initial model $\mathbf{x}^0$.  Each node $i \in [2,S]$ receives the model from its own node, evaluates the $(i-1)$ received model and then takes one step of model update.  For node $i \in [S+1,N]$,  each node will have $S$ models to evaluate, and the time  recursion follows (44). 

The time recursions, for the remaining $ \tau$ rounds, where the training is assumed to happen over $\tau$ rounds, are  given by  

\begin{align} 
    &E_1^{(k+1)} =  E_{n}^{(k)}+ T_{\text{com}}+ T_{\text{comp}}   \text{ for } k \in [\tau-1] \label{41-3} \\
    &E_n^{(k)} =  E_{n-1}^{(k)}+ T_{\text{com}}+ T_{\text{comp}}  \; \text{ for } n \in [N]  \backslash\{1\}, k \in [\tau] \label{eq-Basil_recur} \\
    &E_1^{(\tau+1)} =  E_{n}^{(\tau)}+T_{\text{com}}+ S T_{\text{perf-based}}  \label{41-4}.
\end{align}
  By telescoping (42)-(47),  we get the upper bound   in \eqref{trainig_time}.

The   training time of  \texttt{Basil}+ in  \eqref{trainig_time+}  can be proven by computing the time of each stage of the algorithm: In Stage 1, all groups in parallel apply \texttt{Basil}  algorithm within its group, where the training is carried out sequentially. This results in a training time of $T_{\text{stage1}} \leq n \tau  T_{\text{perf-based}} + n \tau  T_{\text{comm}} + n \tau T_{\text{SGD}}$. The time of the robust circular aggregation stage  is given by   $T_{\text{stage2}} =  G T_{\text{perf-based}} + S G T_{\text{comm}}$. Here,    $ S T_{\text{perf-based}} $ in the first term 
comes from the fact that each node in the set $\mathcal{S}_g$ in parallel evaluates $S$ models received from the nodes in $\mathcal{S}_{g-1}$.    The second term in $T_{\text{stage2}}$  comes from the fact that each node in  $\mathcal{S}_{g}$  receives $S$ models from the nodes in  $\mathcal{S}_{g-1}$.   The term $G$ in stage 2 results from the sequential aggregation over the $G$ groups. The  time of the final stage (multicasting stage) is given by $T_{\text{stage3}} =   T_{\text{perf-based}} + (G-1) T_{\text{comm}}$, where the first term from the fact all nodes in the set $\{\mathcal{U}_1, \mathcal{U}_2, \dots, \mathcal{U}_{G-1} \}$ evaluates the $S$ robust average model in parallel, while the second term follows from  the time needed  to receive    the $S$  robust average model by each  corresponding node  in the remaining groups. By combining the time of the three stages, we get the training time given in \eqref{trainig_time+}. 

\section{Proof of Proposition 5  }\label{prop4}
\noindent \textbf{Proposition 5.}\textit{ The connectivity parameter $S$ in \texttt{Basil+}  can be relaxed to $S<n-1$ while guaranteeing the success of the algorithm (benign local/global subgraph connectivity)  with high probability. The failure probability of \texttt{Basil+} is given by  
\begin{align}
\mathbb {P}(F) \leq   G \sum_{i = 0}^{\min{ (b,n)}}   \left(  \prod_{s=0 }^{S-1} \frac{ \max \left(i-s , 0 \right)}{(N-s) } n  \right) \frac{\binom{b}{i}\binom{N-b}{n-i}}{\binom{N}{n}}, 
\end{align}
where $N$, $n$, $G$, $S$ and $b$ are the number of total nodes, number of nodes in each group, number of groups, the connectivity parameter, and the number of Byzantine nodes.  }

\textit{Proof.} At a high level, \texttt{Basil+} would fail if   at least one group out of the $G$ groups failed (the set of  models $\mathcal{L}_g$ sent from the set $\mathcal{S}_g$ in any particular group $g$ to the  group $g+1$  are faulty). According to the discussion in the proof of   proposition 2, group $g$ fails, when we $S$ Byzantine nodes comes in a row. 

Now, we formally prove the failure probability of \texttt{Basil+}. We start  our proof by  defining    the failure event of \texttt{Basil+} by 
\begin{equation}
F = \bigcup_{g = 1}^{G} F_g, 
\end{equation}
where $F_g$ is the failure event of group $g$ and $G$ is the number of groups.  The failure probability of group $g$ is given by 
\begin{equation}\label{condi}
\mathbb {P}(F_g) =  \sum_{i = 0}^{\min{ (b,n)}}   \mathbb {P} ( F_g \given[\Big] 
 b_g = i)  \mathbb {P} (b_g = i),
\end{equation}
where $b_g$ is the number of Byzantine nodes in group $g$.  Equation \eqref{condi} follows the law of total probability. The conditional probability in  \eqref{condi} represents the failure probability of group $g $ given $i$ nodes in that group are Byzantine nodes. This conditional group failure probability     can be derived similarly to the failure probability  in \eqref{Failure} in Proposition 2. In particular, the conditional probability is formally given by 
\begin{equation}
 \mathbb {P} ( F_g \given[\Big] 
 b_g = i)  \leq  \sum_{j=1}^{n}\mathbb {P}(A_j\given[\Big]  b_g = i )  \label{Failure2}, 
\end{equation}
 where      $A_j$  is  the failure event in which $S$ Byzantine nodes come in a row, where $j$ is the starting node of these $S$ nodes given that there are $i$ Byzantine nodes in that group. The probability $\mathbb {P}(A_j\given[\Big]  b_g = i )$  is given as follows 
\begin{equation}
\mathbb {P}(A_j\given[\Big]  b_g = i )=  \prod_{s=0 }^{S-1} \frac{ \max \left(i-s , 0 \right)}{(N-s) }\label{Prob2}, 
\end{equation}
 where $i$ is the number of Byzantine nodes in group $g$ and $S$ is the connectivity parameter in that group. By combining \eqref{Prob2}   with \eqref{Failure2}, we get the  conditional probability  in the first term in \eqref{condi} which is given as  follows 
 \begin{equation}
 \mathbb {P} ( F_g \given[\Big] 
 b_g = i)  \leq  \sum_{j=1}^{n}\mathbb {P}(A_j\given[\Big]  b_g = i  ) =   \prod_{s=0 }^{S-1} \frac{ \max \left(i-s , 0 \right)}{(N-s) } n.  \label{Failure3}
\end{equation}

 The  probability   $\mathbb {P} (b_g = i)$ in the second term of \eqref{condi} follows a Hypergeometric distribution with parameter $(N,b,n)$ where $N$, $b$, and $n$ are the total number of nodes, total number of Byzantine nodes,  number of nodes in each group, respectively. This probability is   given by 
 \begin{equation} \label{hyb}
 \mathbb {P} (b_g = i) = \frac{\binom{b}{i}\binom{N-b}{n-i}}{\binom{N}{n}}. 
 \end{equation}
 
By substituting  \eqref{hyb} and   \eqref{Failure3} in \eqref{condi}, we get the failure probability of one group in  \texttt{Basil+}, which is given as follows 

\begin{equation}\label{condi1}
\mathbb {P}(F_g) \leq \sum_{i = 0}^{\min{ (b,n)}}   \left(  \prod_{s=0 }^{S-1} \frac{ \max \left(i-s , 0 \right)}{(N-s) } n  \right) \frac{\binom{b}{i}\binom{N-b}{n-i}}{\binom{N}{n}}.
\end{equation}

Finally, the failure probability of \texttt{Basil+}
is given  by 
\begin{align}
\label{Failt}
\mathbb {P}(F) &= \mathbb {P}(\bigcup_{g = 1}^{G} F_g) \nonumber \\ & \overset{a}\leq \sum_{g=1}^{G}  \mathbb {P}(F_g) \nonumber \\ & =   G \sum_{i = 0}^{\min{ (b,n)}}   \left(  \prod_{s=0 }^{S-1} \frac{ \max \left(i-s , 0 \right)}{(N-s) } n  \right) \frac{\binom{b}{i}\binom{N-b}{n-i}}{\binom{N}{n}}, 
\end{align}
where (a) follows the union bound. $\hfill \square$

\section{UBAR}\label{UBAR1}
In this section, we describe UBAR \cite{guo2020byzantineresilient}, the SOTA   Byzantine resilient approach for parallel  decentralized training.  
\subsection{Algorithm}
This decentralized training setup   is defined over  undirected graph: $\mathcal{G} = (V, E)$, where $V$ denotes a set of $N$   nodes and   $E$ denotes a set of edges representing communication links.  Filtering Byzantine nodes is done over two stages for each training iteration. At the first stage, each benign node performs a distance-based strategy to select a candidate pool of potential benign nodes from its neighbors. This selection is performed by   comparing the Euclidean distance of its own model  with  the model  from its neighbors. In the second stage, each benign node performs a performance-based strategy to pick the final nodes from the candidate pool resulted from stage 1.  It reuses the training sample as the validation data to compute the loss function value  of each model. It selects the models whose loss values are smaller than the value of its own model, and calculates the average of those models as the final updated value. 
Formally, the update rule in UBAR is given by 
\begin{equation}
\mathbf{x}^{(i)}_{k+1} = \alpha \mathbf{x}^{(i)}_{k}+ (1-\alpha) \mathcal{R}_{\text{UBAR}}(
\mathbf{x}^{(j)}_{k}, j \in \mathcal{N}_i ) - \eta  {\nabla} f_i( {\mathbf{x}}^{(i)}_{k}), 
\end{equation}
where $\mathcal{N}_i $ is the set of neighbors of Node $i$, ${\nabla} f_i( {\mathbf{x}}^{(i)}_{k})$ is the  local gradient of node $i$ evaluated on a random sample from the local dataset of node $i$ while using its own model,     $k$  is the training round, and  $\mathcal{R}_{\text{UBAR}}$ is given as follows: 
\begin{equation}\label{UBAR}
\mathcal{R}_{\text{UBAR}}=\begin{cases}  \frac{1}{\mathcal{N}^r_{i, k} } \sum_{j \in \mathcal{N}^r_{i, k} }  \mathbf{x}^{(j)}_{k} & \text{ if  }  \mathcal{N}^r_{i, k} \neq \phi \\ 
 \mathbf{x}^{j^*}_{k}  & \text{ Otherwise,} \end{cases}
\end{equation}
where there are two stages of filtering:
\begin{align*}
\text{ (1) }  \; &  \mathcal{N}^s_{i, k} {=}  \arg\min_{ \substack{ \mathcal{N}^* \subset   \mathcal{N}_i, \\ \mathcal{N}^*  = \rho _i |\mathcal{N}_i|} }  \sum_{j \in  \mathcal{N}_i} || \mathbf{x}^{(j)}_{k} - \mathbf{x}^{(i)}_{k} ||,\\ 
\text{ (2) }  \; & \mathcal{N}^r_{i, k} {=} \hspace{-4pt}\bigcup_{ \substack{ j \in  \mathcal{N}^s_{i, k} \\ {\ell}_i(\mathbf{x}^{(j)}_{k} ) \leq   {\ell}_i(\mathbf{x}^{(i)}_{k} )} } j, \text{ and } j^* {=} \arg \min_{j \in  \mathcal{N}^s_{i, k}} {\ell}_i(\mathbf{x}^{(i)}_{k} ).
\end{align*}
 \subsection{ Time Analysis for UBAR}
 The  training time of UBAR is divided into two parts; computation time and communication time. We start by discussing the communication time. For modeling the communication time, we assume that each node in parallel can multicast its model to its neighbors and receive the models from the neighbor nodes, where each node is assumed to be connected to $S$ neighbors. Hence, the time to multicast $S$ models can be  calculated as $\frac{d32}{R}$, where $d$ is the model size and each element of the model is represented by $32$ bits and $R$ is the communication BW in b/s. On the other hand, the time to receives $S$ different models from  the $S$ neighbor nodes is given by $\frac{S*d*32}{R}$. We assume that in UBAR, each node starts the model evaluations and model update after receiving the  $S$ models (the same assumption is used when computing the training time for  \texttt{Basil}).  Therefore, given that each node starts the  training procedure only when it receives the $S$ models while all the communications happen in parallel in the graph, the communication time in one parallel round in UBAR is given as follows:
\begin{equation}
    T_{\text{UBAR-communication}} = \frac{Sd32}{R}.
\end{equation} 
The computation time of UBAR is given by 
\begin{equation}
T_{\text{UBAR-computation}}=T_{\text{dist-based}} +  T_{\text{perf-based}} + T_{\text{agg}} + T_{\text{SGD}}, 
\end{equation}  
where $T_{\text{dist-based}}$, $T_{\text{perf-based}}$, $T_{\text{agg}}$ and  $T_{\text{SGD}}$ are respectively the times to apply the distance-based strategy, the performance-based strategy, the aggregation, and one step of SGD model update. Hence, the total training time  when using UBAR  for $K$ training rounds is given by 
\begin{equation}
T_{\text{UBAR}}= K(T_{\text{dist-based}} +  T_{\text{perf-based}} + T_{\text{agg}} + T_{\text{SGD}} + S  T_{\text{comm}}  ),
\end{equation}
where $ T_{\text{comm}} = \frac{d32}{R}$.

\section{}\label{expe}
In this section, we provide the details of the neural networks used in our experiments, some key details regarding the UBAR implementation, and multiple additional experiments to further demonstrate the superiority of our proposed algorithms. We start in Section \ref{model} by describing the model that is used in our experiments in Section VI and the additional experiments  given in this section.    In Section \ref{impUBAR},  we discuss the implementation of  UBAR. After that,   we run additional experiments by using  MNIST dataset \cite{MNIST} in Section  \ref{exp_MNIST}.  In Section \ref{wall-clock}, we study the computation time of \texttt{Basil}  compared to UBAR, and the training performance of \texttt{Basil}  and UBAR with respect to the training time. Finally, we study the performance of \texttt{Basil}  and ACDS for CIFAR100 dataset with non-IID data distribution in Section H-E, and the performance comparison between \texttt{Basil}  and \texttt{Basil}+ in Section H-F. 
% hyperparameters, schemes, data partitioning, and Byzantine attacks described in the main paper. We also set the total number of nodes to be 100 in which 67  of them are benign . 

\begin{figure*}[ht]
  \centering
  \subfigure[No Attack]{\includegraphics[scale=0.28]{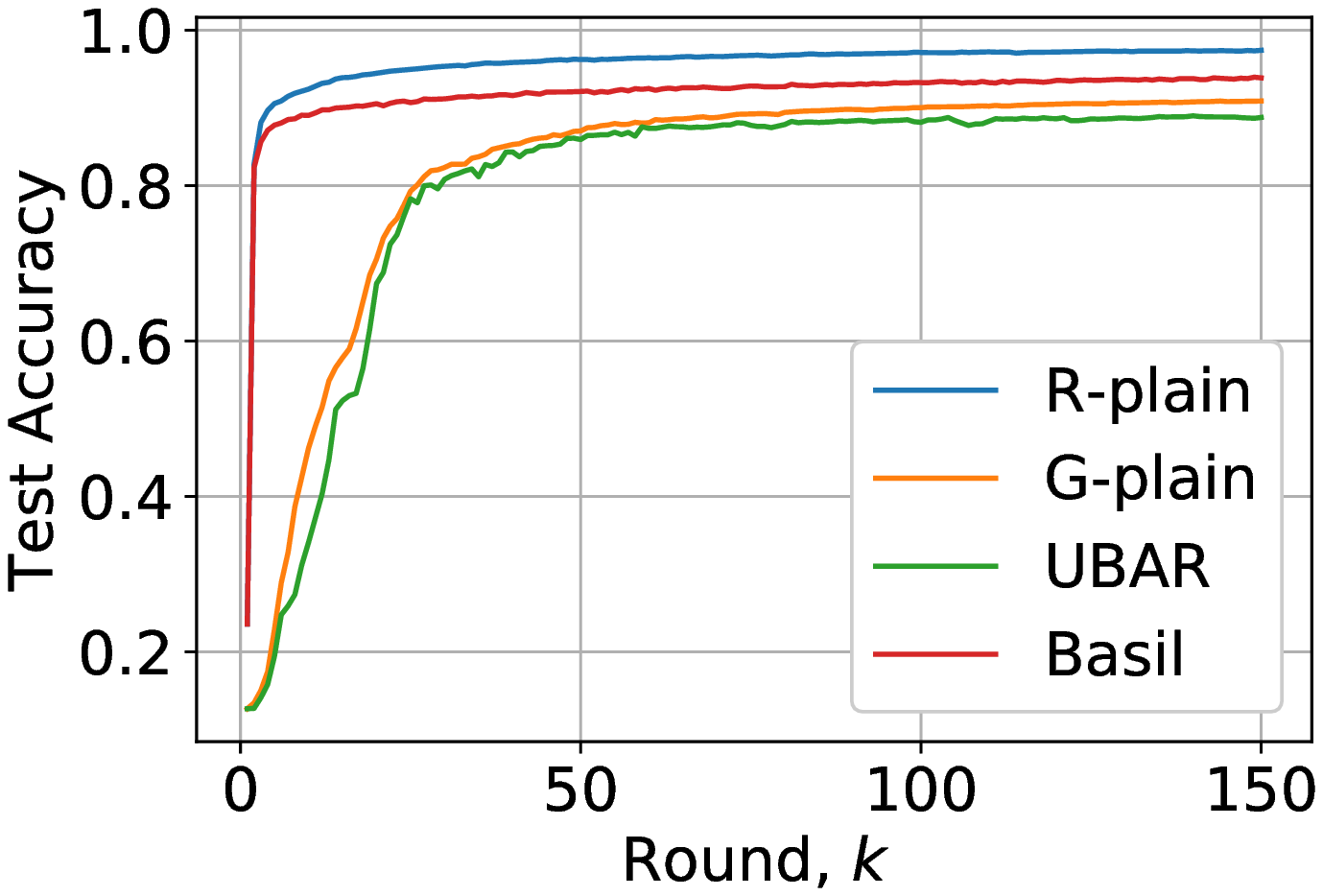}
  }%\quad
  \subfigure[Gaussian Attack]{\includegraphics[scale=0.28]{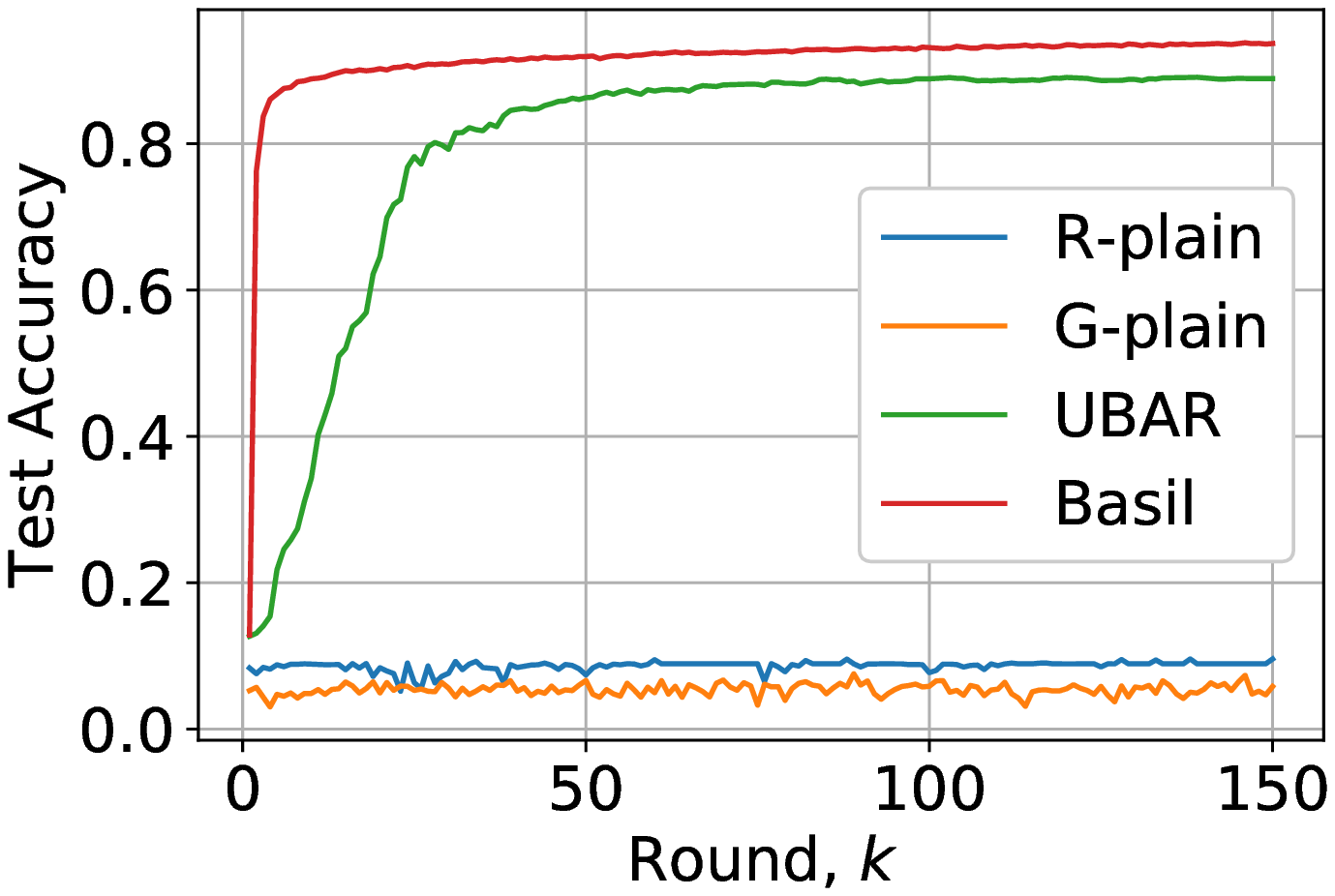}
  }%\quad
  \subfigure[Random Sign Flip Attack]{\includegraphics[scale=0.28]{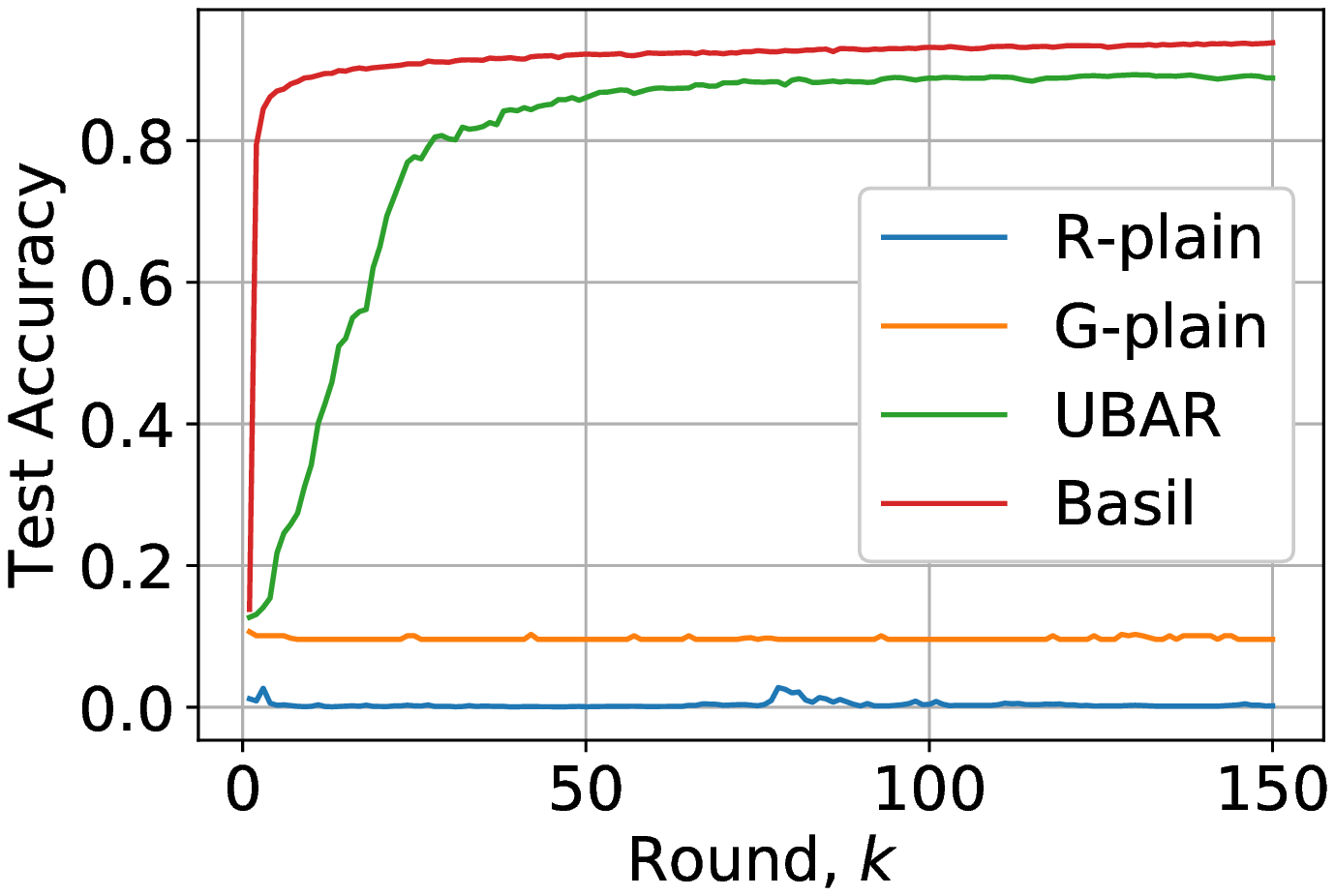}
  }
  \subfigure[Hidden Attack]{\includegraphics[scale=0.28]{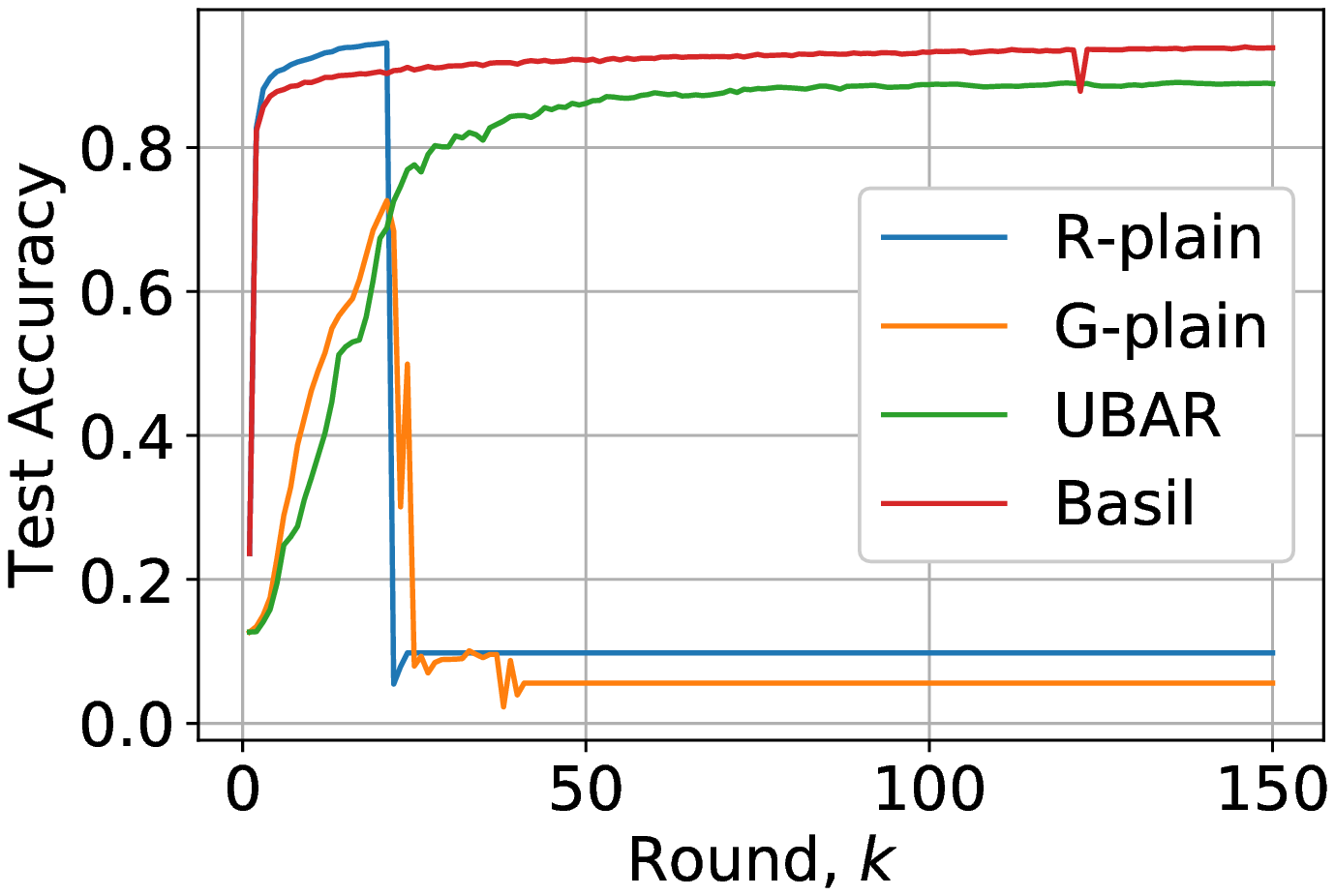}
  }%\quad
  \caption{Illustrating the results for MNIST under IID data distribution setting.}
  \label{fig:results_mnist0}
 \end{figure*}
 
 \begin{figure*}[t]
  \centering
  \subfigure[No Attack]{\includegraphics[scale=0.28]{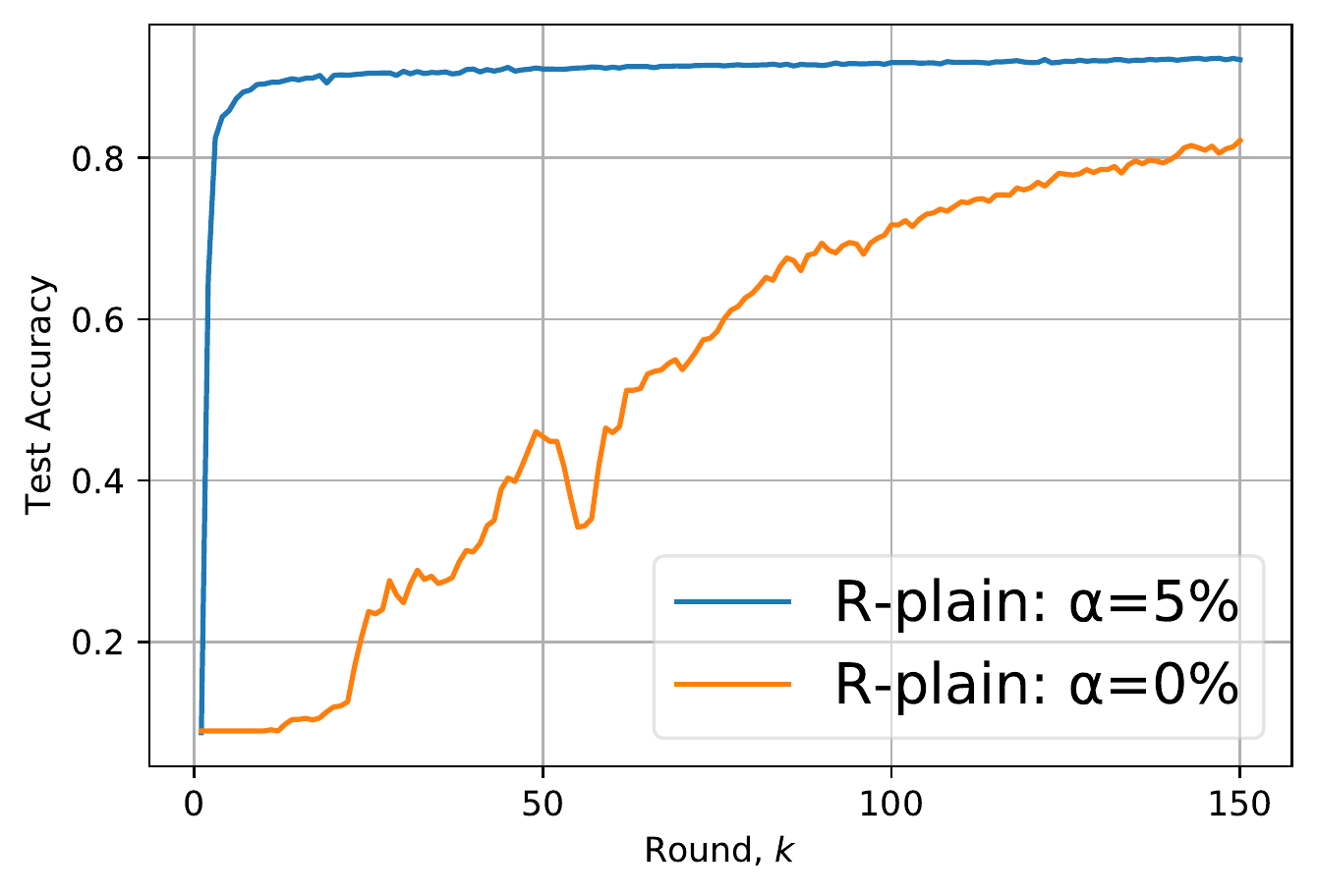}
  }%\quad
  \subfigure[No Attack]{\includegraphics[scale=0.28]{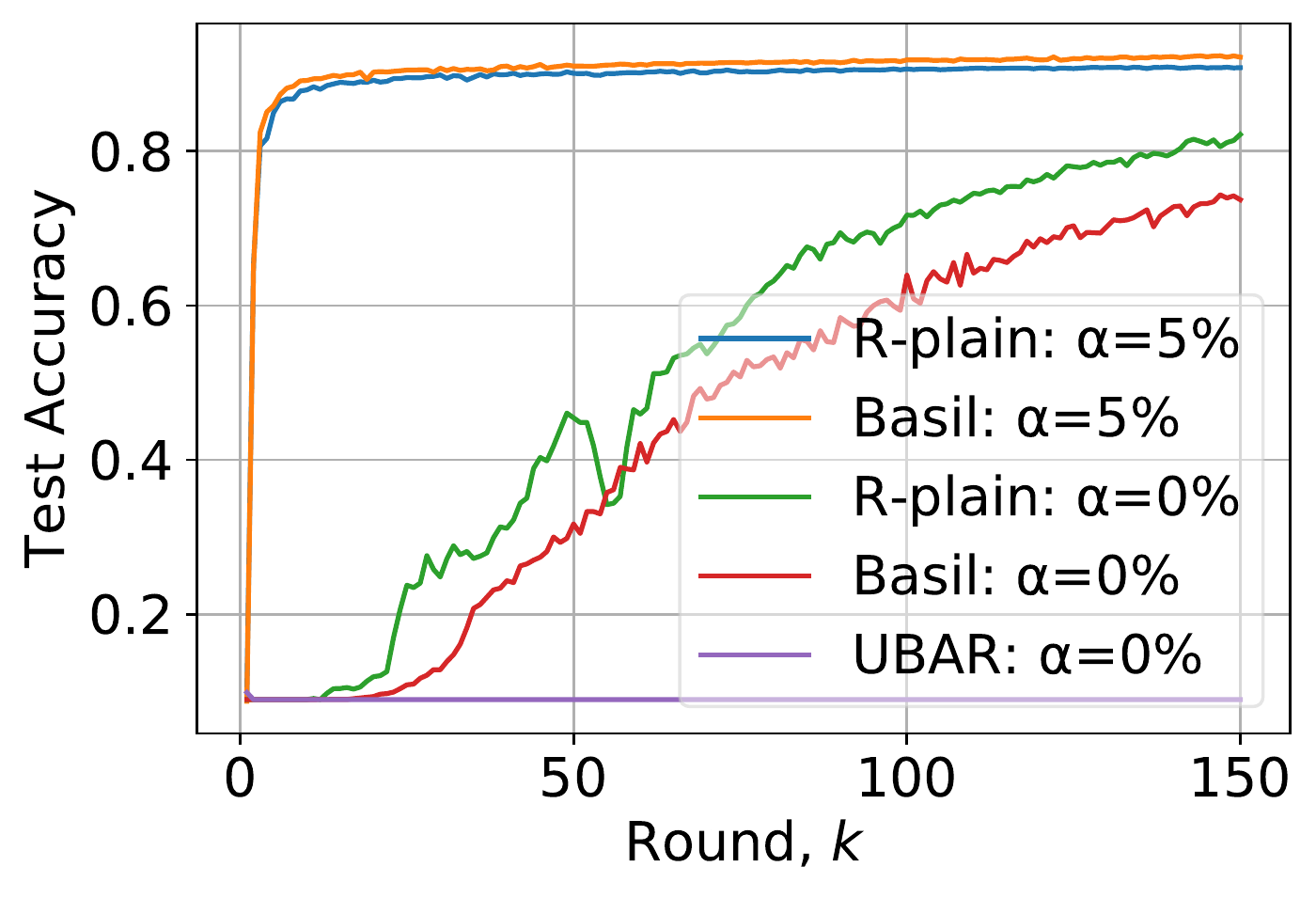}
  }%\quad
  \subfigure[Gaussian Attack]{\includegraphics[scale=0.28]{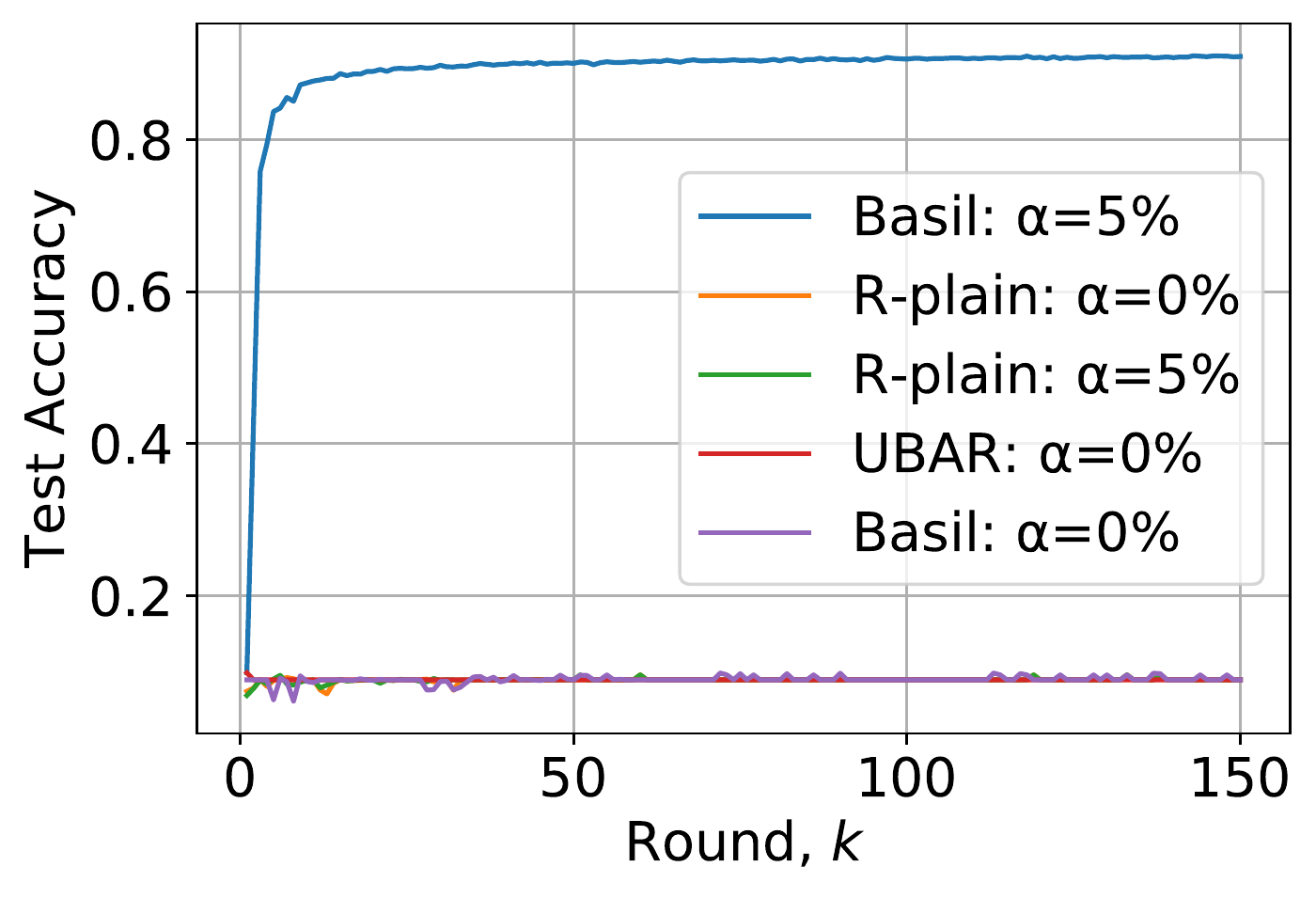}
  }
  \subfigure[Random Sign Flip Attack]{\includegraphics[scale=0.28]{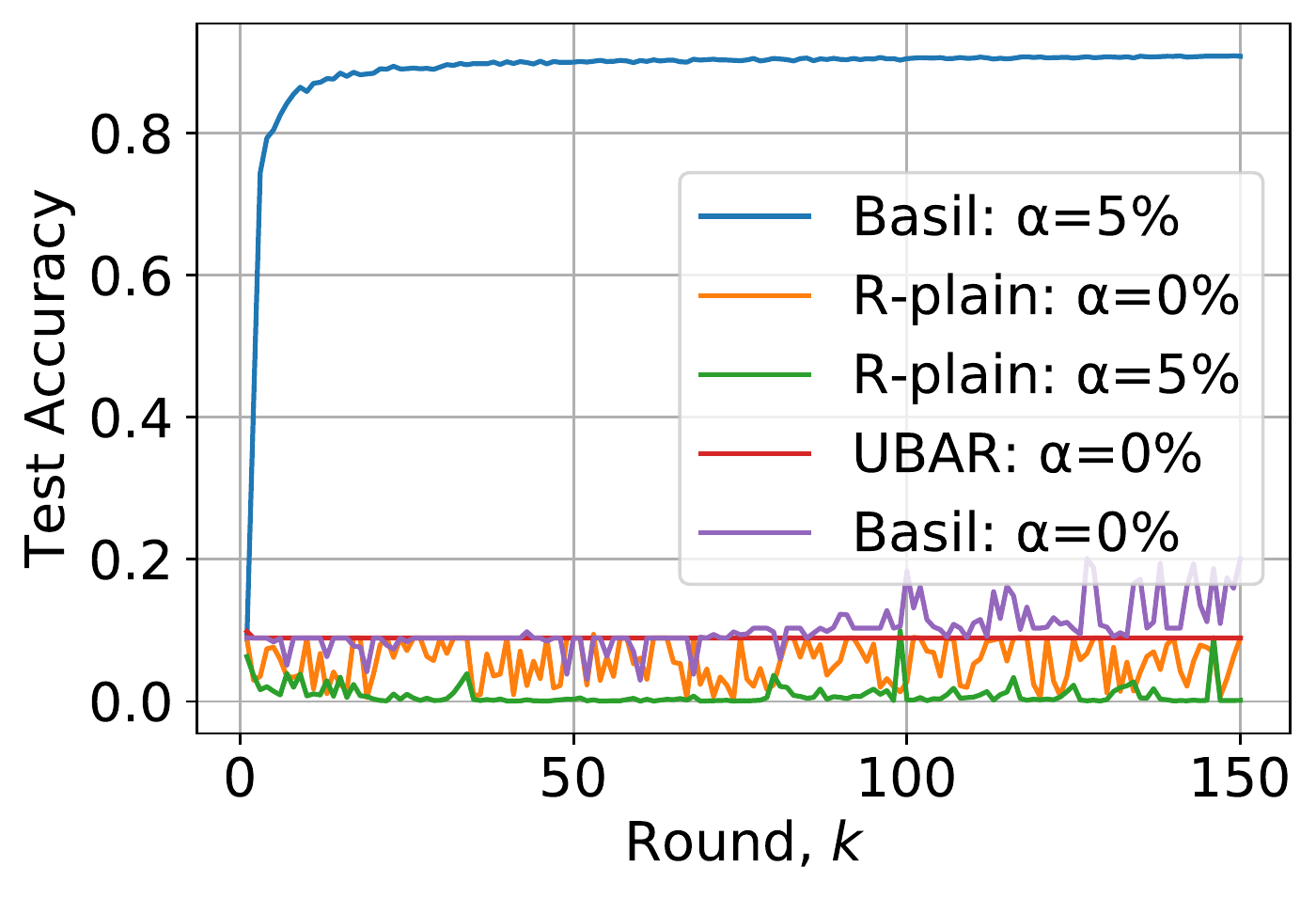}
  }%\quad
  \caption{Illustrating the results for MNIST under non-IID data distribution setting.}
  \label{fig:results_mnist20}
 \end{figure*}
 
  \begin{figure*}[t]
  \centering
  \subfigure[No Attack]{\includegraphics[scale=0.33]{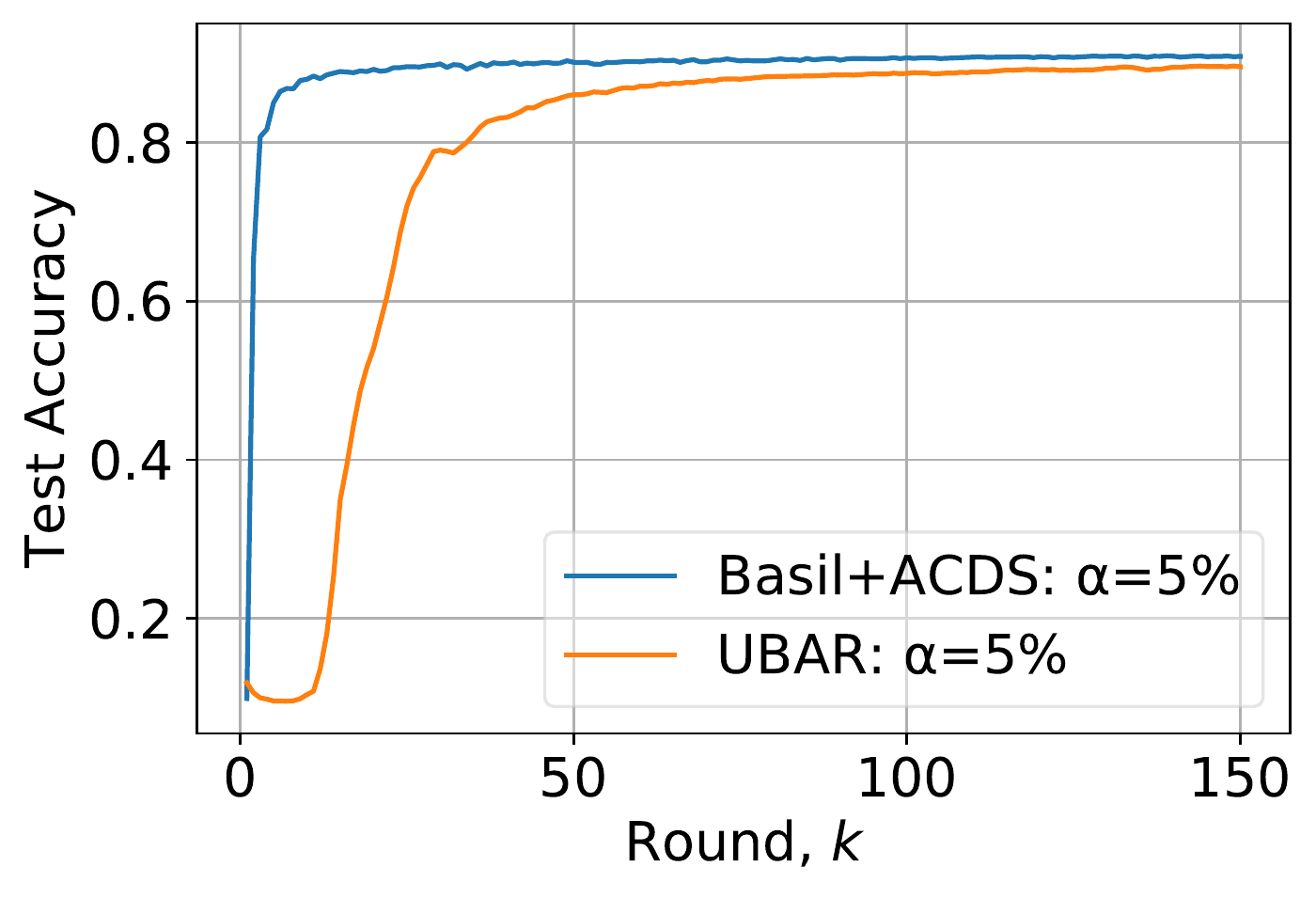}
  }%\quad
  \subfigure[Gaussian Attack]{\includegraphics[scale=0.33]{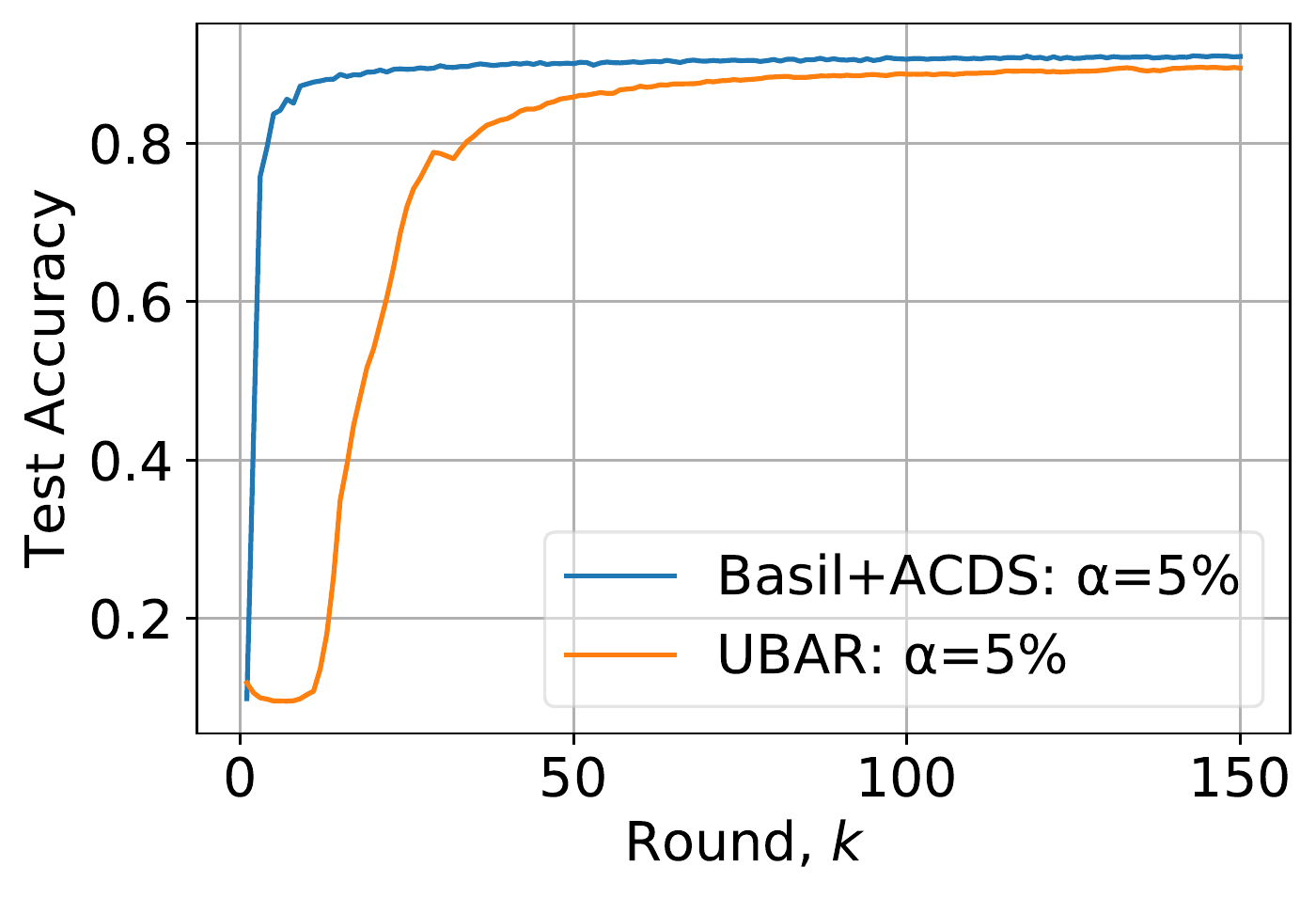}
  }%\quad
  \subfigure[Random Sign Flip Attack]{\includegraphics[scale=0.33]{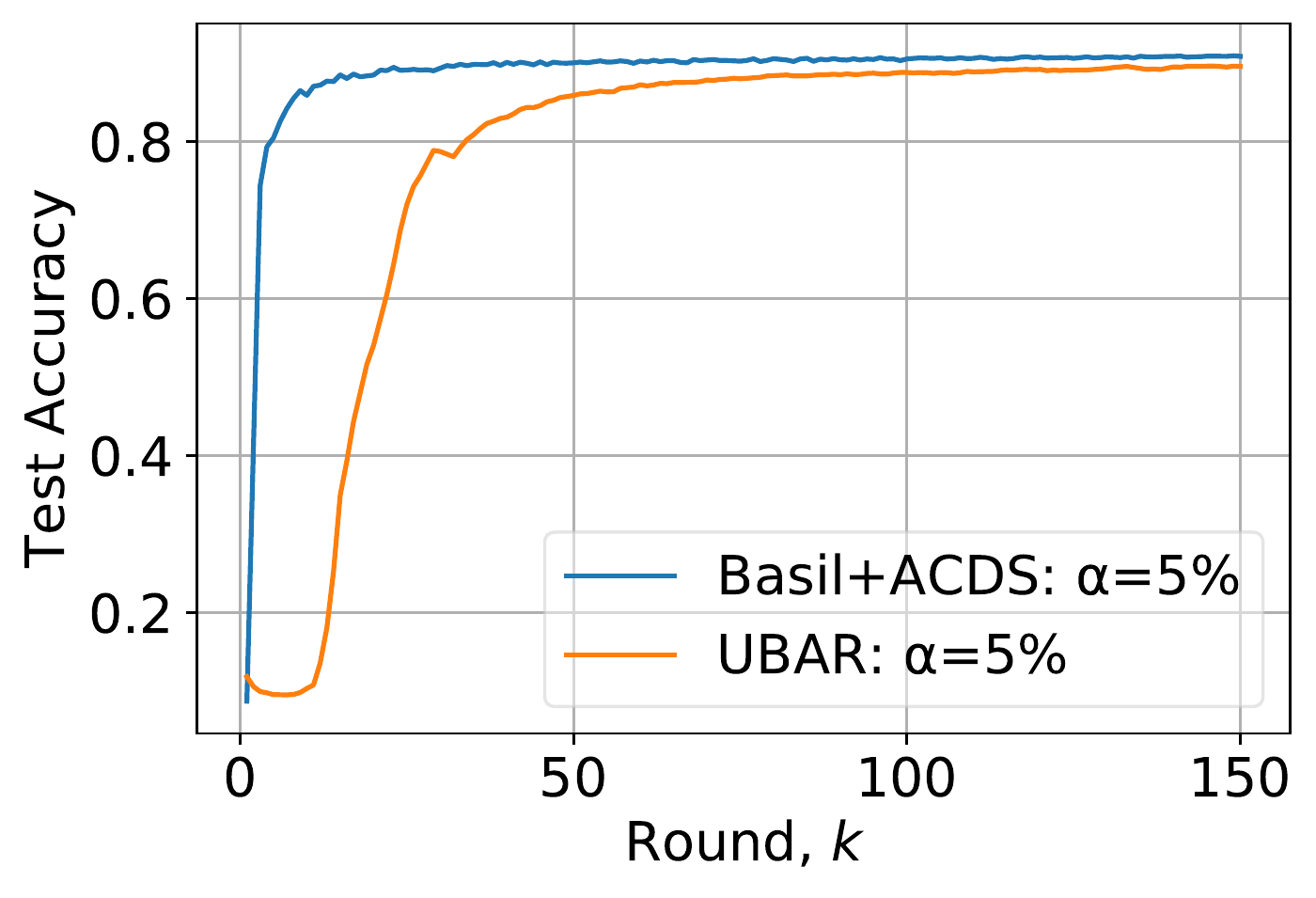}
  }
  \caption{Illustrating the performance of \texttt{Basil} compared with UBAR  for MNIST under non-IID data distribution setting with $ \alpha = 5\%$ data sharing.}
  \label{fig:Basil VS Mozi with data sharing1}
 \end{figure*}

\subsection{Models}\label{model}
We provide the details of the neural network architectures used in our experiments. For MNIST, we use a model with three fully connected layers, and the details for the same are provided in Table \ref{tab:1}. Each of the first two fully connected layers is followed by ReLU, while softmax is used at the output of the third one fully connected layer.

\begin{table}[htb!]
\caption{Details of the parameters in the architecture of the neural network used in our MNIST experiments.}\label{tab:1}
\centering
\begin{tabular}{|l|l|}
\hline
\textbf{Parameter} & \textbf{Shape}\\ \hline
fc1& $784\times100$\\ \hline
fc2& $100\times100$\\ \hline
fc3& $100\times10$\\ \hline
\end{tabular}
\end{table}

For CIFAR10 experiments in the main paper, we consider a neural network with two convolutional layers, and three fully connected layers, and the specific details of these layers are provided in Table \ref{tab:2}. ReLU and maxpool is applied on the convolutional layers. The first maxpool has a kernel size $3\times3$ and a stride of $3$ and the second maxpool has a kernel size of $4\times4$ and a stride of $4$. Each of the first two fully connected layers is followed by ReLU, while softmax is used at the output of the third one fully connected layer.

We initialize all biases to $0$. Furthermore, for weights in convolutional layers, we use Glorot uniform initializer, while for weights in fully connected layers, we use the default Pytorch initialization.
\begin{table}[htb!]
\caption{Details of the parameters in the architecture of the neural network used in our CIFAR10 experiments.}\label{tab:2}
\centering
\begin{tabular}{|l|l|}
\hline
\textbf{Parameter} & \textbf{Shape}\\ \hline
conv1& $3\times16\times3\times3$ \\ \hline
conv2& $16\times64\times4\times4$ \\ \hline
fc1& $64\times384$\\ \hline
fc2& $384\times192$\\ \hline
fc3& $192\times10$\\ \hline
\end{tabular}
\end{table}

\subsection{Implementing UBAR}\label{impUBAR}
 We follow a similar approach as described in the experiments in \cite{ guo2020byzantineresilient}. Specifically, we first assign connections randomly among benign nodes with a probability of $0.4$ unless otherwise specified, and then randomly assign connections from the benign nodes to the Byzantine nodes, with a probability of $0.4$ unless otherwise specified. Furthermore, we set the Byzantine ratio for benign nodes as $\rho=0.33$.
\subsection{Performance of Basil on MNIST} \label{exp_MNIST}

 We present the results for MNIST in Fig. \ref{fig:results_mnist0}  and Fig. \ref{fig:results_mnist20}   under  IID and non-IID data distribution settings, respectively. As can be seen from Fig. \ref{fig:results_mnist0}  and  Fig. \ref{fig:results_mnist20}  that  using   \texttt{Basil} leads to the same conclusions   shown  in CIFAR10 dataset in the main paper in terms of its fast convergence, high test accuracy, and Byzantine robustness compared to the different schemes. In particular, Fig. \ref{fig:results_mnist0} under IID  data setting demonstrates that   \texttt{Basil}  is not only resilient to Byzantine attacks, it maintains its superior convergence performance over UBAR.  Furthermore, Fig. 10(a) and Fig. 10(b) illustrate that the test accuracy when using \texttt{Basil} and R-plain  under  non-IID  data setting increases   when each node shares   $5\%$   of its local data with other nodes in the  absence of Byzantine nodes.  It  can also be  seen from Fig. 10(c) and Fig. 10(d) that 
ACDS with $\alpha = 5 \%$  on the top of \texttt{Basil} provides the same robustness  to software/hardware
faults represented in Gaussian model and random sign flip as concluded in the main paper. Additionally, we observe that both \texttt{Basil} without ACDS and UBAR completely fail in the presence of these faults.

Similar to the results in Fig. \ref{fig:Basil VS Mozi with data sharing} in the main paper,  Fig.   \ref{fig:Basil VS Mozi with data sharing1} shows that even $5\%$  data sharing is done in UBAR, performance remains quite low in comparison to \texttt{Basil}+ACDS.
\subsection{Wall-Clock Time Performance of Basil}\label{wall-clock}
In this section, we show the training performance of \texttt{Basil} and UBAR  with respect to the training time instead of the number of rounds.   To do so, we consider the following setting.

\begin{figure*}
  \centering
  \subfigure[Using a communication Bandwidth of $100$ Mb/s]{\includegraphics[scale=0.33]{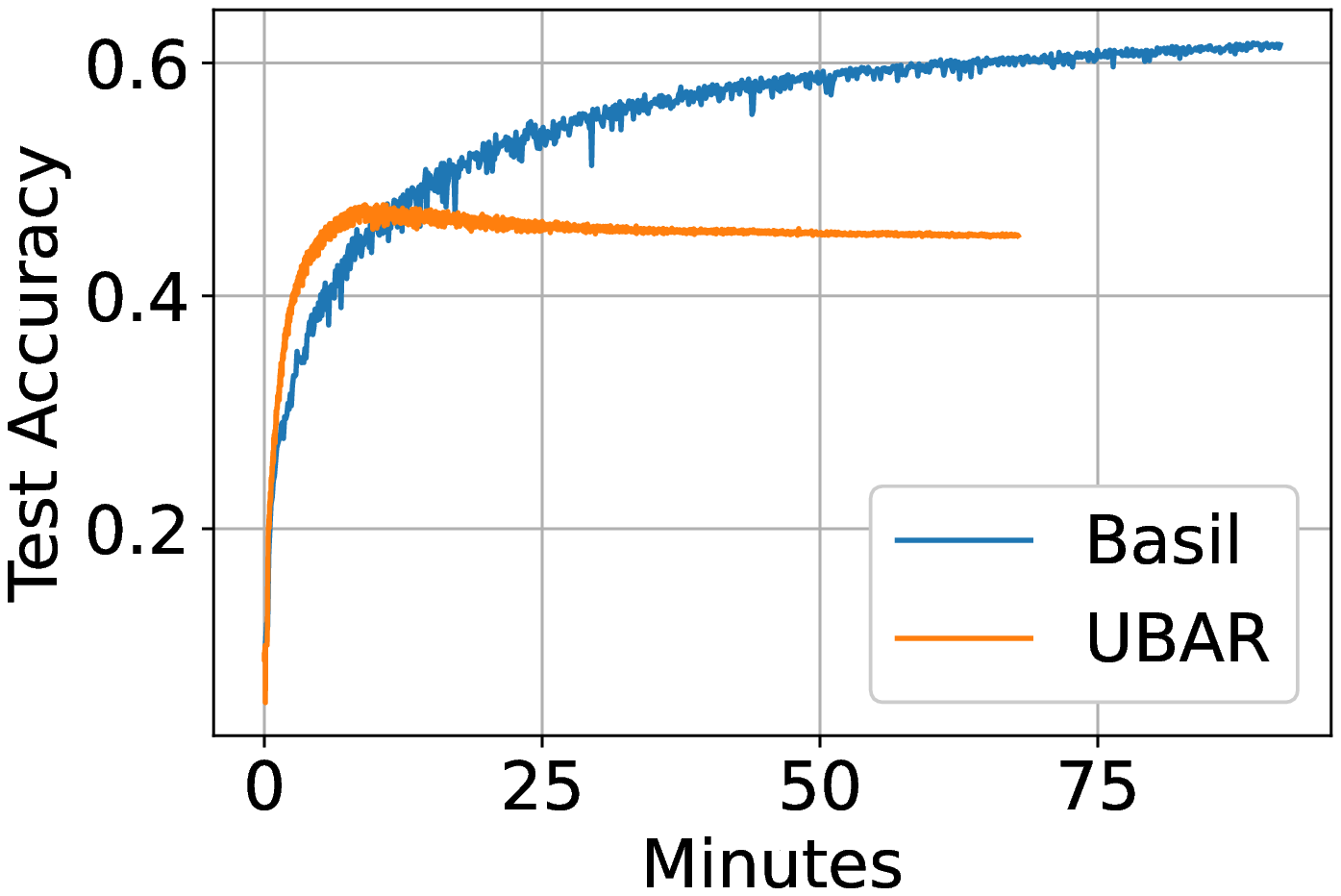}
}  \quad \quad \quad \quad
  \subfigure[Using  a communication Bandwidth of $10$ Mb/s]{\includegraphics[scale=0.33]{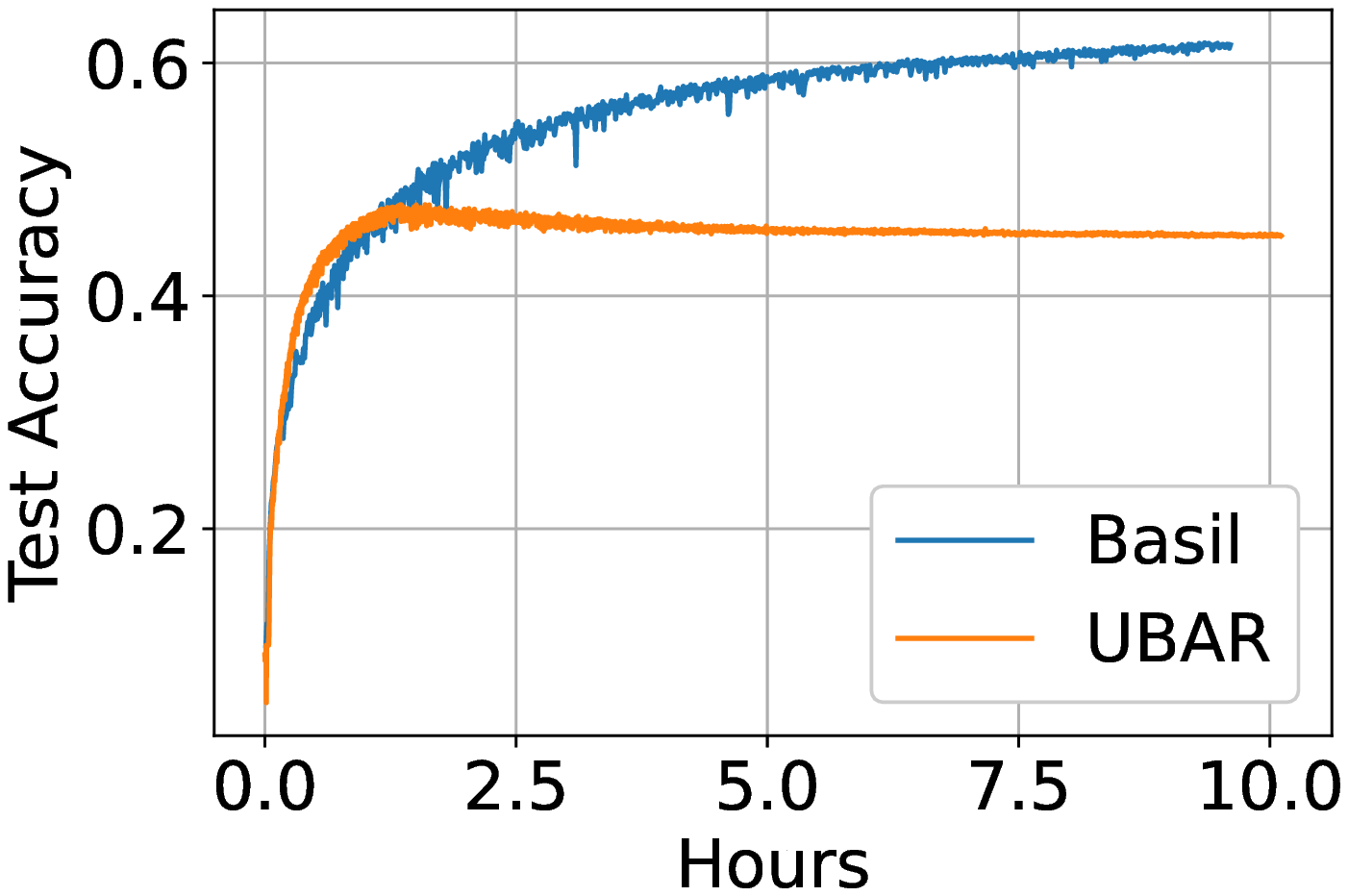}
  }%\quad
  \caption{Illustrating the performance of \texttt{Basil}  using  CIFAR10 dataset under IID data distribution setting   with respect to the training time.}
  \label{fig:latency}
 \end{figure*}

\underline{{Experimental setting}}. We consider the same setting discussed in Section VI-A  in the main paper where there exists a total   of $100$ nodes, in which $67$ are benign.  For the dataset, we use CIFAR10. We also consider the Gaussian attack. We set the connectivity parameter for the two algorithms to be $S = 10$.

Now, we start by giving the computation/communication time of \texttt{Basil}  and UBAR.  

\underline{{Computation time.}} We measured the computation time of \texttt{Basil} and UBAR  on a server with \texttt{AMD EPYC 7502 32-Core} CPU Processor. In particular, in TABLE \ref{table-time_abstract}, we report the average running time of each main component (function) of  UBAR and  \texttt{Basil}. To do that, we take the average computation time   over $10^{3}$ runs of each component in the mitigation strategy  on each training round  for  $100$ rounds.  These functions (components of the mitigation strategy) are the performance-based evaluation for \texttt{Basil}  given in Section III, while for UBAR these functions are  the performance-based evaluation and the  distance-based evaluation along with the model  aggregation for UBAR given in Appendix G.   We can see from TABLE \ref{table-time_abstract} that the average time each benign node in UBAR takes to evaluate the received set of models and take one step of model update using SGD is $\sim 2 \times$  the one in \text{Basil}. The reason is that each benign node in UBAR performs two extra stages before taking the model update step: (1) distance-based evaluation and (2) model aggregation. The distance-based stage includes a comparison between the   model of each benign node   and the received set of models which is a time-consuming operation compared to the performance-based as shown in Table \ref{table-time_abstract}.

  \begin{table*}[]
  \caption{The breakdown of the average computation time per node  for  \texttt{Basil} and UBAR. }
 \centering
\begin{tabular}{|c|cc|c|c|c|}
\hline
Algorithm        & \multicolumn{2}{c|}{Average evaluation
time per node (s)}                                                                                                                                                   & \begin{tabular}[c]{@{}c@{}}Aggregation (s)\\ $T_{\text{agg}}$\end{tabular} & \begin{tabular}[c]{@{}c@{}}SGD step (s)\\ $T_{\text{SGD}}$\end{tabular} & \begin{tabular}[c]{@{}c@{}}Total computation time\\  per node (s)\end{tabular} \\ \hline
                 & \multicolumn{1}{c|}{\begin{tabular}[c]{@{}c@{}}Performance-based\\ $T_{\text{per-based}}$\end{tabular}} & \begin{tabular}[c]{@{}c@{}}Distance-based\\ $T_{\text{dis-based}}$\end{tabular} &                                                                         &                                                                      &                                                                            \\ \hline
$\texttt{Basil}$ & \multicolumn{1}{c|}{$0.019$}                                                                            & -                                                                                & -                                                                       & $0.006$                                                              & $0.025$                                                                    \\ \hline
UBAR             & \multicolumn{1}{c|}{$0.012$}                                                                            & $0.027$                                                                          & $0.002$                                                                 & $0.006$                                                              & $0.047$                                                                    \\ \hline
\end{tabular}
 \label{table-time_abstract}
\end{table*}

\underline{Communication time.}  We consider an idealistic simulation, where the communication time of the trained model  is proportional to the number of elements    of the model. In particular, we simulate the communication time taken to send the model as used in Section VI  and described in Appendix H-(B)  to be given by $ T_{\text{comm}} = \frac{32d}{R}$, where the model size is given by $d = 117706$ parameters where each is represented by $32$ bits and $R$ is the bandwidth in Mb/s. To get the training time after $K$ rounds, we use the per round time result for \texttt{Basil}  and UBAR that are given in Proposition 4 in Section VI and Appendix G-(B), respectively, while considering  $K$ to be  the number of training rounds. 

 \underline{{Results.}}  Fig. \ref{fig:latency} demonstrates the performance of \texttt{Basil}  and UBAR with respect to the training time. As we can observe in Fig. \ref{fig:latency}, the  time it takes for UBAR to reach its maximum accuracy is almost the same as the time for \texttt{Basil}  to reach UBAR's maximum achievable accuracy. We recall from Section VI that   UBAR needs  $5 \times$ more   computation/communication resources  than \texttt{Basil}  to  get $~41\%$ test accuracy.    The performance of \texttt{Basil}  and UBAR  with respect to the training time  is  not surprising as we know that \texttt{Basil}  takes much  less training rounds to reach  the same accuracy that UBAR can reach as shown in Fig. 4. As a result, the latency resulting from the sequential training does not have high impact in comparison to UBAR. Finally, we can see that the communication time is the bottleneck in this setting as when the BW increases from $10$ Mb/s to $100$ Mb/s the training time decreases significantly.  
   
 \subsection{Performance of \texttt{Basil}  for Non-IID Data Distribution using CIFAR100 }
 
To demonstrate the practicality of ACDS  and simulate the scenario where each node   shares data only from  its non-sensitive dataset portion, we  have considered the following experiment. 

\underline{Dataset and hyperparameters. }  We run image classification task on CIFAR100 dataset \cite{krizhevsky2009learning}.  This dataset is similar to CIFAR10 in having the same dimension ($32\times32\times 3$), except it has $100$ classes containing $600$ images each, where each class has its own feature distribution. There are $500$ training images and $100$ testing images per class. The $100$ classes in the CIFAR100 are grouped into $20$ superclasses.  For instance, the superclass Fish includes these five subclasses; 	Aquarium fish, Flatfish, Ray, Shark and Trout. In this experiment, we consider a system of  a total of $100$ nodes, in which $80$ are benign. We have set the connectivity parameter of \texttt{Basil}    to   $S = 5$. We use a decreasing learning rate of $0.03/(1+0.03\,k)$, where $k$  denotes the training round.  For the classification task, we  only consider the    superclasses as the target labels, i.e., we have $20$ labels for our classification task.

\underline {Model architecture. } We use  the same  neural network  that is used for CIFAR10 in the main paper which consists of  $2$ convolutional layers and $3$ fully connected layers, with the modification that the output of the last layer has a dimension of $20$.  The details for this network are included in Appendix H-A. 

\underline{Data distribution. } For simulating the  non-IID  setting,  we first shuffle the data within each superclass,  then partition 
each superclass into $5$ portions, and assign each node one partition. Hence, each node will have only data from one  superclass and includes data from each of the corresponding $5$ subclasses. 
 
\underline{Data sharing. } To simulate the case where  each node  shares data  from the non-sensitive portion of its local data,  we take the advantage of the variation of the  feature distribution across subclasses and simulate the  sensitive and non-sensitive data as per subclasses.  Towards achieving this partition goal,  we define  $\gamma \in (0,1)$ to represent the fraction  of subclasses  that a  node  considers their data    as non-sensitive  out of its available  $5$ subclasses.  For instance, $\gamma  = 1$  implies that all the $5$ subclasses are considered non-sensitive and  nodes can share data from them (e.g., nodes can share data from their entire local data). On the other hand,  $\gamma  = 0.4$  means that all nodes only consider the first two subclasses of their data as  non-sensitive and only share data from them.  We note that for nodes that share the same superclass, we consider  the order of the subclasses are  the same among them (e.g., if node  1 and node  2 have  data   from the same superclass Fish, hence  the  subclass Aquarium fish is the first subclass at both of them). This ensures that for $\gamma  = 0.4$ data in  $3$ subclasses per each superclass will never be shared by any user.  Finally, we allow each node to share $\alpha D$ data points, where $D = 500$ is the local data set size at each node, from the $\gamma$ subclasses, and $\alpha = 5 \%$.

  \begin{figure}
          \centering
          \includegraphics [width=0.3\textwidth]{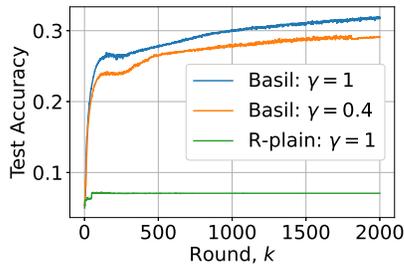}
         \caption{The performance of \texttt{Basil} under different  data sharing scenario in the presence of the Gaussian attack when the data distribution at the nodes is non-IID. Here, $\gamma$  represents the fraction  of subclasses  that   nodes  consider  the  data  from them  as non-sensitive  out of its available  $5$ subclasses.  } 
         \label{fig:partitiong}
      \end{figure}

\underline{Results} Fig. \ref{fig:partitiong} shows the performance of \texttt{Basil}  in the presence of the  Gaussian attack under different $\gamma$ values. As we can see that even when each node shares data from only two subclasses ($\gamma = 0.4)$ out of the five available subclasses, \texttt{Basil}  is giving a comparable performance to the case where each node shares data from its entire dataset ($\gamma = 1$). The intuition behinds getting a good performance in the presence of Byzantine nodes is  the fact that although data from the  three subclasses in each superclass  is never shared, there are nodes in the system originally that have data from these sensitive classes, and when they train the model on their local dataset, the augmented side information from the shared dataset helps to maintain the model performance and resiliency to Byzantine faults. Furthermore, we can see that R-plain fails in the presence of the attack, demonstrating that the ACDS data sharing is quite crucial for good performance when data is non-IID.   

\subsection{Performance Comparison Between Basil and Basil+}\label{comaprsion between basil an}

We compare the performance of \texttt{Basil}  and \texttt{Basil}+ under the following setting:
 \begin{figure}
  \centering
  \subfigure[Reporting the  average test accuracy among the benign nodes in each round.]{\includegraphics[scale=0.38]{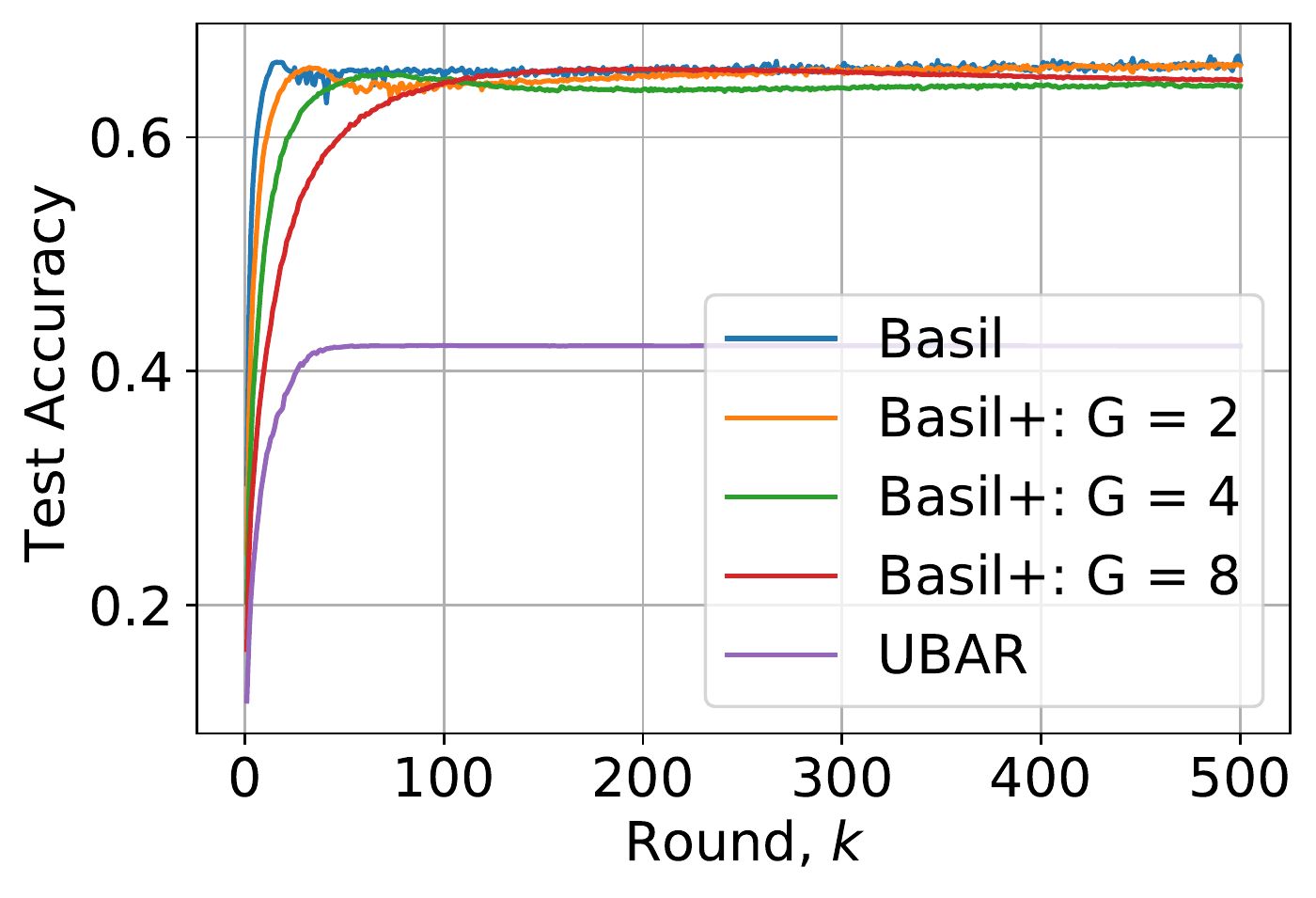}
} \quad  \quad \quad
  \subfigure[Reporting the  worst case test accuracy among the benign nodes in each round.
]{\includegraphics[scale=0.38]{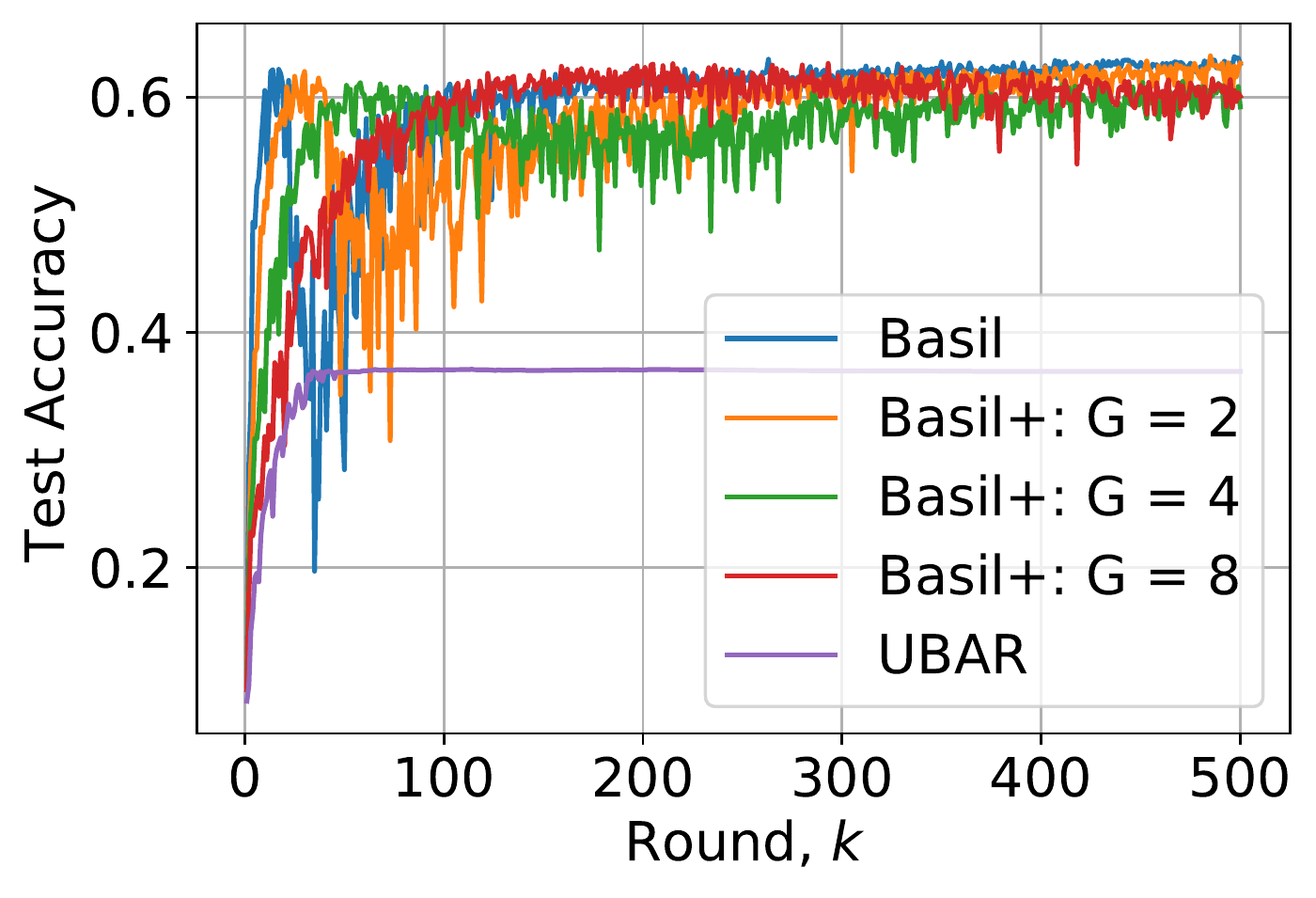}
  }%\quad
  \caption{Illustrating the performance of \texttt{Basil}  and \texttt{Basil}+  for CIFAR10 dataset.}
  \label{fig:Bsail and Basil+}
 \end{figure}
 
\underline{Setting.}
We consider a setting of $400$ nodes in which $80$ of them are Byzantine nodes. For the dataset, we use CIFAR10 dataset, while considering inverse attack for the Byzantine nodes. We use a connectivity parameter of $S = 6$ for both \texttt{Basil} and \texttt{Basil}+, and consider epoch based local training, where
we set the number of epochs to $3$.

\underline{Results:}
As we can see from Fig. \ref{fig:Bsail and Basil+}, \texttt{Basil} and \texttt{Basil}+  for different groups  retain  high test accuracy over UBAR in the  presence of the inverse attack. We can also see from Fig. \ref{fig:Bsail and Basil+}-(b) that when we have a large number of nodes in the system  increasing the number of groups (e.g., $G = 8$) makes the worst case training performance of the benign nodes  more stable  across  the training rounds compared to \texttt{Basil} that has high fluctuation for large setting (e.g., $400$ nodes).

 \end{appendices}

\end{document}